\documentclass[fleqn,usenatbib]{mnras}

\usepackage[T1]{fontenc}

\DeclareRobustCommand{\VAN}[3]{#2}
\let\VANthebibliography\thebibliography
\def\thebibliography{\DeclareRobustCommand{\VAN}[3]{##3}\VANthebibliography}

\usepackage{graphicx}	
\usepackage{amsmath}	
\usepackage{comment}
\usepackage{txfonts}
\usepackage{hyperref}
\usepackage{textgreek}
\usepackage{xargs}
\usepackage{xspace}
\usepackage{booktabs} 
\usepackage[dvipsnames,hyperref]{xcolor}
\usepackage[normalem]{ulem}

\usepackage{hyperref}
\hypersetup{
    colorlinks=true,
    linkcolor=Cerulean,
    citecolor=NavyBlue,
    filecolor=magenta,
    urlcolor=Violet,
    pdftitle= {BLR_profiles},
    pdfpagemode=FullScreen,
    }

\newcommand{\jwst}{\textit{JWST}\xspace}

\newcommand{\Halpha}{\text{H\textalpha}\xspace}

\newcommand{\Hbeta}{\text{H\textbeta}\xspace}
\newcommand{\Hgamma}{\text{H\textgamma}\xspace}

\newcommandx{\permittedEL}[6][1=O,2=III,3=,4=,5=,6=]{\text{{#1}\,{\sc {#2}}{#3}{#4}{#5}{#6}}\xspace}
\newcommandx{\semiforbiddenEL}[6][1=O,2=III,3=,4=,5=,6=]{\text{{#1}\,{\sc {#2}}]{#3}{#4}{#5}{#6}}\xspace}
\newcommandx{\forbiddenEL}[6][1=O,2=III,3=,4=,5=,6=]{\text{[{#1}\,{\sc{#2}}]{#3}{#4}{#5}{#6}}\xspace}
\newcommand{\kms}{km s$^{-1}$\xspace}

\newcommand{\HeIIL}[1][1=1640]{\permittedEL[He][ii][\textlambda][#1]}

\newcommandx{\OIIL}[1][1=3728]{\forbiddenEL[O][ii][\textlambda][#1]}
\newcommand{\OIIall}{\forbiddenEL[O][ii][\textlambda][\textlambda][3727,][30]}
\newcommand{\OIII}{\forbiddenEL[O][iii]}
\newcommandx{\OIIIL}[1][1=5008]{\forbiddenEL[O][iii][\textlambda][#1]}
\newcommand{\OIIIall}{\forbiddenEL[O][iii][\textlambda][\textlambda][5008,][4960]}

\newcommandx{\NIIL}[1][1=6585]{\forbiddenEL[N][ii][\textlambda][#1]}
\newcommand{\NIIall}{\forbiddenEL[N][ii][\textlambda][\textlambda][6550,][85]}

\newcommandx{\NeIIIL}[1][1=3870]{\forbiddenEL[Ne][iii][\textlambda][#1]}

\newcommand{\MBH}{M$_{BH}$\xspace}

\newcommand{\escat}{electron scattering \xspace}
\newcommand{\escats}{e$^{-}$-scattering \xspace}

\newcommand{\MSA}{R1000 - MSA\xspace}

\defcitealias{Rusakov_nature_2026}{R26}

\title[Line profiles in LRDs and LBDs]{
Little Red and Blue Dots: simply stratified Broad Line Regions.}
\author[Jan Scholtz]{\parbox[h]{\textwidth}{
J. Scholtz$^{1,2}$\thanks{E-mail: js2685@cam.ac.uk},
F. D'Eugenio$^{1,2}$,
R. Maiolino$^{1,2,3}$,
M. Brazzini$^{4,5}$,
H. \"{U}bler$^{6}$,
X. Ji$^{1,2}$,
M. Perna$^{7}$,
F. Sun$^{8}$,
G. Brocchi$^{9}$,
S. Carniani$^{10}$,
G. Cresci$^{11,12}$,
A. Harshan$^{1,2}$,
L. R. Ivey$^{1,2}$,
I. Juodžbalis$^{1,2}$,
A. Marconi$^{11,12}$,
G. Mazzolari$^{6}$,
G. Risaliti$^{11,12}$,
B. Trefoloni$^{10,12}$
}\vspace{0.4cm}
\\
$^{1}$Kavli Institute for Cosmology, University of Cambridge, Madingley Road, Cambridge, 
CB3 0HA, UK\\
$^{2}$Cavendish Laboratory, University of Cambridge, 19 JJ Thomson Avenue, Cambridge CB3 0HE, UK\\
$^{3}$Department of Physics and Astronomy, University College London, Gower Street, London WC1E 6BT, UK\\
$^{4}$Department of Physics, Astronomy Section, University of Trieste, Via G.B. Tiepolo, 11, I-34143 Trieste, Italy\\
$^{5}$INAF - Osservatorio Astronomico di Trieste, Via G. B. Tiepolo 11, I-34143 Trieste, Italy\\
$^{6}$Max-Planck-Institut f\"ur extraterrestrische Physik, Gie{\ss}enbachstra{\ss}e 1, 85748 Garching, Germany \\ 
$^{7}$Centro de Astrobiolog\'ia (CAB), CSIC--INTA, Cra. de Ajalvir Km.~4, 28850 -- Torrej\'on de Ardoz, Madrid, Spain\\
$^{8}$Center for Astrophysics $|$ Harvard \& Smithsonian, 60 Garden St., Cambridge, MA 02138, USA\\
$^{9}$Háskóli Íslands, Tæknigarður, Dunhagi 5, 107 Reykjavík, Iceland\\
$^{10}$Scuola Normale Superiore, Piazza dei Cavalieri 7, I-56126 Pisa, Italy\\
$^{11}$Dipartimento di Fisica e Astronomia, Università degli Studi di Firenze, Via G. Sansone 1,I-50019, Sesto Fiorentino, Firenze, Italy\\
$^{12}$INAF - Osservatorio Astrofisico di Arcetri, Largo E. Fermi 5, I-50125, Firenze, Italy\\
}

\date{Accepted XXX. Received YYY; in original form ZZZ}

\pubyear{2025}

\begin{document}
\label{firstpage}
\pagerange{\pageref{firstpage}--\pageref{lastpage}}
\maketitle

\begin{abstract}
It has been claimed that a fraction of the so-called Little Red Dots (LRDs) are characterised by exponential broad line profiles, which have been ascribed to broadening from electron scattering by an ionised cocoon.
In this work, we investigate the \Halpha broad line profiles of 32 AGN, including Little Red Dots (LRDs), Little Blue Dots (LBDs), and X-ray detected sources, using high SNR and resolution spectroscopy. We find that while single Gaussian models are statistically rejected, the exponential model is not universally preferred. Lorentzian and multi-Gaussian profiles provide equally good or superior fits for the majority of the sample, with no statistical preference for exponential profiles in $\sim$60\% of cases across all AGN subtypes. There are indications that exponential profiles are preferred more frequently among LBDs, indicating that exponential profiles are not a prerogative of LRDs, which actually seem to more often favour Lorentzian profiles. Furthermore, we demonstrate that exponential wings can emerge naturally from the stratification of BLR clouds in virial motion, without invoking any scattering process. More generally, we also show that stacking multiple broad lines (either from multiple objects, as done in previous works, or from different BLR components within the same object) generally yields an exponential profile, even if none of the individual profiles are exponential.
Explaining the exponential profiles in terms of BLR stratification solves various observational tensions with the electron scattering interpretation. While electron scattering may play a role, there is no evidence that it dominates the line profiles and that it significantly affects the inferred black hole masses.
\end{abstract}




\section{Introduction} 

Thanks to the James Webb Space Telescope (\jwst), we are now able to study active galactic nuclei (AGN) at high redshifts (z$>$4) with bolometric luminosities much lower than the population of previously known quasars, i.e. in the range $\log L_{\mathrm{bol}} [\mathrm{erg\,s^{-1}}]\sim 42 - 46$ \citep[e.g.,][]{harikane_jwstnirspec_2023, adamo_first_2025, juodzbalis_jades_2025, mazzolari_scholtz_narrow-line_2025}. Unexpectedly, the newly discovered \jwst population of AGN is neither a scaled-down version of the luminous quasar population previously detected at high-$z$, nor a typical AGN at lower redshifts.

The high-quality rest-frame optical and UV spectra from the NIRSpec instrument \citep{jakobsen_near-infrared_2022} have revealed a different population of AGN with the following characteristics: i) absence of high-ionisation nebular emission lines \citep{harikane_jwstnirspec_2023, ubler_ga-nifs_2023, maiolino_jades_2024, juodzbalis_jades_2025, zucchi_black_2025}; ii) lack of the optical Fe\,{\sc ii} bump \citep{trefoloni_missing_2025}, iii) weak variability \citep[e.g.][]{kokubo_challenging_2025, ji_blackthunder_2025, furtak_investigating_2025, naidu_black_2025, zhang_var_2025}; iv) X-ray weakness \citep{lambrides_case_2024, yue_stacking_2024, maiolino_jwst_2025, ananna_x-ray_2024}; v) radio weakness \citep{mazzolari_radio_2024, mazzolari_scholtz_narrow-line_2025} and vi) overmassive black holes \citep[e.g. ][]{harikane_jwstnirspec_2023, juodzbalis_dormant_2024, juodzbalis_jades_2026}. As a result, there has been a proliferation of new, increasingly complex models invoking high gas and dust obscuration \citep{inayoshi_extremely_2025,juodzbalis_jades_2024, ji_blackthunder_2025, Madau2026}, often super-Eddington accretion \citep{pacucci_mildly_2024,madau_x-ray_2024,king_joining_2025}, and dense-gas cocoons fully enshrouding a rapidly accreting black hole, `quasi-stars' \citep{begelman_little_2026} and `black-hole stars' \citep{naidu_black_2025,de_graaff_little_2025}.

The type 1 population of \jwst-discovered AGN (i.e. those with broad Balmer lines) can be broadly split into two subgroups: Little Blue Dots (LBDs) and Little Red Dots (LRDs), which share the majority of the telltale signs of \jwst AGN described above. 
However, LRDs have a characteristic v-shaped continuum \citep{matthee_little_2024, kocevski_spectroscopic_2024}, pivoting at the Balmer limit \citep{setton_little_2025} or close to it \citep{de_graaff_little_2025}, with a compact source at optical wavelengths.
In addition to broad Balmer lines, LRDs often also exhibit Balmer absorption in their broad line profile \citep{juodzbalis_jades_2024,matthee_little_2023, deugenio_irony_2025,deugenio_jades_2026,deugenio_blackthunder_2025,lin_discovery_2025}, which, together with the smooth Balmer breaks described above, have been interpreted as absorption by dense circumnuclear gas along the line of sight \citep{juodzbalis_jades_2024,inayoshi_extremely_2024,
inayoshi_weakness_2025,ji_blackthunder_2025,naidu_black_2025}.

However, LRDs constitute less than 30\% of the AGN identified by JWST at redshift $z\sim5$ and at bolometric luminosities below the quasar regime \citep{hainline_investigation_2025,kocevski_rise_2025,taylor_broad-line_2025}, while, as mentioned above, the rest of the spectroscopically confirmed type-1 AGN show typical blue optical and UV colours (LBDs), similar to those of standard type-1 AGN and star-forming galaxies. Similarly to LRDs, LBDs are also generally compact \citep[e.g.][who find that 60\% of JWST-discovered broad-line AGN are unresolved in NIRCam]{hainline_investigation_2025,juodzbalis_jades_2025}. It is worth stressing that, similarly to LRDs, the presence of a dominant point source does not rule out the existence of an extended photometric component, e.g.\ a host galaxy \citep{chen_host_2025, rinaldi_not_2025, baggen_2026}. 

The high sensitivity of \jwst/NIRSpec instrument has allowed a detailed investigation of the broad line profiles in the Hydrogen and Helium lines \citep[e.g. ][]{Rusakov_nature_2026, brazzini_ruling_2025, deugenio_irony_2025,deugenio_jades_2026,deugenio_blackthunder_2025, chang_impact_2025}, finding that the profiles are generally non-Gaussian, with significant wings, often exponential \citep{Rusakov_nature_2026, torralba_warm_2025}. The exponential wings have been interpreted as evidence of electron scattering, first shownby \citet{laor_evidence_2006} for a local AGN and then applied by \citet[][hereafter: \citetalias{Rusakov_nature_2026}]{Rusakov_nature_2026} for a sample of high-z AGN found by JWST. According to this scenario, a large column of hot (Te $\sim$ several to tens of thousands K), free electrons exists in the BLR \citep{laor_evidence_2006} or a hypothesized ``cocoon'' \citep[e.g. ][]{de_graaff_little_2025,sneppen_inside_2026, Matthee_2026}, enabling Thomson scattering which significantly broadens the emission-line profiles.  We note that \citet{laor_evidence_2006} assumes optically-thin electron scattering and it is not sufficient to shape entirely the broad line region profile. On the other hand, the models of \citetalias{Rusakov_nature_2026} and \citet{Matthee_2026} obtain an optically thick result where the line profile is dominated by scattering. This model implies that the intrinsic widths of the broad lines could be 5--10 times narrower than currently measured, potentially leading to systematic overestimates of black hole masses by up to two orders of magnitude, when using standard virial relations (since $M_{\rm BH;~virial}\propto {\rm FHWM_{broad~line}^2}$). However, the electron scattering scenario has been recently challenged by a detailed analysis of multiple hydrogen and helium emission lines in two sources with deep observations \citep[28074 and GS-3073;][]{brazzini_ruling_2025, Brazzini_rosettas_2026}. Additionally, \citet{juodzbalis_direct_2025} obtained a direct, kinematic measurement of the black hole mass in the prototypical LRD Abell2744-QSO1 at z=7.04 with NIRSpec/IFU observations, showing that the independently measured kinematic mass agrees with the single-epoch virial measurement \citep{deugenio_blackthunder_2025}, while two orders of magnitude larger than what is inferred from the electron scattering scenario. Similarly, for the prototypical local example of exponential wings, NGC4395 \citep{laor_evidence_2006}, the black hole mass inferred from the virial relations \citep{Lira1999} matches very well the mass inferred from reverberation mapping \citep{Peterson_2005} and from direct dynamical measurements \citep{denbrok_2015}.

However, in addition to the case of NGC4395, non-Gaussian BLR profiles have been observed in AGN and quasars well before the launch of \jwst. Studies have utilised exponential, double-Gaussian, Lorentzian, or broken power-law models to model \Halpha and \Hbeta BLR profiles \citep{Netzer_1990, nagao_gas_2006, cano-diaz_observational_2012, kollatschny_shape_2013, scholtz_impact_2021, santos_spectroscopic_2025, laor_evidence_2006, kollatschny_2018}, while a single Gaussian model was often used for the BLR profile due to the limited SNR of ground-based near-infrared observations \citep[e.g.][]{scholtz_kashz_2020, kakkad_super_2020}.

Previous work have investigated single objects \citep{brazzini_ruling_2025, deugenio_irony_2025,deugenio_jades_2026,deugenio_blackthunder_2025, chang_impact_2025, torralba_weak_2025} or a small sample \citep{Rusakov_nature_2026}. In this work, we build on this by compiling a large sample of 32 high SNR \Halpha type-1 AGN, carefully distinguishing between Little Blue and Red dots as well as more usual X-ray AGN. Furthermore, we investigate the interpretation of exponential profiles and their effect on the black hole mass (\MBH) measurement. 

In \S~\ref{s.obs} we present our sample selection and data reduction, in \S~\ref{s.eml_fit} we describe our emission line fitting, in \S~\ref{s.fits} we present our results, and \S~\ref{s.discuss} we discuss the implications of the line profiles. Finally, in \S~\ref{s.conclusion} we summarise our results.
Throughout this work, we adopt a flat $\Lambda$CDM cosmology: H$_0$= 67.4 km s$^{-1}$ Mpc$^{-1}$, $\Omega_\mathrm{m}$ = 0.315, and $\Omega_\Lambda$ = 0.685 \citep{2020A&A...641A...6P}. We use vacuum wavelengths for the emission lines throughout the paper.

\section{Sample selection and data reduction}
\label{s.obs}

\subsection{Sample selection}\label{s.sample_selection}

The aim of this work is to investigate the \Halpha emission line of Little Blue Dots (LBDs), Little Red Dots (LRDs) and typical AGN (see \S~\ref{s.classification}). To this end, we compiled a sample of \Halpha spectra observed with NIRSpec at medium (R$\sim$1000) or high (R$\sim$2700) spectral resolution. In order to distinguish between the models, we require a high SNR for the \Halpha line (\Halpha SNR$>$20) and a detected BLR in the \Halpha. For such, we compile targets from the following source catalogues:

\begin{enumerate}
    \item LBDs, LRDs and an X-ray source from \citet{Rusakov_nature_2026} compiled by searching for high-SNR spectra in the DJA archive. In this sample, we also include the high-SNR observation of the Irony LRD \citet[][and references therein]{deugenio_irony_2025}. We note that, while \citet{Rusakov_nature_2026} makes claims about the association of LRDs with exponential profiles, only half of their sample is made of LRDs, the rest are LBDs and X-ray AGN.
    \item LBDs and LRDs from the NIRSpec/IFS programmes BlackThunder (PIs: H. Ubler \& R. Maiolino, PID: 5015) and GO: 5664 programme (PI J. Matthee), including Abell2744-QSO1 \citep[QSO1;][]{furtak_high_2024,ji_blackthunder_2025,deugenio_blackthunder_2025}, ID 159717 \citep{deugenio_jades_2025-2} and GN-9771 \citep{torralba_warm_2025}. Full sample from GO 5664 has also been presented in \citet{Matthee_2026}.
    \item NIRSpec/IFS observations of the extreme LRD, Cliff \citep{de_graaff_remarkable_2025}, from DDT programme 9433 (PIs. Maiolino \& D'Eugenio; see Ivey et al. 2026).  
    \item X-ray sources detected with NIRSpec from (Yixiao Liu et al. in prep), compiled by cross-matching the X-ray catalogues \citep{luo_chandra_2017, xue_2_2016, civano_chandra_2016} with the DJA NIRSpec archive. The NIRSpec-MSA data for these objects were observed as part of WIDE survey \citep{maseda_nirspec_2024} and BlueJay \citep{belli_star_2024}.
    \item X-ray AGN XID2028 from the NIRSpec/IFS ERS Q3D programme (PI: Wylezalek, PID: 1335) from \citet{cresci_bubbles_2023, veilleux_q3d_2023}.
    \item LRD (28074) and LBD (GS-3073) ``Rosetta stones'' from \citet{Brazzini_rosettas_2026}, originally published in \citet{juodzbalis_jades_2024} and \citet{ubler_ga-nifs_2023}, respectively. 
    \item We also added the extreme LRD, ``Uncover Monster'' (Uncover Abell2744-45924), with NIRCam/Slitless spectroscopy \citep[PI: Naidu, Matthee; PID: 3516;][]{torralba_weak_2025} and NIRSpec/PRISM observations from \citet{greene_uncover_2024}.
\end{enumerate}

\begin{figure*}
    \centering
    \includegraphics[width=0.42\paperwidth]{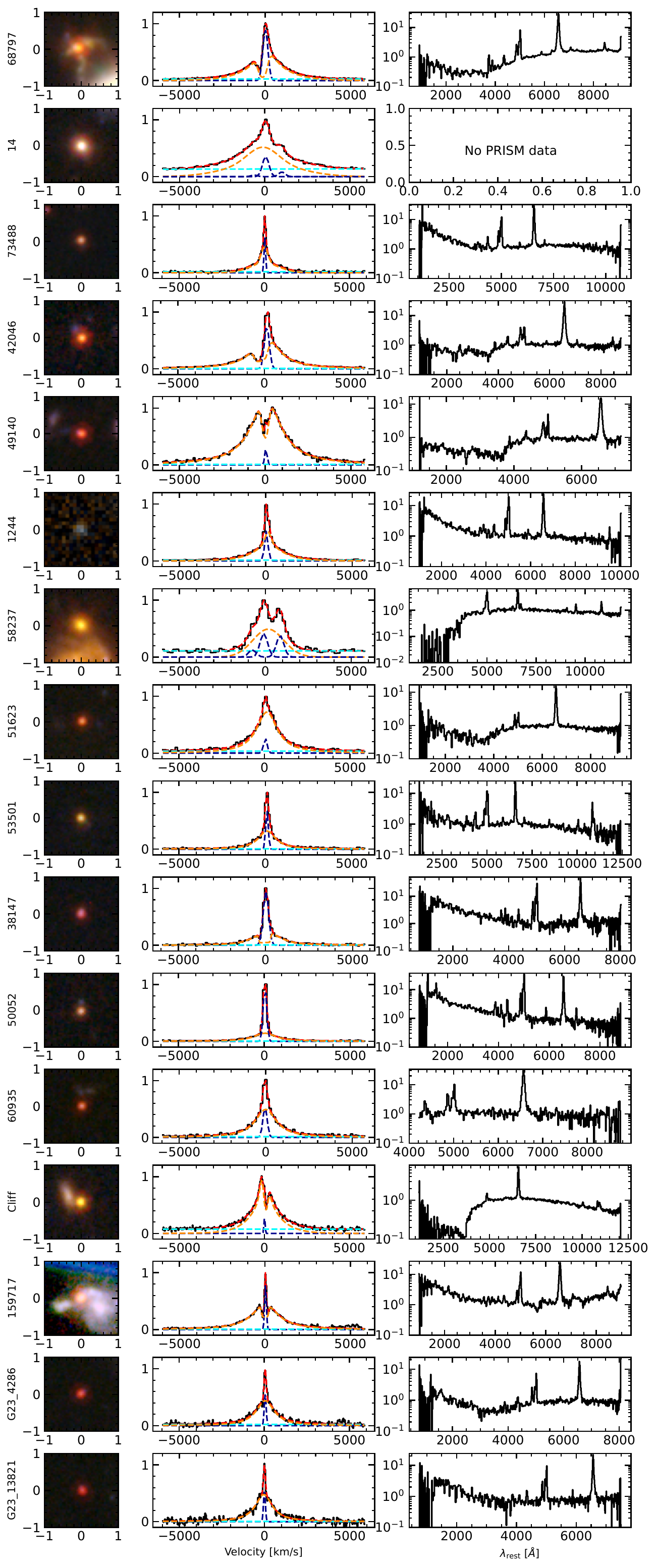}
    \includegraphics[width=0.42\paperwidth]{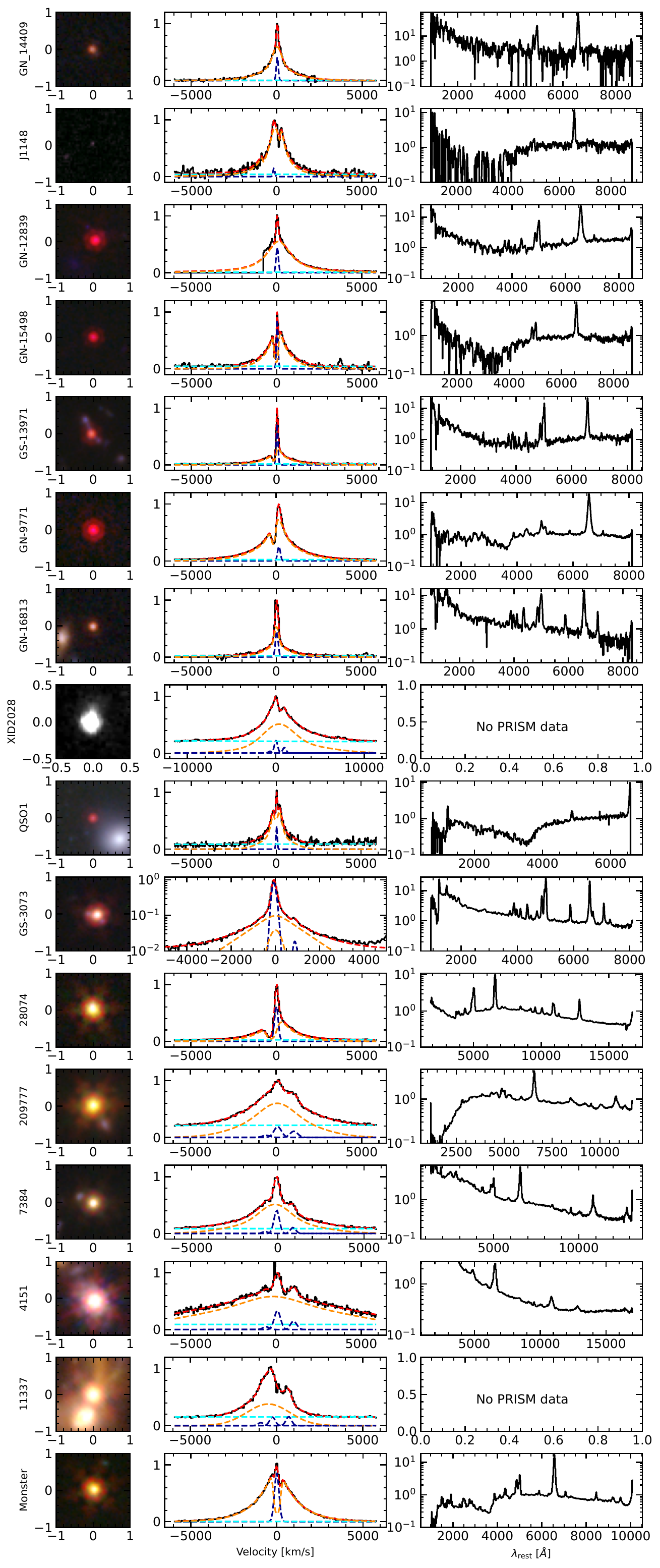}
    \caption{Overview of the sample used in this work. Left: \jwst/NIRCam RGB image (R -- F444W, G -- F277W, B -- F090W); the stamp size is $1\times1$ arcseconds. For XID2028 we used an HST I-band image as no NIRCam imaging is available, while for 1244, we used \textit{HST} I, J and H-band images. Middle panel: Higher- spectral-resolution spectrum of \Halpha. The best fit is shown as a red dashed line, the BLR model as an orange dashed line, and the narrow component (\Halpha or \NIIall) as a blue dashed line. Right panel: The PRISM spectrum when available.
    }
    \label{fig.sample}
\end{figure*}

We show the full overview of the sources with NIRCam images, \Halpha emission lines and PRISM spectra when available in Fig.~\ref{fig.sample}, and we present basic properties of the sample in Table~\ref{tab.sample}.

\begin{table*}
    \caption{List of targets in our sample with basic properties: ID, ID in other works, Survey, coordinates, redshift, presence of Absorption, AGN type and UV and optical slopes.}
    \centering
    \begin{tabular}{@{}lcccccccccc@{}} 
\hline 
\hline 
ID & Other Names & Survey & RA & Dec & z  & Absorber & AGN type & Resolution & $\beta_{opt}$ & $\beta_{UV}$ \\
    &      &  & deg& deg &    &          &     &            &                  & \\
\hline 
68797 & A$^{*}$ & JADES & 189.2291 & 62.1462 & 5.04 & Yes & LRD  & R2700  & 2.7$\pm$0.2 & -0.6 $\pm$0.1\\
14 & B$^{*}$ & 2674 & 189.1998 & 62.1615 & 5.18 & No & X-ray  & R1000  & -1.3$\pm$0.2 & -1.9$\pm$0.1 \\
73488 & C$^{*}$ & JADES & 189.1974 & 62.1772 & 4.13 & No & LRD  & R2700  & 0.7$\pm$0.1 & -1.7 $\pm$0.1\\
42046 & D$^{*}$ & Rubies & 214.7954 & 52.7888 & 5.28 & Yes & LRD  & R1000  & 0.7$\pm$0.2 & -0.3 $\pm$0.1\\
49140 & E$^{*}$ & Rubies & 214.8922 & 52.8774 & 6.68 & Yes & LRD  & R1000  & 0.7$\pm$0.1 & -0.7 $\pm$0.1\\
1244 & F$^{*}$ & CEERS & 215.2407 & 53.0360 & 4.48 & No & LBD  & R1000  & -0.1$\pm$0.1 & -1.8 $\pm$0.1\\
58237 & G$^{*}$ & Rubies & 214.8506 & 52.8660 & 3.65 & No & LBD  & R1000  & 1.0$\pm$0.1 & 1.8 $\pm$1.3\\
51623 & H$^{*}$ & 4106 & 214.8868 & 52.8554 & 4.95 & No & LRD  & R1000  & 1.0$\pm$0.1 & -1.2 $\pm$0.1\\
53501 & I$^{*}$ & JADES & 189.2951 & 62.1936 & 3.43 & No & LRD  & R1000  & 0.6$\pm$0.2 & -1.3 $\pm$0.1\\
38147 & J$^{*}$ & JADES & 189.2707 & 62.1484 & 5.87 & Yes & LRD  & R1000  & 0.4$\pm$0.2 & -1.5 $\pm$0.1\\
50052 & K$^{*}$ & Rubies & 214.8235 & 52.8303 & 5.24 & No & LBD  & R1000  & -0.8$\pm$0.1 & -1.7 $\pm$0.1\\
60935 & L$^{*}$ & Rubies & 214.9234 & 52.9256 & 5.29 & No & LBD  & R1000  & 0.2$\pm$0.2 & 0.1 $\pm$0.3\\
The Cliff & -- & DDT & 34.4107 & -5.1297 & 3.55 & Yes & LRD  & R2700  & 1.0$\pm$0.1 & -0.9 $\pm$0.3\\
159717 & -- & BlackThunder & 53.0975 & -27.9013 & 5.08 & Yes & LRD  & R2700  & 0.4$\pm$0.1 & -1.5 $\pm$0.1\\
G23\_4286 & -- & BlackThunder & 3.6192 & -30.4233 & 5.83 & No & LRD  & R2700  & 1.1$\pm$0.1 & -1.5 $\pm$0.1\\
G23\_13821 & -- & BlackThunder & 3.6206 & -30.4000 & 6.35 & No & LRD  & R2700  & 0.8$\pm$0.1 & -1.7 $\pm$0.1\\
GN\_14409 & -- & BlackThunder & 189.0721 & 62.2734 & 5.15 & No & LBD  & R2700  & -0.4$\pm$0.2 & -2.4 $\pm$0.1\\
J1148-18404  & -- & BlackThunder & 177.0580 & 52.8628 & 5.01 & Yes & LRD  & R2700  & 1.4$\pm$0.1 & -2.6 $\pm$0.4\\
GN-12839 & -- & Matthee-IFS & -170.6552 & 62.2631 & 5.24 & No & LRD  & R2700  & 1.9$\pm$0.2 & -1.6 $\pm$0.1\\
GN-15498 & -- & Matthee-IFS & -170.7145 & 62.2808 & 5.08 & Yes & LRD  & R2700  & 1.2$\pm$0.1 & -2.0 $\pm$0.1\\
GS-13971 & -- & Matthee-IFS & 53.1386 & -27.7903 & 5.48 & Yes & LRD  & R2700  & 1.4$\pm$0.1 & -1.9 $\pm$0.1\\
GN-9771 & -- & Matthee-IFS & -170.7190 & 62.2473 & 5.53 & Yes & LRD  & R2700  & 0.7$\pm$0.1 & -0.8 $\pm$0.1\\
GN-16813 & -- & Matthee-IFS & -170.8207 & 62.2925 & 5.36 & No & LBD  & R2700  & -1.0$\pm$0.1 & -2.2 $\pm$0.1\\
XID2028 & -- & Q3D & 150.5470 & 1.6185 & 1.59 & No & X-ray  & R2700  & - & - \\
QSO1 & Abell2744-QSO1$^{+}$ & BlackThunder & 3.5835 & -30.3967 & 7.04 & Yes & LRD  & R2700  & 1.6$\pm$0.1 & -1.2 $\pm$0.2\\
GS-3073 & Blue Rosetta$^{\times}$ & GA-NIFS & 53.0789 & -27.8842 & 5.55 & No & LBD  & R2700  & -0.8$\pm$0.1 & -1.8 $\pm$0.1\\
28074 & Red Rosetta$\dagger$ & JADES & 189.0646 & 62.2738 & 2.26 & Yes & LRD  & R2700  & 1.0$\pm$0.3 & -1.5 $\pm$0.1\\
209777 & -- & WIDE & 53.1585 & -27.7740 & 3.71 & No & X-ray  & R1000  & -0.4$\pm$0.1 & 2.5 $\pm$0.1\\
7384 & -- & WIDE & 53.1785 & -27.7841 & 3.19 & No & X-ray  & R1000  & -1.3$\pm$0.2 & -0.9 $\pm$0.1\\
4151 & -- & WIDE & 189.3246 & 62.3155 & 2.24 & No & X-ray  & R2700  & -1.9$\pm$0.1 & -2.0 $\pm$0.1\\
11337 & -- & BlueJay & 150.1195 & 2.2958 & 2.10 & No & X-ray  & R1000  & - & - \\
Monster & Abell2744-45924$^{o}$ & ALT & 3.5848 & -30.3436 & 4.47 & Yes & LRD  & R1600  & 0.6$\pm$0.2 & -0.5 $\pm$0.1\\
\hline 
\end{tabular}  
    \par $^{*}$\citetalias{Rusakov_nature_2026}; $^{+}$ \citet{furtak_constraining_2022,ji_blackthunder_2025, deugenio_jades_2026}; $^{\times}$ \citet{Brazzini_rosettas_2026}, $\dagger$ \citet{juodzbalis_jades_2024, brazzini_ruling_2025}; $^{o}$\citet{greene_uncover_2024, torralba_weak_2025}.
    \label{tab.sample}
\end{table*}

\subsection{Identifying LRDs, LBDs and Typical AGN}\label{s.classification}

We split our sample into three separate classes Little Red Dots, Little Blue Dots and X-ray sources. In order to identify the X-ray AGN, we cross-matched our sample with the latest X-ray catalogues in the GOOD-S, COSMOS and UDS fields  \citep[][see more details in Liu et al., in prep.]{luo_chandra_2017, xue_2_2016, civano_chandra_2016}. For X-ray undetected, broad line sources 
we split our sample into Little Red Dots (LRDs) and Little Blue Dots (LBDs). To do this, we employ the selection criteria from \citet{de_graaff_little_2025} based on the optical and UV slopes. Specifically, we fitted a power-law model to the PRISM spectra (or NIRCam photometry when no PRISM observations are available) in the form of $f_{\lambda} = \alpha(\lambda/3650)$ in two rest-frame wavelength ranges: 1200--3600 $\text{\AA}$ and 3700--6600 $\text{\AA}$; masking a $\pm50\,\text{\AA}$ region around all significant bright emission lines (\OIIall, \Hgamma, \Hbeta, \OIIIall, \Halpha). We identified the LRDs and LBDs from the X-ray undetected sources in our sample based on their optical and UV spectral slopes. To identify an object as an LRD, we search for a V-shaped continuum using the following criteria:

\begin{enumerate}
    \item $\beta_{\rm opt}$ > 0 
    \item $\beta_{\rm UV}$ < -0.2
    \item $\beta_{\rm UV}$ - $\beta_{\rm opt}  <$ 0.5
\end{enumerate}

\begin{figure}
    \centering
    \includegraphics[width=0.99\columnwidth]{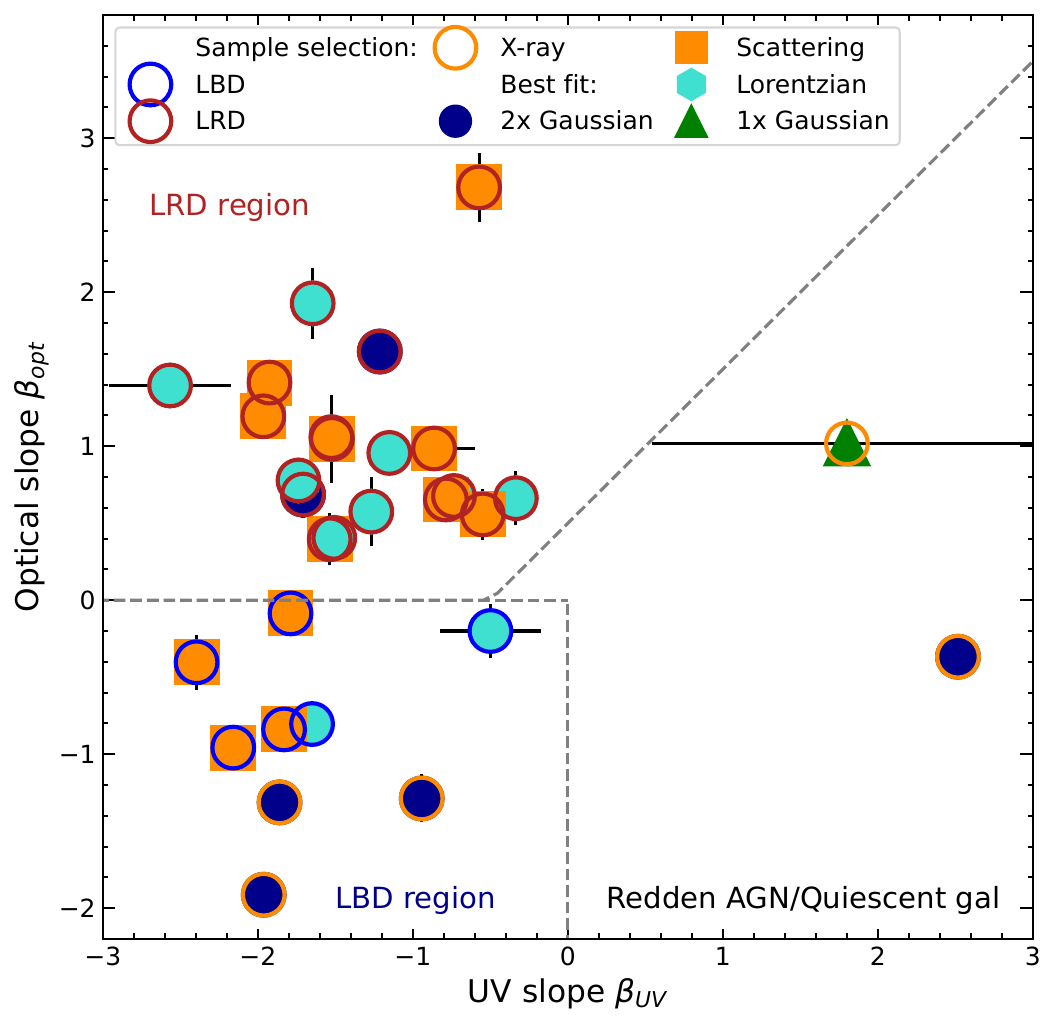}
    \caption{Plot of $\beta_{\rm opt}$ vs $\beta_{\rm UV}$ for selecting LRDs in the sample. The dashed lines show the different regions for selecting LRDs and LBDs. We highlight the sources selected as LRDs, LBDs and X-ray AGN as red, blue and orange circles, respectively. We also show the best fits for each object with different symbols.}
    \label{fig.beta}
\end{figure}

We show the plot of $\beta_{\rm opt}$ vs $\beta_{\rm UV}$ in Fig.~\ref{fig.beta}. 
The LBDs were selected as objects with blue $\beta_{\rm opt}$ and $\beta_{\rm UV}$ slopes without any detections in the X-rays. We note that our LRDs and LBDs are also satisfying the compactness criterion as described in \citet{Brazzini_rosettas_2026}.

Overall, in our sample, we identified seven typical AGN (22\%), 19 LRDs (60\%) and six LBDs (19\%). We note that our sample is not representative of the overall population of LBDs and LRDs, as LRDs are believed to constitute $\sim$30\% of the overall AGN population \citep[e.g.][]{hviding_rubies_2025, juodzbalis_jades_2026}. However, due to the relatively straightforward pre-selection of LRDs based on NIRCam imaging and the community's interest in them, it is possible that LRDs dominate NIRSpec/MSA observations. We note that our X-ray sample is biased towards sources with powerful outflows as they were either targetted NIRSpec programmes to study the large scale outflow properties in the X-ray AGN.

\subsection{NIRSpec-IFS Data}\label{s.IFS_obs}

The IFS data in this work are from the Large programme BlackThunder (PI: H. Ubler \& R. Maiolino),  GO: 5664 (PI: J. Matthee) and Q3D (PI: Wylezalek; PID 1315). While GO 5664 and BlackThunder utilise the G395H/F290LP and PRISM/CLEAR grating/filter combination, the Q3D observations use the G140H/F100LP grating/filter combination.

Raw data files of these observations were downloaded from the Barbara A.~Mikulski Archive for Space Telescopes (MAST) and then processed with the {\it JWST} Science Calibration pipeline\footnote{\url{https://jwst-pipeline.readthedocs.io/en/stable/jwst/introduction.html}} version 1.11.1 under the Calibration Reference Data System (CRDS) context jwst\_1149.pmap. We made several modifications to the default reduction steps to increase data quality, which are described in detail by \citet{perna_ga-nifs_2023} and are briefly summarised here. Count-rate frames were corrected for $1/f$ noise through a polynomial fit. Furthermore, we removed regions affected by failed open MSA shutters during calibration in Stage 2. We also removed regions with strong cosmic ray residuals in several exposures. Any remaining outliers were flagged in individual exposures using an algorithm similar to {\sc lacosmic} \citep{van_dokkum_cosmic-ray_2001}: we calculated the derivative of the count-rate maps along the dispersion direction, normalised them by the local flux (or by three times the rms noise, whichever was highest), and rejected the 95\textsuperscript{th} percentile of the resulting distribution \citep[see][for details]{deugenio_fast-rotator_2024}. The final cubes were combined using the `drizzle' method. The main analysis in this paper is based on the combined cube with a pixel scale of $0.05''$.

Prior to our analysis of the emission line cube of the PRISM or R2700 observations, we perform a number of data preparatory steps: 1) background subtraction; 2) masking of any outlier pixels that may have been missed by the pipeline; 3) flux uncertainty verification. For these tasks and the rest of the analysis, we use \texttt{QubeSpec} (see e.g. \citealt{scholtz_ga-nifs_2025}), an analysis pipeline written for NIRSpec/IFS data.

For both the R2700 and PRISM IFS observations, we need to subtract the strong background affecting our observations. We follow the procedure described in \citet{scholtz_ga-nifs_2025}. Briefly, we mask the location of the source based on its \Halpha emission ($2\sigma$ SNR contours), and we estimate the background using \texttt{astropy.photutils.background.Background2D} (2D background estimator). To reduce noise in the estimated background, we smoothed the background in spectral space using a median filter with a width of 25 channels. The final estimated background is subtracted from the flux data cube. We extracted the integrated spectrum of the AGN by summing pixels within a circular aperture of 0.2 arcsec and correcting for any aperture losses following the procedure described in \citet{Jones_2026}.

We additionally mask any major pixel outliers that were not flagged by the data reduction pipeline. To identify the residual outliers not flagged by the pipeline, we used the error extension of the data cube. We flagged any pixels whose error is 10$\times$ above the median error value of the cube. 

\citet{ubler_ga-nifs_2023, scholtz_ga-nifs_2025} reported that the uncertainties on the flux measurements in the \texttt{ERR} extension of the data cubes are underestimated compared to the noise estimated from the rms of the spectrum, calculated inside a spectral window free from emission lines. However, the error extension still carries information about the relative uncertainties between pixels and flags outliers. Therefore, when extracting each spectrum, we first retrieve the uncertainty from the error extension. We then scale it so that its median uncertainty matches the spectrum's sigma-clipped rms in emission-line-free regions. This scaling is performed across both detectors independently, without wavelength dependence.

\subsection{NIRSpec-MSA and NIRCam/Slitless data}\label{s.MSA_obs}

This study makes use of public JWST data collected as part of several observational programmes with the NIRSpec/MSA spectrograph with PIDs: 1345 (CEERS), 1180, 1181, 1210, 3215 (JADES), 4106 (PI: E. Nelson), 4233 (RUBIES), 2674 (PI A. Haro), WIDE survey (PID: 1211, 1212, 1214) and BlueJay survey (PID 1812). These observations have been uniformly reduced and published as part of the DAWN JWST Archive (https://dawn-cph.github.io/dja) (DJA) using the v4.4 data reduction. We verified that the data reduction does not influence our conclusions by performing the same analysis on JADES DR4 spectra \citep[]{scholtz_jades_2025} for sources from the JADES survey.

The Uncover Monster NIRCam WFSS data was reduced following the routine outlined by \citet{SunF_2023}, which includes a large number of customised steps and calibration files different from those used by the standard \jwst pipelines. For transparency and reproducibility, the code and calibration files are publicly available\footnote{\url{https://github.com/fengwusun/nircam_grism}.}.

\section{Emission line fitting}\label{s.eml_fit}

The aim of this work is to model the profile of the broad component of \Halpha. We model the broad \Halpha profiles using common models used in the literature:

\begin{enumerate}
    \item Single Gaussian profile -- The broad H$\alpha$ profile is modelled as a single broad Gaussian profile with its redshift, FWHM and flux as the free parameters, for a total of three free parameters. 
    \item Double Gaussian profile -- The broad H$\alpha$ profile is modelled as two separate Gaussian profiles, with each Gaussian profile having independent FWHM and flux but a tied redshift for a total of five free parameters. For XID2028, we allow a velocity shift of up to 500 \kms between the two Gaussian profiles. 
    This model can represent an extreme simplification of the BLR velocity structure.
    \item Lorentzian profile -- The broad H$\alpha$ is modelled as a Lorentzian profile with its redshift, FWHM and flux as the free parameters, for a total of three free parameters. Compared to the single Gaussian profile, the Lorentzian profile has additional flux in the wings of the emission line profile, often attributed to turbulence in the BLR or outflows.
    \item Electron scattering exponential profile -- A model comprising an intrinsic Gaussian profile (ascribed to the intrinsic BLR) with a fraction of it being scattered by electrons. The electron scattering contribution is modelled as Gaussian component convolved by a symmetric exponential  \citep{laor_evidence_2006} in the form: 
    \begin{equation}
        E(\lambda_0, W; \lambda) \propto e^{- \frac{|\lambda - \lambda_0 |}{W}},
    \end{equation}
    where $\lambda_0$ is the central wavelength (assumed to be the same as that of the Gaussian BLR), and $W$ is the exponential width, which is allowed to vary in the range 500--10{,}000 \kms. 
    The full profile for each broad line is therefore given by: 
    \begin{equation}
        f_\text{scatt} E(\lambda)*G_\text{BLR}(\lambda) + (1 - f_\text{scatt}) G_\text{BLR}(\lambda),
    \end{equation}
    where $f_\text{scatt}$ is the fraction of scattered light. $f_\text{scatt}$ can be related to the optical depth of the scattering medium as $(1-e^{-\tau_{\rm thom}})$ and the exponential decay scale is related to $\tau_{\rm thom}$, and the temperature of the scattering gas ($T$) as  $W=(428 \times \tau_{\rm thom} + 370.) \times (T/10^{4}K)^{0.5}$ \kms. 
\end{enumerate}

\begin{figure*}
    \centering
    \includegraphics[width=0.8\paperwidth]{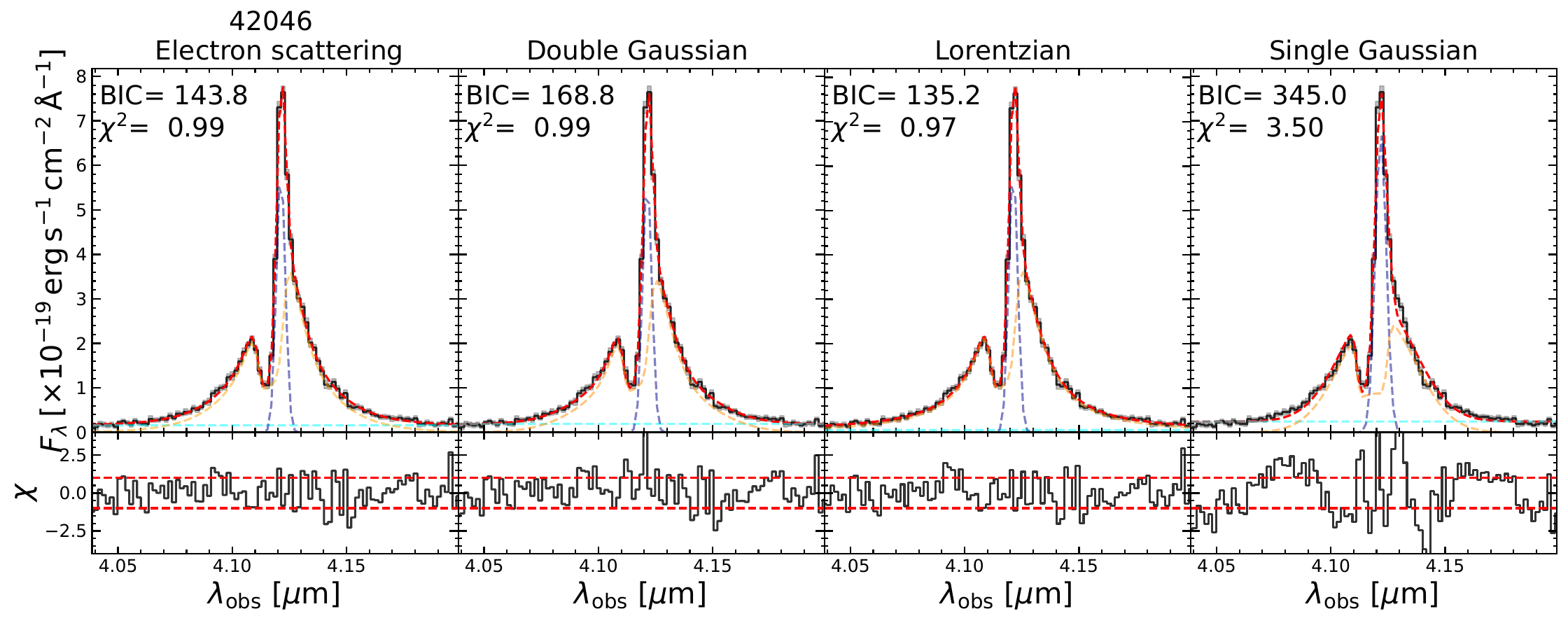}
    \includegraphics[width=0.8\paperwidth]{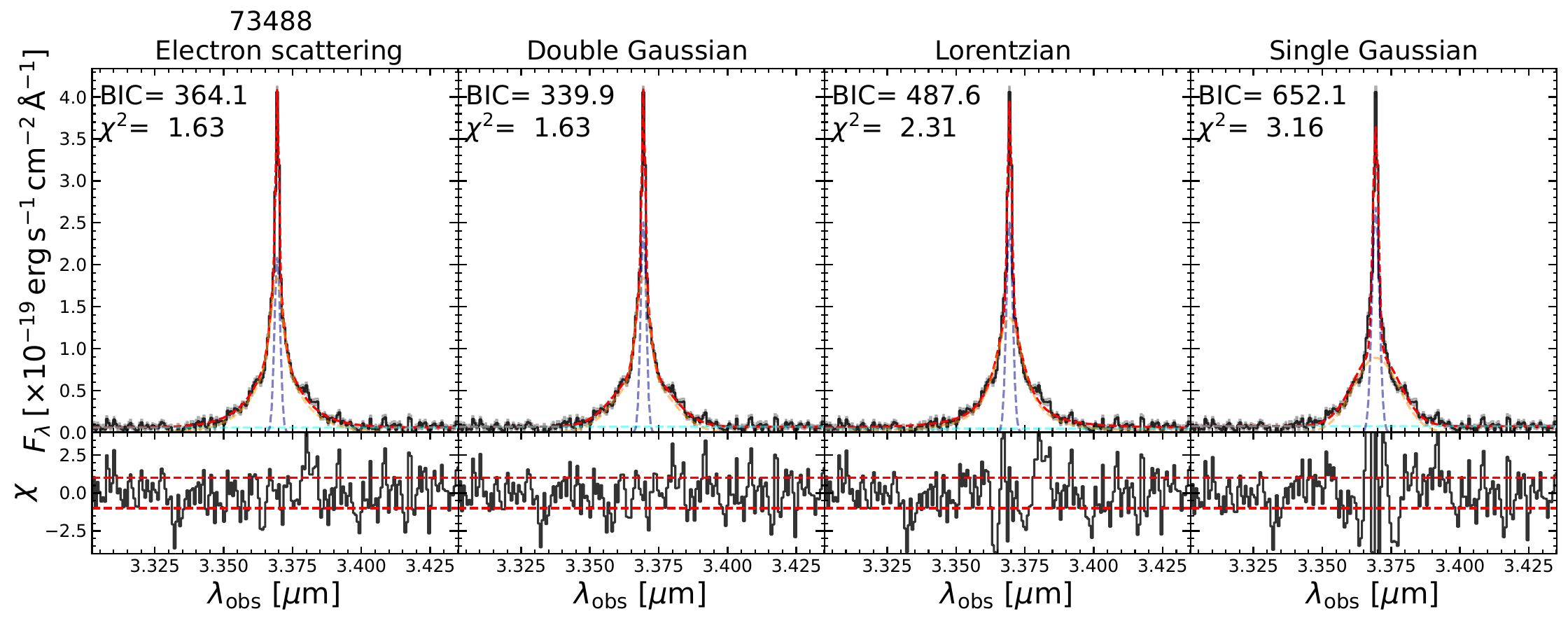}
    \caption{Examples of the model fitting procedure used in this work. The top and bottom panels show examples of targets with and without and absorption in the Balmer line. From left to right: Electron scattering model, double Gaussian model, Lorentzian model and single Gaussian model. The data are shown as a black solid line with the best-fit model shown as a red dashed line. The BLR model is shown as an orange line and the narrow \Halpha component as a blue dashed line. We show the $\chi$ residuals in the bottom panel for each model. For 42046, the best-fit model is the Lorentzian profile.}
    \label{fig.fit_example}
\end{figure*}

We show each of the profile fits for ID-42046 and ID-38147 in Fig.~\ref{fig.fit_example}. Furthermore, for sources with detected \Halpha absorption, we add a dense hydrogen absorber with a common covering factor
$C_f$, FWHM$_{\rm abs}$, and independent velocity offsets $v_{\rm abs}$ for \Halpha (the velocity offset is relative to the redshift of the narrow lines). The residual intensity at wavelength $\lambda$ is given by
\begin{equation}\label{eq.residual}
\begin{split}
    I(\lambda)/I_0(\lambda) &= 1 - C_f + C_f \cdot \exp \left(- \tau(k;\,\lambda) \right)\\
    \tau(k;\,\lambda) &= \tau_0(k) \cdot f[v(\lambda)],
\end{split}
\end{equation}
where $I_0(\lambda)$ is the spectral flux density before absorption, $\tau_0(\Halpha)$ is the optical depth at the centre of the \Halpha line and $f[v(\lambda)]$ is the velocity distribution of the absorbing atoms, assumed to be a Gaussian probability distribution. $I_0(\lambda)$ represents the component being absorbed, in this case, the BLR emission and continuum models. 

The continuum is modelled as a power-law, while the narrow \Halpha is modelled as a simple Gaussian. For sources with \NIIall emission, we model the doublet with two Gaussian profiles with a flux ratio of 3 \citep{dimitrijevic_flux_2007}. The redshift and FWHM of the \NIIall component are tied to the narrow \Halpha emission line.

As 28 out of 32 sources in our sample also have observations of the \OIIIL[5008] emission line, this allows us to independently measure the FWHM of the narrow lines, reducing the degeneracies between the broad and narrow \Halpha emission. However, tying the narrow widths of \OIIIall and \Halpha emission lines is challenging for the \MSA observations, which suffer from large uncertainties in the line spread function (LSF) due to the size of the narrow emission. For these reasons, we only tie the narrow \Halpha emission to the measured \OIIIall widths when R2700 observations of the \OIIIall are available (15 sources in total). We verified that tying the narrow \Halpha width to that of the \OIIIall does not affect our conclusions. 

A few sources in our sample also exhibit outflows in the \OIIIall (see Fig. A\ref{fig.OIII_spec}). For these sources, we also include the outflow component in the narrow \Halpha and \NIIall (when present), with outflow velocity and FWHM of the outflow tied to those determined from the \OIIIall fitting \citep[for a description of the \OIIIall fitting, see][]{scholtz_kashz_2020}.

Finally, the model is convolved with the LSF of the NIRSpec instrument. For the NIRSpec-MSA observations, we use the LSF from \citet{de_graaff_early_2024}, which is estimated for point sources observed with the MSA. For the NIRSpec-IFS mode, we use the nominal LSF from the JDOCS\footnote{Available at \href{https://jwst-docs.stsci.edu/jwst-near-infrared-spectrograph/nirspec-instrumentation/nirspec-dispersers-and-filters}{jwst-docs website}.} multiplied by a factor 0.7 \citep[e.g.][]{Shajib_nirspec, de_graaff_ionised_2024} as the in-flight characteristics of the spectrograph are better than the original instrument simulations. For the NIRCam/Slitless observations, we used the description of LSF from \citet{danhaive_dawn_2025}.

We note that GS-3073 and 28074 have previously been fitted in \citet{Brazzini_rosettas_2026}, and we adopt their fitting results for GS-3073. For 28074, we refit this object using the G140M observations that have been extended to 2.7 $\mu$m (DJA v4.4 or also \citealt{scholtz_jades_2025}) and provide higher spectral resolution than the already published G235M observations. Our results are consistent with those from \citet{brazzini_ruling_2025}.

The posterior probability distribution is estimated using the Markov-Chain Monte-Carlo (MCMC) ensemble sampler \texttt{emcee} \citep{foreman-mackey_emcee_2013}. To initialise the chains, we first identify the minimum $\chi^{2}$ solution as initial conditions for the MCMC chains. The fiducial model parameters of our best-fit models are estimated with a Bayesian approach. For each free parameter, we define a prior for the MCMC integration. The prior on the redshift of each spectrum is set as a truncated Gaussian distribution, centred on the systemic redshift (taken from \OIIIall or narrow \Halpha line) of the galaxy with a sigma of 300 km s$^{-1}$ and boundaries of $\pm 1000$ km s$^{-1}$ (the truncation is for chains finding unphysical local minima). The prior on the intrinsic FWHM of the narrow-line component is set as a uniform distribution between 30--300 km s$^{-1}$, while the prior on the amplitude of the lines is set as a uniform distribution in log-space between $0.5\times$rms of the spectrum and twice the maximum of the flux density in the spectrum. The FWHM of the BLR profile is set as a uniform distribution between 300--5000 km s$^{-1}$. For the double Gaussian profile, we also only allow models where the second Gaussian profile is broader than the first, to avoid degeneracies in the fit.

The final best-fit parameters and their uncertainties are calculated as the median value and 68\% confidence interval of the posterior distributions. We note that all the quantities derived from our spectral fitting (e.g.\ metallicities) are calculated from the posterior distribution to account for any correlated uncertainties in the spectrum.

We use the Bayesian Information Criterion \citep[BIC;][]{schwarz_estimating_1978} to distinguish between the fitted models. The BIC is defined as $\chi^{2} + k \log(N)$, where $N$ is the number of data points and $k$ is the number of free parameters for each model. We use a $\delta$BIC $<10$ (often used in the literature for spectroscopy; e.g. \citealt{scholtz_ga-nifs_2025}, \citetalias{Rusakov_nature_2026}, \citealt{brazzini_ruling_2025}) to choose whether the fit needs a second narrow component (using $\delta$BIC $>10$ as a boundary for choosing a more complex model). We summarise the BIC values for each fit with respect to the best fit in Table~\ref{tab.bics}.

\begin{table}
    \caption{$\delta$BIC with respect to the best-fit value for each model. All models with $\delta$BIC of the best fit are highlighted in bold.}
    \centering
    \begin{tabular}{@{}lccccc@{}} 
\hline 
\hline 
ID & Best fit & $\delta$BIC$_{1G}$ & $\delta$BIC$_{2G}$ & $\delta$BIC$_{Lorentz}$ & $\delta$BIC$_{Scatt}$ \\
   \\
\hline 
68797 & \escats & 1152.2 & 128.8 & 16.8 & \textbf{0.0}\\
14 & 2G & 143.8 & \textbf{0.0} & 25.7 & \textbf{6.9}\\
73488 & 2G & 312.2 & \textbf{0.0} & 147.7 & 24.2\\
42046 & Lorentzian & 209.8 & 33.6 & \textbf{0.0} & \textbf{8.6}\\
49140 & Lorentzian & 60.3 & 13.2 & \textbf{0.0} & 13.9\\
1244 & \escats & 89.3 & 15.9 & 15.5 & \textbf{0.0}\\
58237 & 1G & \textbf{0.0} & \textbf{8.8} & 22.8 & \textbf{9.7}\\
51623 & Lorentzian & 35.0 & \textbf{8.0} & \textbf{0.0} & \textbf{1.7}\\
53501 & Lorentzian & 21.3 & 21.7 & \textbf{0.0} & 10.9\\
38147 & \escats & \textbf{6.3} & 40.8 & \textbf{5.0} & \textbf{0.0}\\
50052 & Lorentzian & 11.3 & \textbf{0.3} & \textbf{0.0} & \textbf{0.3}\\
60935 & Lorentzian & 17.4 & \textbf{5.5} & \textbf{0.0} & 11.8\\
Cliff & \escats & 375.4 & 34.1 & 23.8 & \textbf{0.0}\\
159717 & Lorentzian & 131.7 & \textbf{9.2} & \textbf{0.0} & \textbf{3.5}\\
G23\_4286 & Lorentzian & 24.0 & 29.6 & \textbf{0.0} & \textbf{6.2}\\
G23\_13821 & Lorentzian & 48.2 & 21.6 & \textbf{0.0} & \textbf{9.6}\\
GN\_14409 & \escats & 77.3 & 16.0 & \textbf{0.0} & \textbf{0.0}\\
J1148 & Lorentzian & 104.4 & 11.8 & \textbf{0.0} & \textbf{2.9}\\
GN-12839 & Lorentzian & 46.0 & 12.3 & \textbf{0.0} & \textbf{6.8}\\
GN-15498 & \escats & 219.1 & \textbf{8.5} & \textbf{2.0} & \textbf{0.0}\\
GS-13971 & \escats & 342.4 & 19.6 & 38.4 & \textbf{0.0}\\
GN-9771 & \escats & 5005.7 & 321.4 & 1221.9 & \textbf{0.0}\\
GN-16813 & \escats & 46.7 & \textbf{0.2} & \textbf{5.3} & \textbf{0.0}\\
XID2028 & 2G & 1656.8 & \textbf{0.0} & 1001.0 & 623.1\\
QSO1 & 2G & 145.8 & \textbf{0.0} & 12.4 & \textbf{6.8}\\
GS-3073 & \escats & 139.0 & 14.0 & 274.0 & \textbf{0.0}\\
28074 & \escats & 1204.3 & 95.9 & 95.0 & \textbf{0.0}\\
209777 & 2G & 115.9 & \textbf{0.0} & 107.3 & \textbf{4.2}\\
7384 & 2G & 264.3 & \textbf{0.0} & 14.2 & \textbf{6.1}\\
4151 & 2G & \textbf{9.4} & \textbf{0.0} & 36.3 & 40.5\\
11337 & 2G & 362.4 & \textbf{0.0} & 873.3 & 373.2\\
Monster & \escats & 1078.9 & 109.0 & 1068.0 & \textbf{0.0}\\
\hline 
\end{tabular}
    \label{tab.bics}
\end{table}

\section{Results of the broad \Halpha modelling}\label{s.profiles}

\begin{figure*}
    \centering
    \includegraphics[width=0.7\paperwidth]{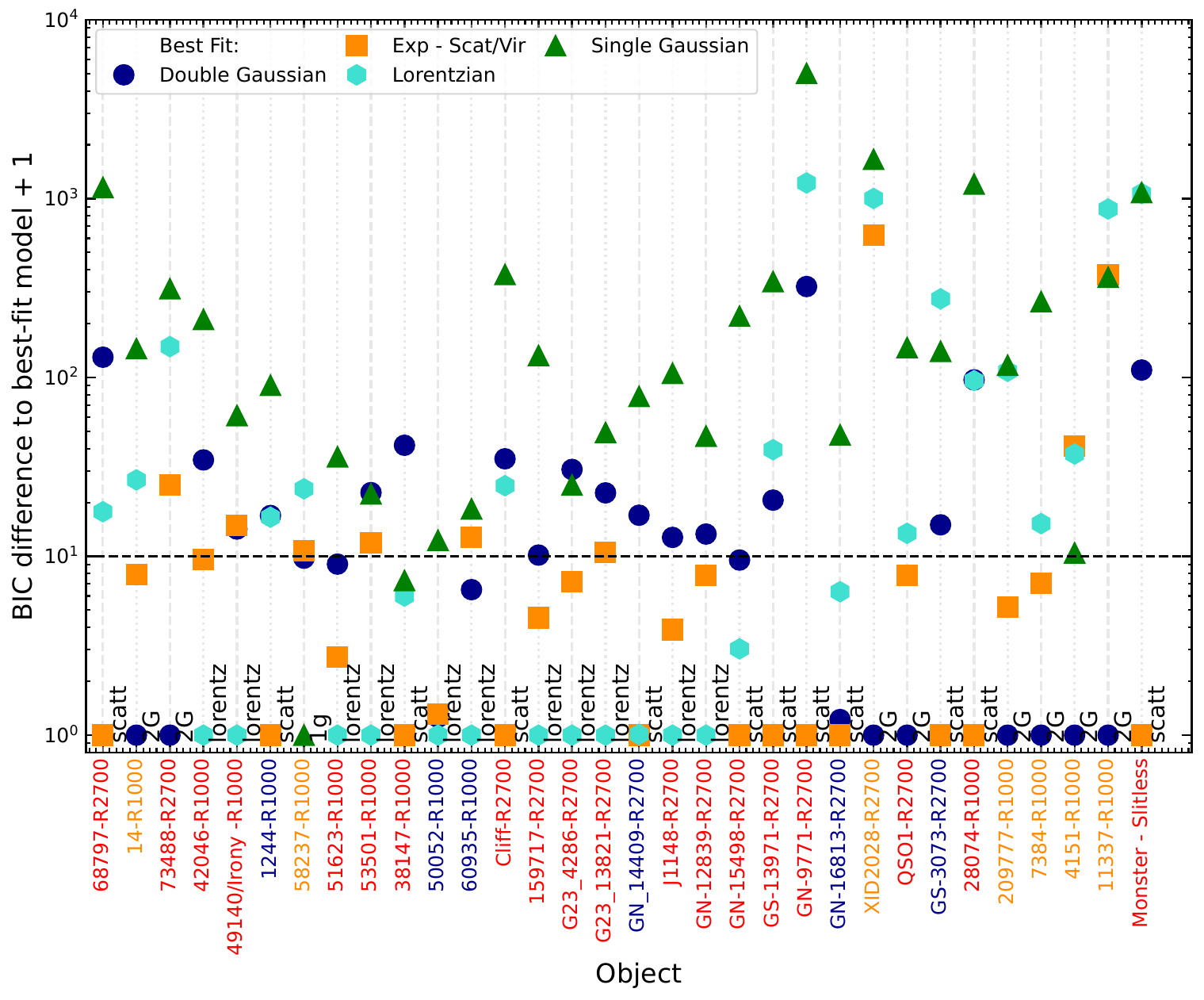}
    \caption{Comparison of the fitted models in this work for each source. The horizontal dashed line shows $\delta$BIC of 10 from the best model. The colour of the labels shows the type of the target: LRD (red), LBD (blue) and X-ray AGN (orange).}
    \label{fig.model_comp}
\end{figure*}

We show the performance of each model discussed in Sect.~\ref{s.eml_fit} for our targets in Fig.~\ref{fig.model_comp}. However, we note that in 48\% of our sample, there are two or more models with $\delta$BIC $< 10$, showing no statistical preference for any single model, resulting in substantial uncertainties in the best-fit model fractions. 

\begin{figure}
    \centering
    \includegraphics[width=0.99\columnwidth]{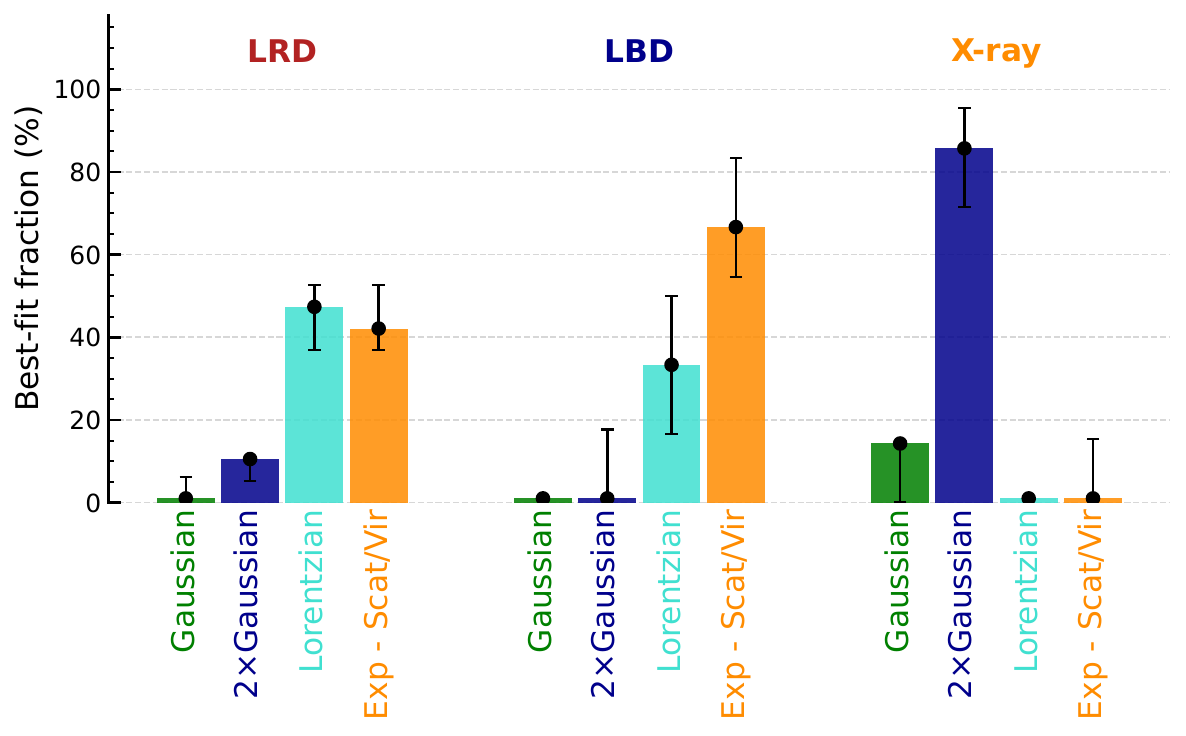}
    \caption{Overview of the best fit profile, with error bars (1$\sigma$) considering all of the broad H$\alpha$ profiles with $\delta$BIC$<$10. We see no evidence that LRDs or LBDs statistically prefer the \escat model. The X-ray AGN prefer the 2 Gaussian model.}
    \label{fig.frac_overview}
\end{figure}

While estimating the fraction of objects with best-fit broad profiles, we have to account for multiple broad line profiles that are within $\delta$BIC $< 10$. To estimate the uncertainties, we bootstrapped errors by randomly selecting a fit for each object from the profiles with $\delta$BIC $< 10$. Each broad line profile with $\delta$BIC $< 10$ has the same probability of being selected. We do this in total of 1000 times for the sample. The final quoted uncertainty of best-fits for each broad line profile is 16th and 84th percentile of the distribution. We summarise the best-fit fractions for each category and the BIC values in Tables~\ref{tab.bics}\&~\ref{tab.res} and in Fig.~\ref{fig.frac_overview}. 

Overall, we see that a single Gaussian profile is the least preferred model, being the best fit for only a single AGN in a quiescent galaxy (ID-58237), similarly to the results presented in \citetalias{Rusakov_nature_2026}. However, we find no evidence that the exponential profiles are ubiquitously preferred, especially in the case of LRDs. This may appear in contrast with the claim by
\citetalias{Rusakov_nature_2026}, but it is actually perfectly consistent with their finding -- indeed, only about half of their sample is made of LRDs (seven of twelve targets), and eight sources of the full sample are best fitted with \escat or exponential model.

Lorentzian and double Gaussian profiles are equally good fits for twenty sources, and these profiles are strongly preferred relative to the exponential in a significant fraction of LRDs and LBDs. The exponential profile is a statistically acceptable model for LRDs,  LBDs and X-ray AGN in 38.9$^{+6.2}_{-6.2}$\%, 71.4$^{+14.3}_{-28.6}$\% and 0$^{+14.3}_{-0.0}$\% of cases, respectively (see Fig.~\ref{fig.model_comp}).

We find a very marginal indication that Lorentzian profiles are preferred in the LRDs, while exponential profiles are preferred in the LBDs, contrary to the findings by \citetalias{Rusakov_nature_2026}. However, given the uncertainties associated with selecting the best fit, we see no evidence for any category of AGN preferring the \escat profile, or for the \escat profile being overall preferred for high-$z$ AGN.

\begin{table}
    \caption{Overview of the best fits.}
    \centering
    \begin{tabular}{@{}lcccc@{}} 
    \hline
    \hline
    Model & All & LRDs & LBDs & X-rays \\
    \hline
    Gaussian & $3.1^{+3.1}_{-3.1}\%$ & $0.0^{+0.0}_{-0.0}\%$ & $0.0^{+14.3}_{-0.0}\%$ & $14.3^{+0.0}_{-14.3}\%$ \\[6pt]
    2$\times$Gaussian & $25.0^{+6.2}_{-6.2}\%$ & $11.1^{+11.1}_{-0.0}\%$ & $0.0^{+14.3}_{-0.0}\%$ & $85.7^{+14.3}_{-14.3}\%$ \\[6pt]
    Lorentzian & $34.4^{+6.2}_{-3.1}\%$ & $50.0^{+5.6}_{-11.1}\%$ & $28.6^{+14.3}_{-14.3}\%$ & $0.0^{+0.0}_{-0.0}\%$ \\[6pt]
    \escats & $37.5^{+6.2}_{-6.2}\%$ & $38.9^{+5.6}_{-11.1}\%$ & $71.4^{+14.3}_{-28.6}\%$ & $0.0^{+14.3}_{-0.0}\%$ \\[6pt]
    \hline
    \hline    
    \end{tabular}
    \label{tab.res}
\end{table}

While previous studies have focused on highlighting exponential wings for the broad Balmer lines of LRDs, we have shown here that non-Gaussian profiles, including exponential wings, are also common across LBDs and in X-ray sources. In this context, we remark that even in the `standard' scenario where the broad lines arise from virial motions, there is no theoretical prescription for why the integrated BLR spectrum should have a Gaussian shape \citep[e.g.][and see \S~\ref{s.exponential_creation}]{kollatschny_shape_2013}. Before the launch of \jwst, studies had already utilised exponential, double-Gaussian, Lorentzian, or broken power-law models to model the \Halpha and \Hbeta BLR profile \citep{nagao_evolution_2006, nagao_gas_2006, cano-diaz_observational_2012, kollatschny_shape_2013, scholtz_impact_2021, santos_spectroscopic_2025}. Therefore, we conclude that exponential wings and non-Gaussian BLR profiles should not be interpreted as a defining characteristic of LRDs or high-$z$ AGN found by \jwst. 

\begin{figure}
    \centering
    \includegraphics[width=0.99\columnwidth]{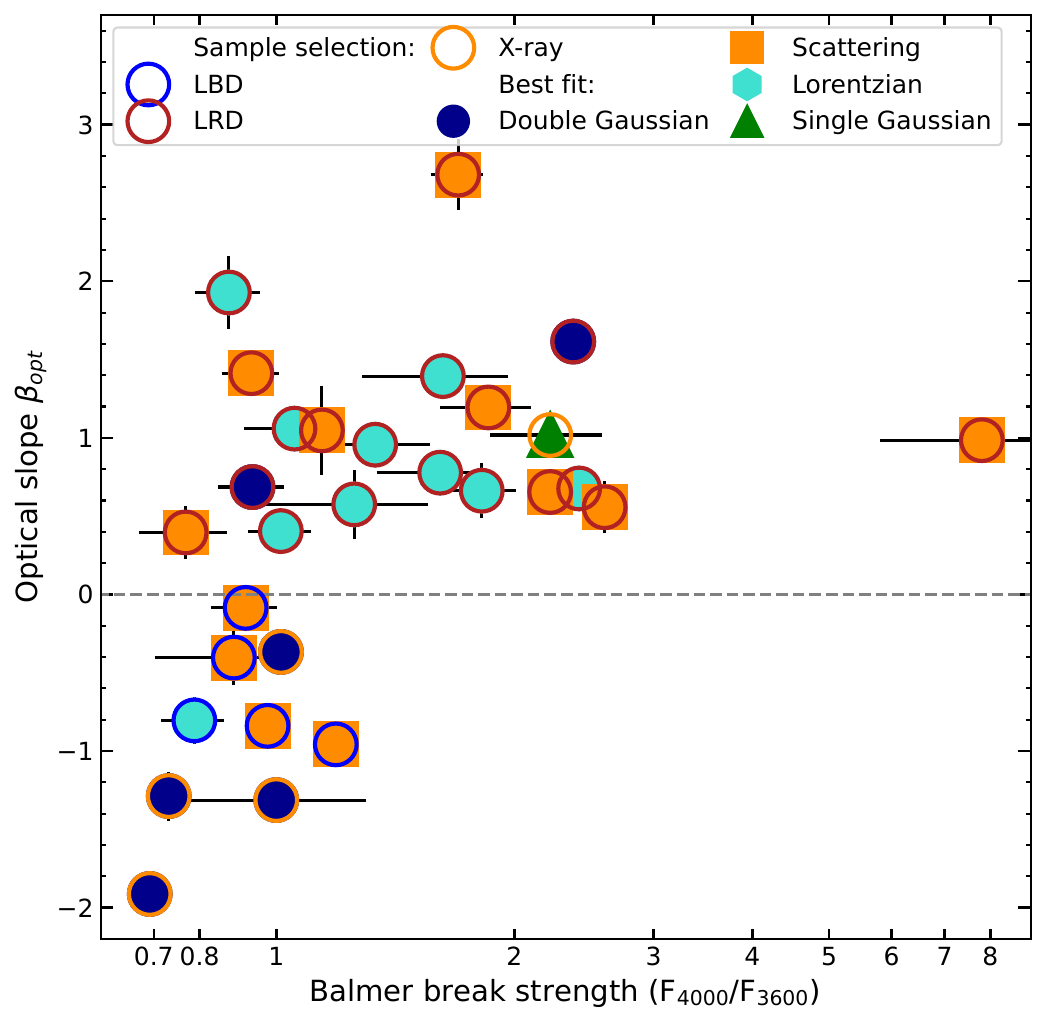}
    \caption{Plot of $\beta_{\rm opt}$ vs Balmer Break strength. We highlight the sources selected as LRDs, LBDs and X-ray AGN as red, blue and orange circles, with each symbol indicating the best-fit model. We see no correlation between the spectral break strength and preferred line profile.}
    \label{fig.break}
\end{figure}

We further investigate the evolution of the broad H$\alpha$ profiles with the spectral break around the Balmer limit (see Fig.~\ref{fig.break}). We estimated the Balmer break strength as a median flux ratio between 4000 and 3600 $\AA$.  We see no evidence for an increase in the fraction of \escat exponential profiles with the spectral break strength in LRDs, given that we see the same number of \escat profiles in objects (and LRDs specifically) with low and high spectral break strength. We further discuss this result in \S~\ref{s.exponential_creation}.

\section{Discussion}\label{s.discuss}


\subsection{Non-scattering scenario of exponential profiles}\label{s.exponential_creation}

We have shown that exponential profiles are not a widely preferred model for high-$z$ \jwst AGN, especially for LRDs, while there is only a slight preference for LBDs. However, even if these exponential cases are not the majority, it is important to discuss the origin of such profiles. These are generally attributed to electron scattering broadening (see e.g. \citetalias{Rusakov_nature_2026} and \citealt{Matthee_2026}). In the scenario proposed these recent works, a shell of warm ionised gas surrounds the broad line region and scatters the light coming from the BLR, broadening the overall profile.
We note that electron scattering can also naturally emerge from the ionized gas in the BLR clodus, as originally proposed by \citet{laor_evidence_2006}, without having to invoke new exotic structures, and also by surrounding ionized gas responsible for the polarization of the nuclear spectrum in normal AGN.
However, in this section, we investigate whether other effects can reproduce exponential profiles.

Based on reverberation mapping studies, it is well known that the BLR has a stratified structure, with clouds closer to the black hole contributing to the larger velocities of the broad lines, and clouds further away contributing more to the core of the broad line \citep[e.g.][]{Zetzl_2018, Feng_2025, gravity_collaboration_spatially_2025}.
Therefore, one wonders whether the observed exponential profiles could actually be reproduced by the superposition of a continuous distribution of Gaussians, each of them providing the contribution of a different layer of the BLR (we will discuss in the next section that an exponential profile is obtained also in the case of combining other non-Gaussian profiles). This scenario holds the advantage of leveraging a well-known structure, without having to invoke a new exotic scenario (a putative gas cocoon).
It is indeed possible to reproduce a broken exponential profile -- or Laplacian distribution -- as a sum of Gaussian profiles $G(0, \sigma; v)$ with common centre $v=0$, with the same integral (assumed to be unity), and with the linewidth $\sigma$ of each Gaussian randomly drawn from a Rayleigh probability distribution
\begin{equation}
  p(\sigma) \coloneqq \displaystyle\dfrac{\sigma}{W^2} \exp\left\{-\dfrac{1}{2}\left(\dfrac{\sigma}{W}\right)^2 \right\} \quad W>0.
\end{equation}
We now define
\begin{equation}\label{eq.intdef}
  f(v) \coloneqq \displaystyle\int_0^\infty d\sigma p(\sigma) \dfrac{1}{\sqrt{2 \pi} \sigma} \exp\left\{-\dfrac{v^2}{2\,\sigma^2}\right\}.
\end{equation}
After using the result
\begin{equation}
    \displaystyle\int_0^\infty dx \exp\left\{-\dfrac{a}{x^2} - b x^2\right\} = \dfrac{1}{2} \sqrt{\dfrac{\pi}{b}} \exp\left(-2 \sqrt{ab}\right),
\end{equation}
Eq.~\ref{eq.intdef} evaluates to a Laplacian probability distribution
\begin{equation}\label{eq.laplace}
  f(v) = \displaystyle\dfrac{1}{2 W} \exp\left\{- \left| \dfrac{v}{W} \right| \right\}.
\end{equation}
This shows that the scale parameter $W$ is equal to the mode $W$ and proportional to the mean $\sqrt{\pi/2}\,W$ of the Rayleigh distribution.

\begin{figure}
    \centering
    \includegraphics[width=0.99\columnwidth]{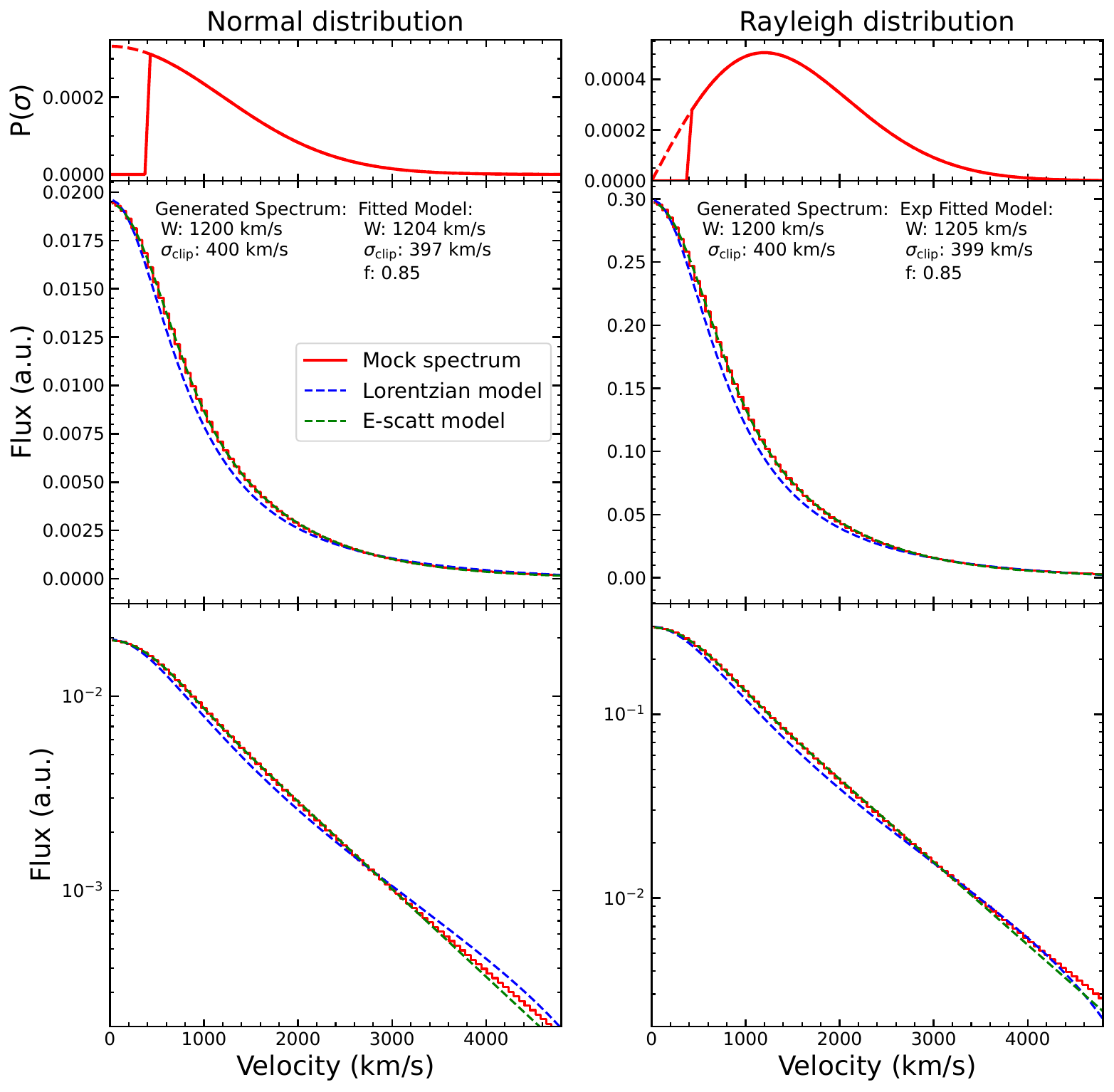}
    \caption{Mock exponential broad line profiles expected from a stratified BLR, generated as a sum of Gaussian profiles with their widths drawn from a clipped normal distribution (left panel) and a Rayleigh distribution (right panel). We show both distributions with and without clipping as solid and dashed red lines, respectively, in the top panels. The bottom panels show the mock BLR profile (red line), best-fit Lorentzian profile (blue dashed line) and \escat model (green dashed line). Regardless of the distributions chosen, we are able to excellently reproduce the exponential profiles without invoking any scattering.}
    \label{fig.expo_model}
\end{figure}

In Fig.~\ref{fig.expo_model}, we show two mock profiles drawn from two separate distributions: a truncated normal distribution (left panel) and a Rayleigh distribution (right panel). We show an example using both normal and Rayleigh distribution to illustrate that our results do not depend on exact shape of the distribution. Each distribution is clipped at a clipping value $\sigma_{\rm clip}$. In both cases, the profile was created by drawing 2000 random $\sigma$ values from each distribution. Clipping at $\sigma_{\rm clip}$ removes the narrowest Gaussian components, causing the natural rounding of the profile around zeroth velocity as seen in the data. The physical interpretation of clipping below $\sigma_{\rm clip}$ is that this represents the velocity distribution of the outer clouds of the BLR, e.g. at the dust sublimation radius or where the clouds start to be disrupted \citep[e.g.][]{maiolino_comets_2010}. We fitted both a Lorentzian profile and an \escat exponential model to the mock profile, and we show that the latter model is an excellent fit, while the Lorentzian profile can broadly reproduce the mock profile as well (see \citealt{Jin_Gauss_2020} for more information about Gaussian transforms). The final profile of the broad H$\alpha$ can therefore be explained by the stratified distribution of BLR clouds \citep[e.g.][]{brazzini_ruling_2025, Brazzini_rosettas_2026, Popovic_2006}.

Crucially, this model shows that \escat is not the only explanation for the exponential profile. The width of each constituent Gaussian reflects the stratification of BLR clouds around the black hole, described by the $W$ parameter, while the $\sigma_{\rm clip}$ parameter describes the largest radius/lowest velocity of the stable BLR clouds before they are either destroyed (e.g. by an outflow or erosion) or once the BLR clouds are beyond the dust sublimation radius \citep[where their contribution to the emission lines would be hampered by extinction or the ionizing flux is preferentially absorbed by dust rather than photo-ionizing the gas][]{netzer_star_2016}.

Summarizing, our simple model shows that the exponential profiles in the BLR can be reproduced without invoking electron or resonant scattering, with the exponential profile arising from the stratification of the BLR and its virial motion around the black hole.


\subsection{The stacking of non-exponential profiles gives an exponential profile}\label{s.stacking}

One of the most powerful tools in astronomy is stacking, which has been used in spectroscopy to detect faint lines or even faint wings lines \citep[e.g.][]{stanley_spectral_2019}. This approach has also been used to increase the S/N to investigate the line profiles, often as a verification of individual fits (e.g. \citetalias{Rusakov_nature_2026}). However, stacking lines with different widths and profiles can generate spurious profiles. In this section, we investigate the effect of stacking on the resulting line profiles and the appropriateness of using the stacked profile to infer information on the profile of the individual lines.

\begin{figure}
    \centering
    \includegraphics[width=0.99\columnwidth]{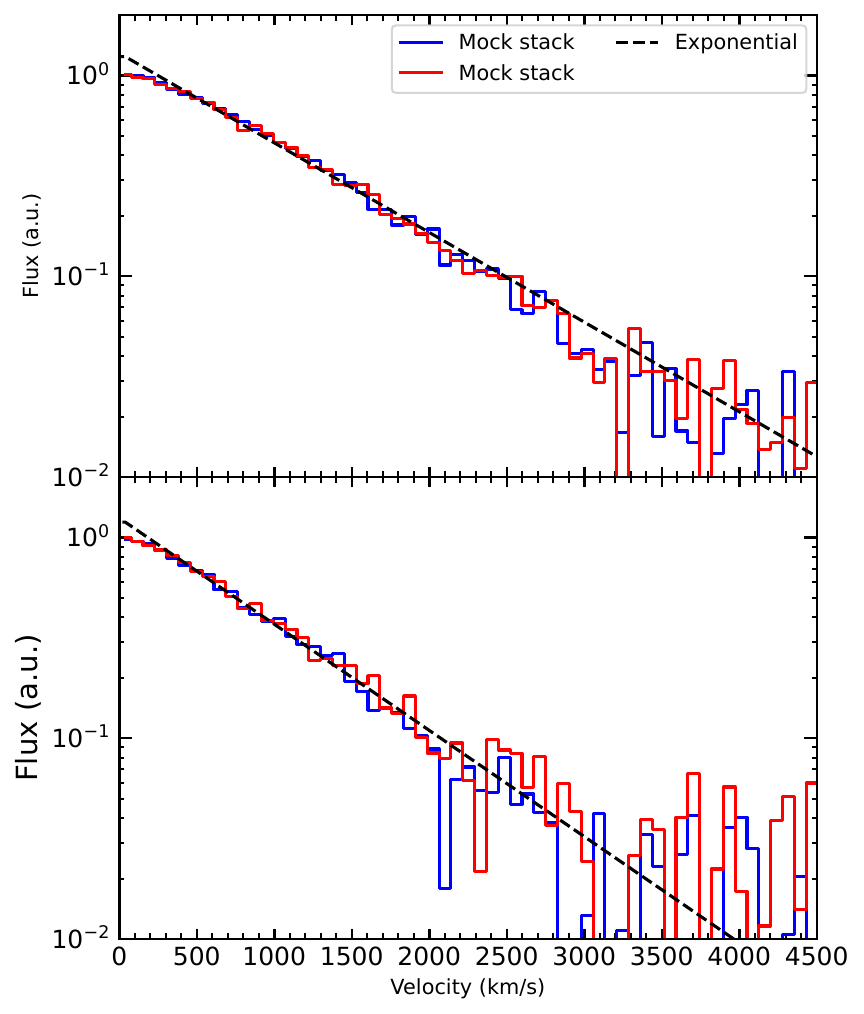}
    \caption{Stacking of mock BLR profiles (single Gaussian, double Gaussian and Lorentzian) to test stacking procedures. Top panel: each profile is normalised to its peak. Bottom panel: Each profile is normalised at 1500 \kms from the peak. The red and blue lines show the red and blue wings of the mock stacked profile, while the dashed black line shows the best broken exponential fit to the data. The exponential fit shows excellent agreement with the data despite none of the individual profiles having no exponential origin.}
    \label{fig.stack_mock}
\end{figure}

We created a set of 20 mock BLR spectra with random profiles (drawn from single Gaussian, double Gaussian and Lorentzian profiles) with FWHM of 2000--4000 \kms. We intentionally excluded exponential or electron scattering profiles from the stack to avoid biasing our experiment. We set the peak SNR of the mock spectra to 20 to simulate the depth of the \jwst/NIRSpec observations. We created a mean stack of the individual profiles with two normalization: i) normalized at the peak of the profile; and ii) normalized at 1,500 \kms \citep[often used in the literature; e.g. ][]{Matthee_2026}. We show the stacked profiles in Fig.~\ref{fig.stack_mock}.

We fit the mock data stacked in velocity space with a simple broken exponential profile, fitting the stacked spectrum outside of $\pm$300 \kms, in order to determine whether the wings are exponential. The fitted simple broken exponential model is shown as a red dashed line. The best exponential fit shows a reduced $\chi^{2}$ of 1.14 and 1.13 for the stacks normalized to the peak and the stacks normalized at 1,500 \kms, respectively. Our simple modelling experiment shows that stacking results into exponential profiles {\it despite none of the individual spectra going into the stack having an exponential or electron scattering profiles}. While one can argue about the details of the stacking set, the outcome of this experiment is rooted in the mathematical foundation of Eq.s~(\ref{eq.intdef}) and~(\ref{eq.laplace}). Therefore regardless if the individual spectra come from separate objects or from individual BLR layers within the same object (see in the \S~\ref{s.exponential_creation}), the combination of the different line profiles will result in an  exponential profile.

This puts severe limitations on studies using stacking to infer the average H$\alpha$ emission line profile of high-$z$ AGN: not only will stacking naturally result in exponential-like profiles -- but the more spectra one stacks, the more the resulting profile will approach the Laplacian distribution. We have seen similar behaviour elsewhere in astronomy; for example, a sum of blackbodies of different temperatures in an accretion disc will result in a power-law spectrum (see below). 

As such, we conclude that it is inappropriate to use stacking analysis to determine the average BLR profile of AGN, LRDs and LBDs, as this will result in an exponential profile even with only a few ($N=10$) stacked spectra, rendering such results unreliable for scientific interpretation. 

Furthermore, as mentioned, the the same reasoning applies to the line profile of the individual objects. In this case, the exponential profile can be the result of the superposition of individual kinematical components in the BLR, none of which are exponential. While in the BLR stratification scenario presented in the previous section the exponential profile arises from the continuous distribution of different BLR layers, the experiment performed in this section illustrates that an exponential profile also naturally originates from the combination of different BLR kinematics components, which can also be disconnected, and individually having a combination of Gaussian and Lorentzian profiles. This further supports, in an even more model-independent way, that exponential profiles in individual objects do not require invoking electron scattering.

\subsection{Implication of the non-scattering scenario}\label{s.disc_profiles}

In \S~\ref{s.exponential_creation}\&~\ref{s.stacking}, we showed that the exponential profile can arise from either stratification of the BLR or multiple separate components inside the BLR. In this section, we discuss the implications of our results on the nature of LRDs and LBDs.

The non-scattering interpretation of the exponential profiles solves several tensions that the scattering scenario has produced. For instance, the direct, kinematic black hole mass measurement of QSO1 at z=7.04 ( \citealt{furtak_supermassive_2023, furtak_high_2024}), results in a value in complete agreement with the virial estimation, and two orders of magnitude higher than the electron scattering scenario \citep{juodzbalis_direct_2025}. Our interpretation of the exponential profiles, still in the context of the virial broadening, would clarify the inconsistency. Similarly, as already mentioned, in the local universe, the prototypical example of exponential wings, NGC4395 \citep{laor_evidence_2006}, has a black hole mass inferred from reverberation mapping (\citealt{Peterson_2005}) and from direct dynamical measurements  (\citealt{denbrok_2015}) fully consistent with the virial estimators \citep{Lira1999}. Indeed, \citet{Popovic_2006} have discussed the complexity of line profiles and BLR kinematics (see their section 3.3). 

\citet{juodzbalis_jades_2025} pointed out that the electron scattering scenario should leave a strong narrow component in the Balmer lines, resulting from the recombination in the scattering medium, which is not observed. Our scenario would obviously solve the tension, as in that case, there is no dominant electron scattering medium.
Furthermore, \citet{brazzini_ruling_2025} showed that the LRD Rosetta Stone has different line profiles across different hydrogen lines; this would naturally fit with a model of stratified emission. While bound-free absorption has also been invoked as an explanation for the different hydrogen line widths \citep{sneppen_inside_2026}, such a scenario would not explain the findings of \citet{Brazzini_rosettas_2026}. In their work, they showed that in the LBD GS-3073, which also displays non-Gaussian broad \Halpha, the \HeIIL[4686] line is much broader than \Halpha (by a factor $\sim3$). This mismatch is in direct contradiction to the scenario of dominant electron scattering, as all lines would display the same broadening \citep[with hydrogen bound-free absorption only able to explain subtle differences;][]{sneppen_inside_2026}.
In contrast, the result by \citet{Brazzini_rosettas_2026} would fit very well in our stratification scenario as the higher ionisation lines are emitted from higher-ionization regions of the BLR compared to the Balmer lines. Since these regions would presumably be closer to the central source than the bulk of the Balmer-emitting regions, the characteristic $W$ of the HeII-emitting clouds would be naturally broader than that of the HII regions, which include lower-ionization gas further away from the black hole. Indeed, it is observationally seen in reverberation mapping studies that HeII is emitted closer to the black hole relative to the Balmer lines.

As described in \S~\ref{s.profiles}, we do not find a correlation between the Balmer break strength and the presence of exponential profiles in our data, especially when LBDs are also considered. This is in contrast with previous studies \citep{Matthee_2026}. However, even if such a correlation is found in future works with higher statistics, we note that our BLR stratification scenario can explain this result. Indeed, the increase in Balmer break strength results from a larger amount of gas along the line of sight, which in the stratified BLR scenario is easily interpreted in terms of a larger covering factor of the BLR clouds and/or larger gas mass in the BLR clouds \citep{inayoshi_extremely_2024,maiolino_jwst_2024}. These additional BLR clouds would add additional kinematical components to the overall line profile, enhancing the exponential profile seen in the data.


Recent works which have focused on the investigation of the emission line profiles of AGN lines \citep[e.g.][]{Kokorev_2025, torralba_warm_2025, Matthee_2026} have observed similar non-Gaussian line profiles and similarity to the exponential profiles. Their interpretation of these features, including emission line ratios, is a clumpy ionised gas cocoon. In this scenario, the Balmer line profiles are shaped by radiative transfer effects in dense gas envelopes. We note that \Halpha and \Hbeta emission lines do not have the same profiles even in typical X-ray AGN or quasars \citep{netzer_accretion_2009, cano-diaz_observational_2012, carniani_fast_2016} and it is not unique to high-z AGN discovered by \jwst. Furthermore, the scenario of a clumpy, partially ionised gas cocoon closely resembles a BLR region with a high covering factor \citep[][Ji et al. in prep.]{inayoshi_weakness_2024}. In this case, we have shown that the Balmer line profiles are explained by the stratification in the BLR \citep{Baskin_2014}, also observed in SDSS AGN \citep{Son_2025}, or as the combination of multiple separate BLR components. Therefore, there is no observational evidence 
requiring
new exotic modes of accretion onto supermassive black holes for the population of AGN found by \jwst. 

We, however, recall that \escat is important in studies of BLR geometry and kinematics based on the results from spectro-polarimetry \citep{Young_2000, Smith_2002, Smith_2005}. In local Seyfert-1 (Sy1), the spectro-polarimetry results are best explained with broad line region rotation by a flattened rotating BLR, surrounded by an equatorial BLR scattering region. We note, though, that if the BLR has a large covering factor as suggested for \jwst AGN \citep{maiolino_jwst_2024, inayoshi_weakness_2024}, the BLR kinematics would no longer be in a flattened disk as in local Sy1 AGN. The electron scattering is present even in these local Sy1 AGN, and it has been successfully used to infer the structure of the BLR on otherwise unresolvable scales.
However, in all these cases, scattering is not the dominant source of the line width. The broadening induced by scattering is not generally large enough to account for the full observed widths of the lines; the kinematics of the emitting gas (largely rotational, with possible additional outflow components) still dominate the line profile \citep{Smith_2005, laor_evidence_2006}. Based on the local AGN results, we would expect some weak presence of electron scattering in high-z AGN, which can smooth the overall BLR profile. However, together with the results presented in this paper, we conclude that scattering does not dominate the line broadening in LBDs and LRDs.

\subsection{Considerations on other properties of LRDs and LBDs}\label{s.other_props}

While the focus of this paper is on the analysis of the profile of the broad component of \Halpha, in this section, we briefly discuss the scenario in which the exponential profiles are simply a result of the BLR stratification. This scenario can also explain other properties of LRDs and LBDs, either in the case of a stratified BLR with large covering factor \citep[e.g.][]{maiolino_jwst_2025}, or a flattened BLR with a super-Eddington accreting black hole \citep[e.g.][]{Madau2026}.

Indeed, the X-ray weakness can be ascribed to either heavy (Compton thick) absorption of the X-rays by the BLR clouds along our line of sight, and/or a cooled corona in the super-Eddington case \citep[][]{maiolino_jwst_2025}. The Balmer absorption features and Balmer breaks can also be ascribed to the dense gas associated with the BLR clouds along our line of sight \citep{juodzbalis_jades_2024,inayoshi_extremely_2025}.

The red optical colours of LRDs can simply be due to reddening by dust along the line of sight. This scenario was initially discarded for LRDs because of initial claims of absence of dust emission in LRDs, however recent studies have shown that most LRDs have clear hot dust emission, i.e. in the vicinity of the AGN, seen in the mid-IR, and with implied dust masses that are far too low (a few tens $M_\odot$)  to be seen with ALMA \citep[e.g.][]{delvecchio_active_2025,lin_discovery_2025,ji_lord_2025,Perez-gonzalez_2026,juodzbalis_jades_2024,Brazzini_rosettas_2026,Madau2026,pacucci_little_2026}. Specifically, \citet{Madau2026} and \citet{pacucci_little_2026} have shown that the spectral shape of LRDs can be easily fit with a dust-reddened, AGN-like powerlaw. Alternatively, the optical continuum may actually be associated with thermal emission at $\sim~5000~K$; however, even in this case, there is no need to invoke a pseudo-atmosphere/cocoon -- the thermal emission can naturally arise from the dense, thermalised BLR clouds \citep{Baldwin2004}

Summarising, the more general spectral properties of LRDs and LBDs can be explained with known physical processes associated with the BLR and BH accretion. A detailed treatment of these spectral properties is beyond the scope of this paper.  

\subsection{Effect of best-fit model on the derived physical properties}\label{s.derived_property}

One of the major implications of the \escat model is its impact on the derived physical properties, especially the black hole masses (\MBH). We use the individual BLR profile fits and their interpretation to constrain the black hole mass using single-epoch virial relations. We employ the virial calibration from  \citet{reines_relations_2015}:
\begin{multline}
\label{eq:virial_mass}
    \log{\frac{M_{\rm BH}}{M_{\odot}}} = \\
    6.60 + 0.47\log{\left(\frac{L_{H\alpha}}{10^{42}\ {\rm erg\,s^{-1}}}\right)+2.06\log\left(\frac{{\rm FWHM}_{H\alpha}}{1000~{\rm km\,s^{-1}}}\right)}, 
\end{multline}
where $L_{H\alpha}$ is the luminosity of the broad \Halpha line and ${\rm FWHM}_{H\alpha}$ its width of the line profile. We estimated the bolometric luminosities of our AGN following the calibrations of \cite{stern_type_2012}, which yield $L_{bol} = 130L_{H\alpha}$. In the electron scattering scenario, the \MBH is estimated by taking the width of the putative intrinsic broad line and the total luminosity of the broad line. We present the estimated \MBH and bolometric luminosities in Table~\ref{tab.prop}.

\begin{figure*}
    \centering
    \includegraphics[width=0.7\paperwidth]{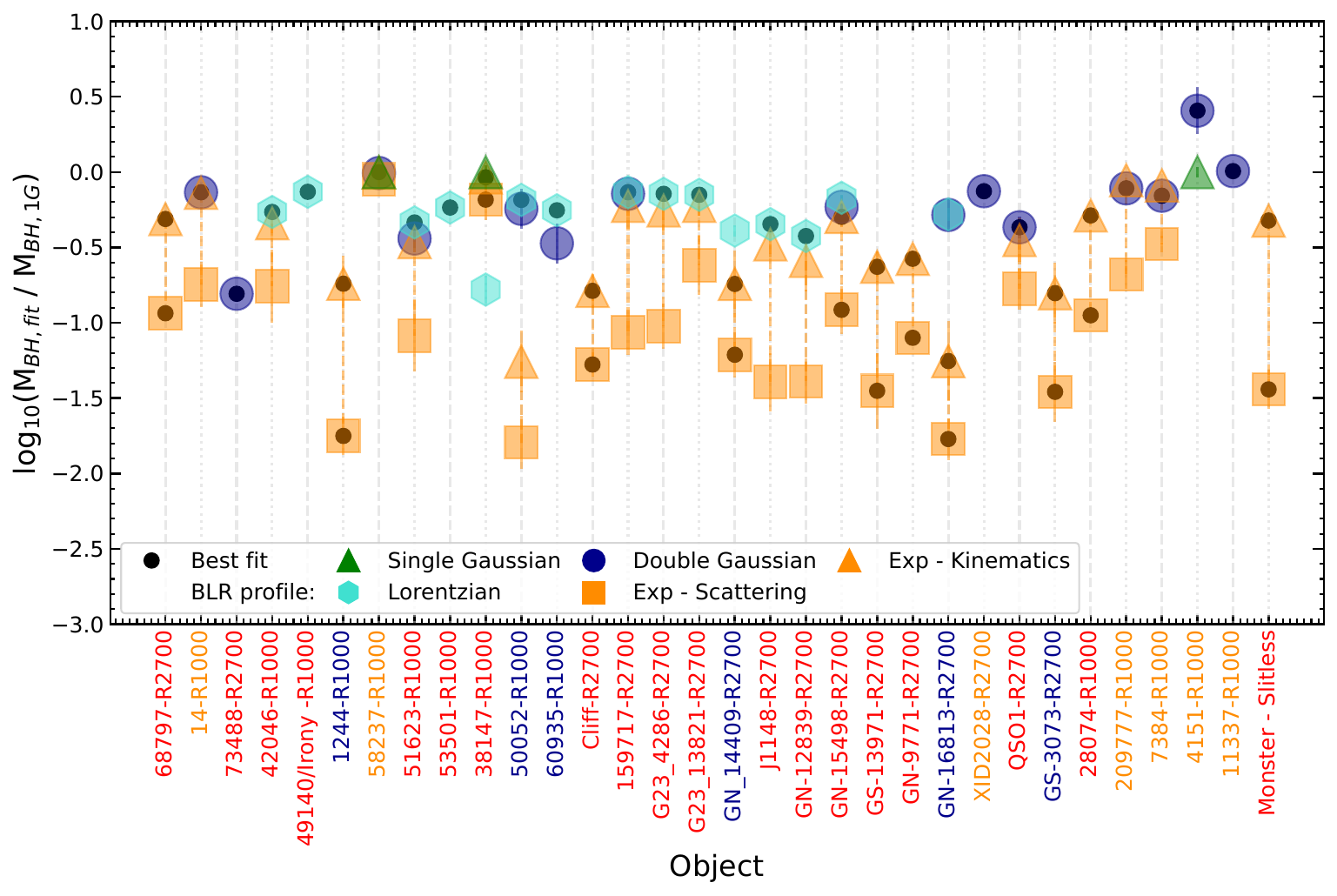}
    \caption{Difference in estimated \MBH from accepted BLR profiles (all with $\delta$BIC$<10$ of the best fit) from a single Gaussian fit. The colour of the labels shows the type of the target: LRD (red), LBD (blue) and X-ray AGN detected (orange). For \escat profile, we show both \escat and BLR kinematics interpretation.}
    \label{fig.MBHcomp}
\end{figure*}

\begin{figure*}
    \centering
    \includegraphics[width=0.7\paperwidth]{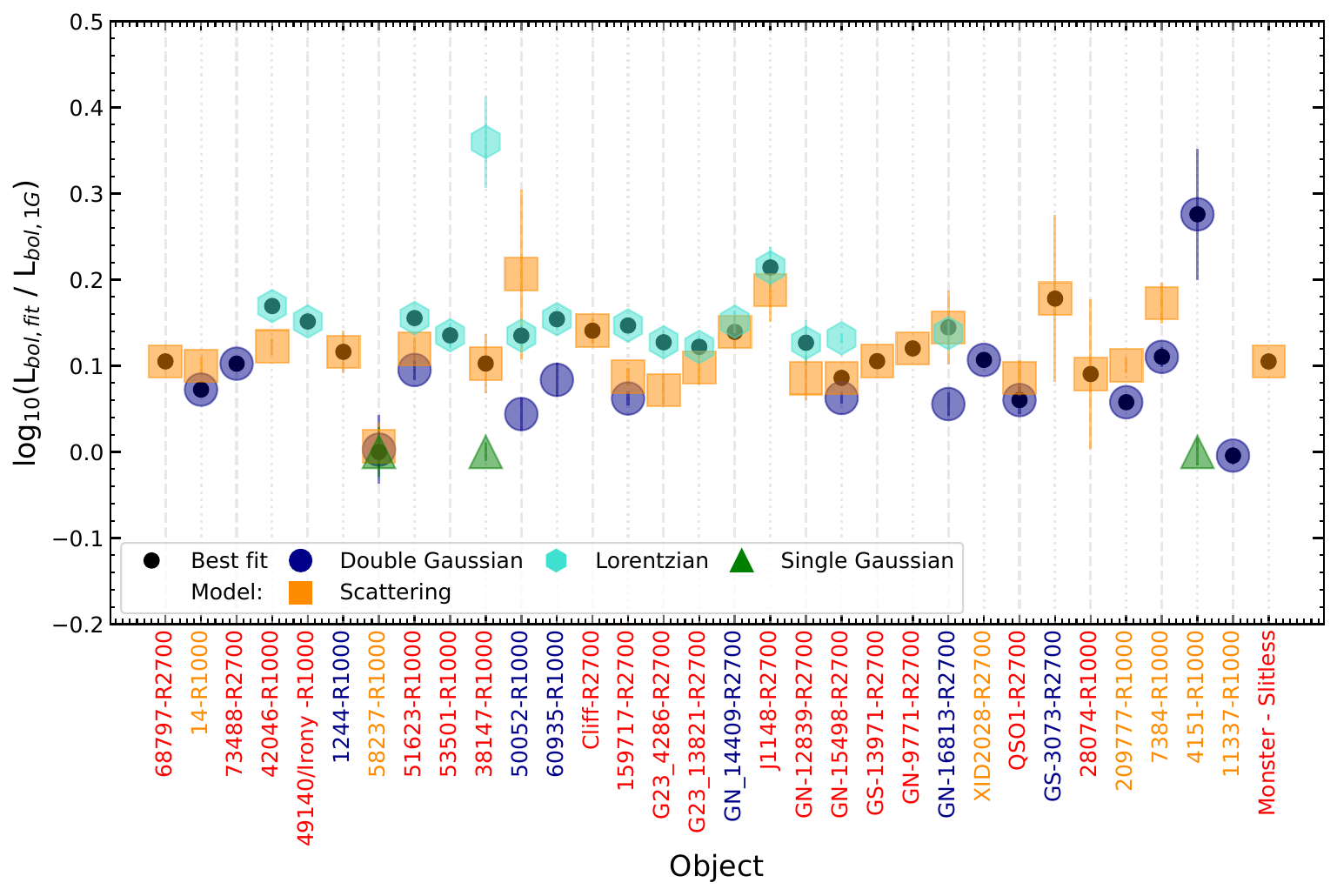}
    \caption{Difference in estimated bolometric luminosity from accepted BLR profiles (within $\delta$BIC$<10$ of the best fit) from a single Gaussian fit. The colour of the labels shows the type of the target: LRD (red), LBD (blue) and X-ray AGN detected (orange).}
    \label{fig.Lbol_comp}
\end{figure*}

In Figs.~\ref{fig.MBHcomp} \& \ref{fig.Lbol_comp}, we compare the \MBH and bolometric luminosities derived from all accepted models to a `simple fit', corresponding to a single Gaussian profile often used in the literature for low-SNR observations, particularly for low-luminosity targets. Furthermore, based on our findings regarding the exponential profile in \S~\ref{s.exponential_creation}, we show both interpretations of the exponential profile -- electron scattering and the kinematics of BLR clouds around a black hole. 

We note that there is a significant discussion in the literature about the use of single epoch virial calibrations and their application to the \jwst AGN \citep[e.g.][]{greene_what_2026}. However, recent results by the \cite{gravity_collaboration_spatially_2025}, which measured the BLR size via interferometric data at z$\sim$4, found the discrepancy between their measurement and \Halpha is "only" a factor of 2.5, within the 0.3~dex calibration uncertainty. This uncertainty is further verified by recent NIRSpec work by \cite{bertemes_jwst_2025}, testing different BH mass constraints for a bright QSO. As mentioned, \citet{juodzbalis_direct_2025} found that the single-epoch measurement of the black hole mass is in excellent agreement with the direct, dynamical measurement. This work stresses the need to investigate more sources with IFS to acquire a direct \MBH measurement.
However, we acknowledge that \MBH measurements at high-$z$ remain a heavily discussed topic, with arguments against the virial calibrations due to the super-Eddington accretion, going in either direction, with studies claiming that virial estimations heavily underestimate them \citep[e.g.][]{marconi_effect_2008, marconi_observed_2009}, while others claiming that they overestimate them  \citep[e.g.][]{lambrides_case_2024, lupi_size_2024}, and also discussing the effects of large covering factors of the BLR clouds \citep{maiolino_jwst_2024}. Similarly, we do not correct for dust obscuration in this work, as these corrections can be very uncertain.

The results of this comparison are presented in Fig.~\ref{fig.MBHcomp}, where the models with a black dot indicate those that resulted in the best fit. In the case of profiles best fit by an exponential, we cannot distinguish between the electron scattering scenario and the BLR stratification scenario, so both of them are marked (and connected with a segment). The comparison shows that the scatter in \MBH for models without \escat interpretation is within 0.5~dex, with the Lorentzian and double Gaussian models agreeing on average within 0.2~dex of each other.

The exponential model with BLR stratification sometimes provides a slightly lower BH mass, as it allows the fit to model part of the narrower emission line profile to attribute more of the low-velocity flux in the line profile to the BLR (see Fig.~A\ref{fig.BLR_best_comp}), which results in a narrower FWHM and hence a lower \MBH. However, in most cases, the exponential model with BLR stratification provides a BH mass very similar to the other models.  
 
Obviously, as already pointed out by other authors (\citetalias{Rusakov_nature_2026}, \citealt{Matthee_2026}), if the exponential wings are interpreted in terms of electron scattering, then the \MBH differences are much more pronounced -- in many cases, the inferred \MBH are lower by about an order of magnitude, and in some cases down to 1.8 dex lower. Furthermore, recent work by \citet{Matthee_2026} argued that the lines are purely based on scattering with no coupling to the \MBH.
However, 1) less than half of the LRDs and LBDs are best fit by an exponential profile, and 2) even in these cases, the interpretation does not need to invoke electron or resonant scattering. As such, there is no significant evidence to claim that the \MBH are indeed 1~dex lower for the entire population of high-$z$ AGN, and hence the black holes at high-$z$ are indeed overmassive.

\subsection{The case of ``X-ray'' AGN}

Here, we discuss the BLR profile of the X-ray AGN, which are overwhelmingly double Gaussian in our sample. We note that our X-ray AGN sample is heavily biased towards objects with strong \OIII outflows. The majority of our X-ray AGN exhibit strong outflows in the \OIIIall emission line (see e.g. \citealt{cresci_bubbles_2023} for description of outflows in XID-2028). We present the \OIIIall fits in Appendix Fig.~A\ref{fig.OIII_spec}. The outflows seen in ionised gas with \OIIIall has an origin at the accretion disc and the BLR \citep[e.g.][]{Wang_2016, Shin_2017}. In this scenario, the double Gaussian profile is likely tracing both the BLR and the launch of the outflow close to the accretion disk \citep{Bon_2009}. This is more extreme in the case of XID 2028, where the \Halpha line is asymmetric with a velocity offset of 250 \kms between the 2 Gaussian Gaussian profiles. Therefore, the prevalence of double Gaussian broad \Halpha profile is most likely driven by the bias towards objects with powerful outflows, rather than a different structure around the black hole. A full analysis of typical AGN from SDSS will be presented in Trefeloni et al. in prep.

\section{Conclusions}\label{s.conclusion}

In this work, we analysed the broad \Halpha profiles of 32 AGN detected with \jwst with high-SNR \Halpha emission, including 19 Little Red Dots, six Little Blue Dots and seven X-ray detected AGN. We fitted single Gaussian, double Gaussian, Lorentzian and \escat exponential models to the broad \Halpha component to determine the most common profile in these classes of objects, and to investigate any evidence of \escat around the BLR of high-$z$ AGN. Our main conclusions are:

\begin{enumerate}
    \item A single Gaussian model is not sufficient to describe the broad \Halpha 
    profile in these high-SNR observations, as already demonstrated in the literature with \jwst and in pre-\jwst studies.

    \item We have found no evidence that \escats exponential models are statistically preferred over Lorentzian or double Gaussian models in LRDs, LBDs or X-ray AGN (see \S~\ref{s.profiles} and Figs.~\ref{fig.model_comp}\&~\ref{fig.frac_overview}). For 48\% of the objects, there are two or more models within $\delta$BIC$<10$, showing there is no statistical evidence to distinguish between the different profiles.
    There is a marginal indication that Lorentzian profiles are preferred in LRDs, while exponential profiles are preferred in LBDs; this opposite of what was claimed by previous works, and clearly indicates that exponential profiles are certainly not a prerogative of LRDs. Finally, we rule out the Lorentzian and \escat profiles as the preferred broad \Halpha profile for the X-ray AGN in our sample.

\item Although not the focus of this paper, we have discussed that various other spectral properties of LRDs and LBDs (X-ray weakness, Balmer absorption features and Balmer break, as well as optical spectral slope) can be explained in the scenario in which the exponential wings are simply due to the stratified BLR.  

    \item We have demonstrated that a sum of Gaussian profiles can produce an exponential profile with no need to invoke electron or resonant scattering. This shows that a scattering scenario is not the only explanation for the exponential line profiles seen in some AGN. The sum of Gaussian profiles can be reasonably interpreted in terms of contributions of different layers of BLR clouds in virial motions around the black hole. Given that BLRs are observationally known to be stratified, this interpretation is probably more plausible than invoking new exotic scenarios.

    \item We have investigated the approach of stacking to determine the average broad line profile of different AGN populations in \S~\ref{s.stacking}. We have created a mock stack of 20 random profiles (single Gaussian, double Gaussian and Lorentzian) with peak SNR$=20$, intentionally excluding any exponential profile. We have shown that stacking of as few as 20 spectra always produces exponential wings (see Fig.~\ref{fig.stack_mock}), despite no exponential profiles being present in the original mock data. We conclude that stacking analysis should not be used to investigate broad line profiles, as it will always produce spurious exponential results. In connection to the previous point, we note that the same consideration applies to individual objects -- exponential profiles seen in individual objects can result from the superposition of various kinematic components of a normal BLR, none of which are exponential.

    \item We have not found statistical evidence that objects with stronger Balmer break prefer exponential profiles (see Fig.~\ref{fig.break}); however, we have discussed that, should future works find such a correlation, this would fit well in a BLR stratification scenario (see \S~\ref{s.profiles}).

    \item We have investigated the effect of the choice of BLR model on the derived physical properties in \S~\ref{s.derived_property}. For models that do not invoke \escat, the scatter in \MBH is within 0.2~dex (see Fig.~\ref{fig.MBHcomp}). The bolometric luminosity is even more robust, with an average model-to-model scatter of only 0.1~dex (see Fig.~\ref{fig.Lbol_comp}). The \escat model systematically shifts flux from the wings into a narrow intrinsic Gaussian, yielding FWHM values -and hence \MBH estimates that can differ by more than 1~dex from those of the other accepted models in individual cases. However, as we have demonstrated in \S~\ref{s.profiles} \& \ref{s.exponential_creation}, the exponential profile can be reproduced with more physical scenarios, such as a stratified BLR, without having to invoke \escat in new exotic scenarios. We therefore conclude that the presence of exponential wings does not imply that \MBH estimates from single-epoch virial calibrations are
    systematically biased.

    \item Our findings provide a simple solution to the tensions that had emerged between the \escat model, the direct black hole mass measurements, and other inconsistencies with multiple line profiles. Furthermore, we have shown that there is no widespread evidence for electron scattering being the dominant driver of the broad \Halpha line profile, and we do not require a large reservoir of ionised gas surrounding the black hole (so-called cocoon). We note that electron scattering can occur on a small scale in the broad line region, as seen in local AGN (see \S~\ref{s.exponential_creation}). 
    
\end{enumerate}

Our work shows that the broad line profiles of high-z AGN are as complex of those of pre-\jwst AGN. There is certainly the need to expand the sample of LBDs and LRDs with direct measurement of black hole masses from gas or stellar kinematics at high redshift, to verify that the broad line profiles are indeed tracing the black hole mass and hence arise from  typical broad line region. This would require a large sample of LBDs and LRDs observed with deep high resolution integral field spectroscopy.

\section*{Acknowledgements}
This work is based on observations made with the NASA/ESA/CSA James Webb Space Telescope. The data were obtained from the Mikulski Archive for Space Telescopes at the Space Telescope Science Institute, which is operated by the Association of Universities for Research in Astronomy, Inc., under NASA contract NAS 5-03127 for JWST. These observations are associated with program 1180, 1181, 1210, 1211, 1212, 1214, 1812, 1287, 2674, 3215, 3516, 4106, 4233,  5015, 5664, 9433.
JS, RM, FDE, XJ, GCJ, A.H and L.R.I. acknowledge support by the Science and Technology Facilities Council (STFC), ERC Advanced Grant 695671 ``QUENCH'' and the UKRI Frontier Research grant RISEandFALL. RM also acknowledges funding from a research professorship from the Royal Society.
S.C acknowledges support from the European Union (ERC, WINGS,101040227)
MP acknowledges support through the grants PID2021-127718NB-I00, PID2024-159902NA-I00, and RYC2023-044853-I, funded by the Spain Ministry of Science and Innovation/State Agency of Research MCIN/AEI/10.13039/501100011033 and El Fondo Social Europeo Plus FSE+.
H\"U and G.M. acknowledge funding by the European Union (ERC APEX, 101164796). Views and opinions expressed are however those of the authors only and do not necessarily reflect those of the European Union or the European Research Council Executive Agency.
GC acknowledges support from the INAF GO grant 2024 “A JWST/MIRI MIRACLE: MidIR Activity of Circumnuclear Line Emission”
\section*{Data Availability}

The datasets were derived from sources in the public domain: JWST/NIRSpec MSA and JWST/NIRCam data from MAST portal -- \url{https://mast.stsci.edu/portal/Mashup/Clients/Mast/Portal.html} and from Dawn JWST archive -- \url{https://dawn-cph.github.io/dja/}.


\bibliographystyle{mnras}
\bibliography{mybib,mybib_add}


\appendix

\section*{Appendix A: Results of the fitting}\label{s.fits}

We show our fits for each of the profiles in Figs.~\ref{fig.a1_obj}, \ref{fig.a2_obj} and \ref{fig.a3_obj}. Furthermore, for comparison, we also show the broad line profiles of each fit within $\delta$BIC$<$10 of the best fit. s

\begin{figure*}
    \centering
    \includegraphics[width=0.42\paperwidth]{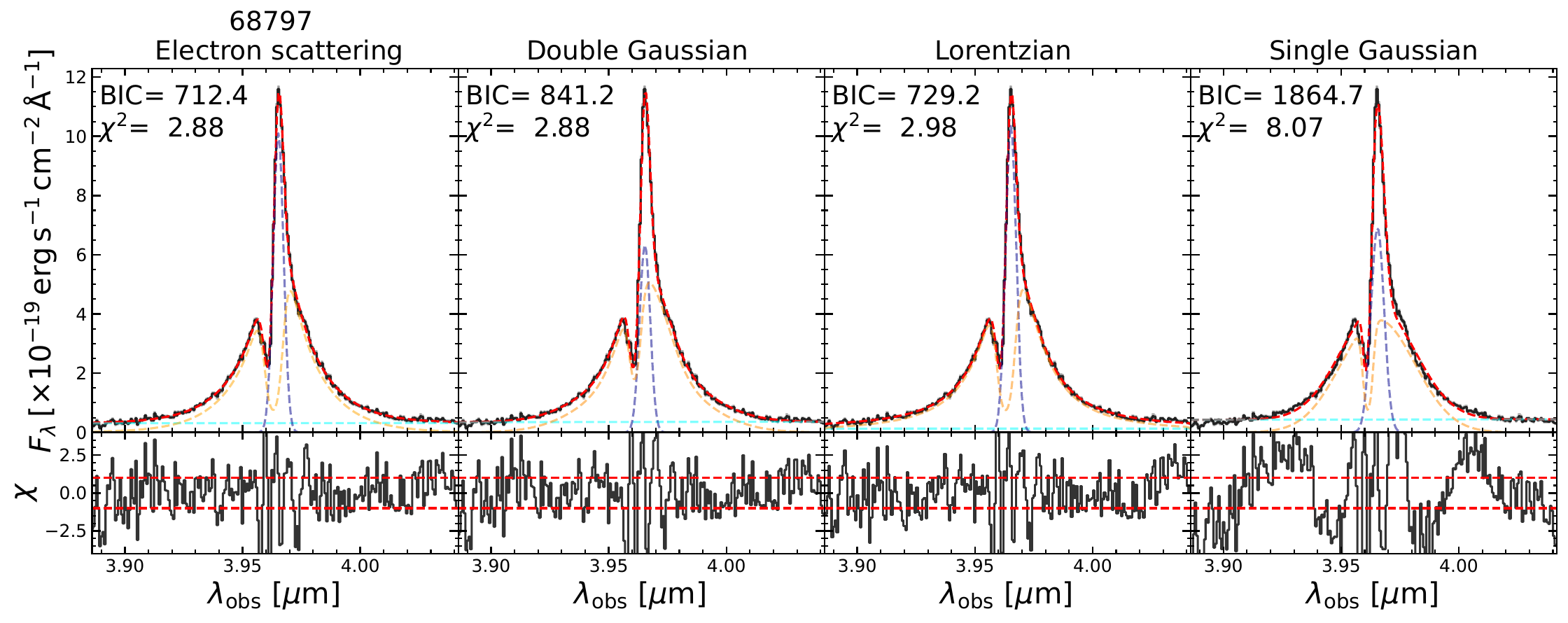}
    \includegraphics[width=0.42\paperwidth]{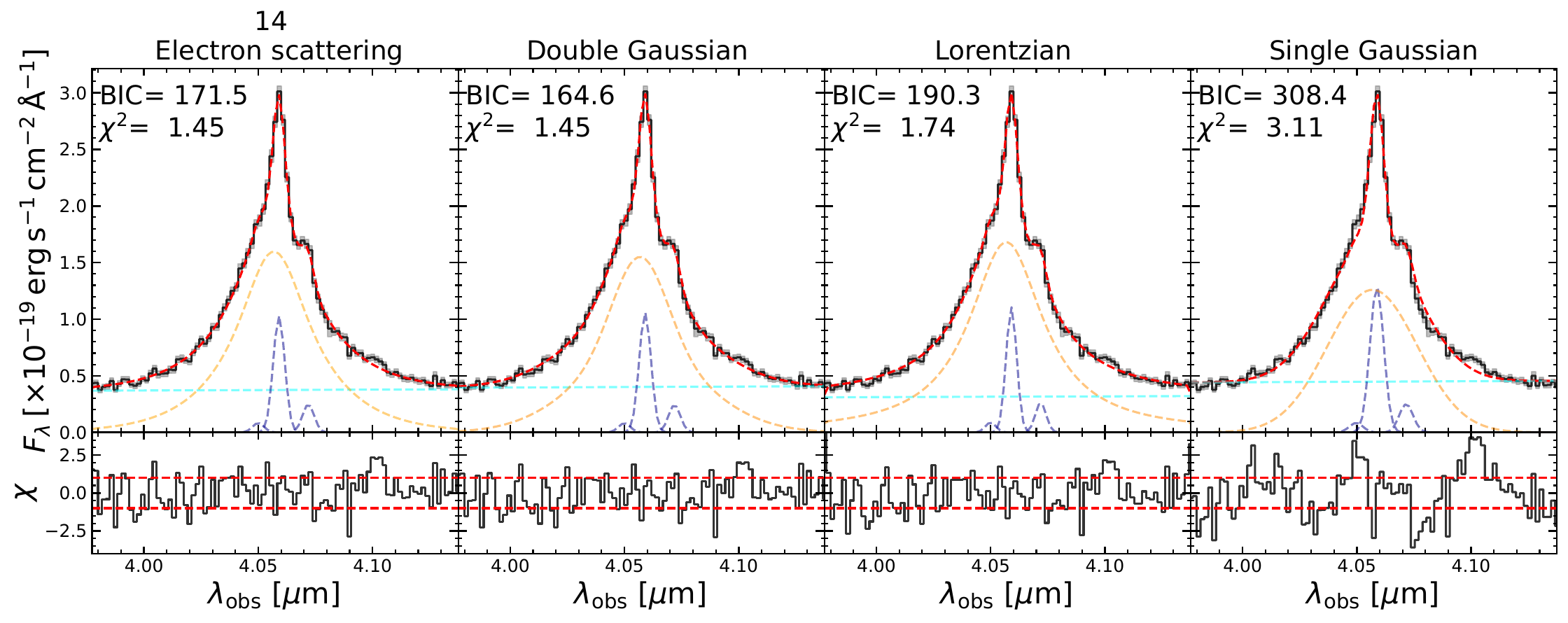}
    \includegraphics[width=0.42\paperwidth]{Graphs/objects/73488_comp.pdf}
    \includegraphics[width=0.42\paperwidth]{Graphs/objects/42046_comp.pdf} 
    \includegraphics[width=0.42\paperwidth]{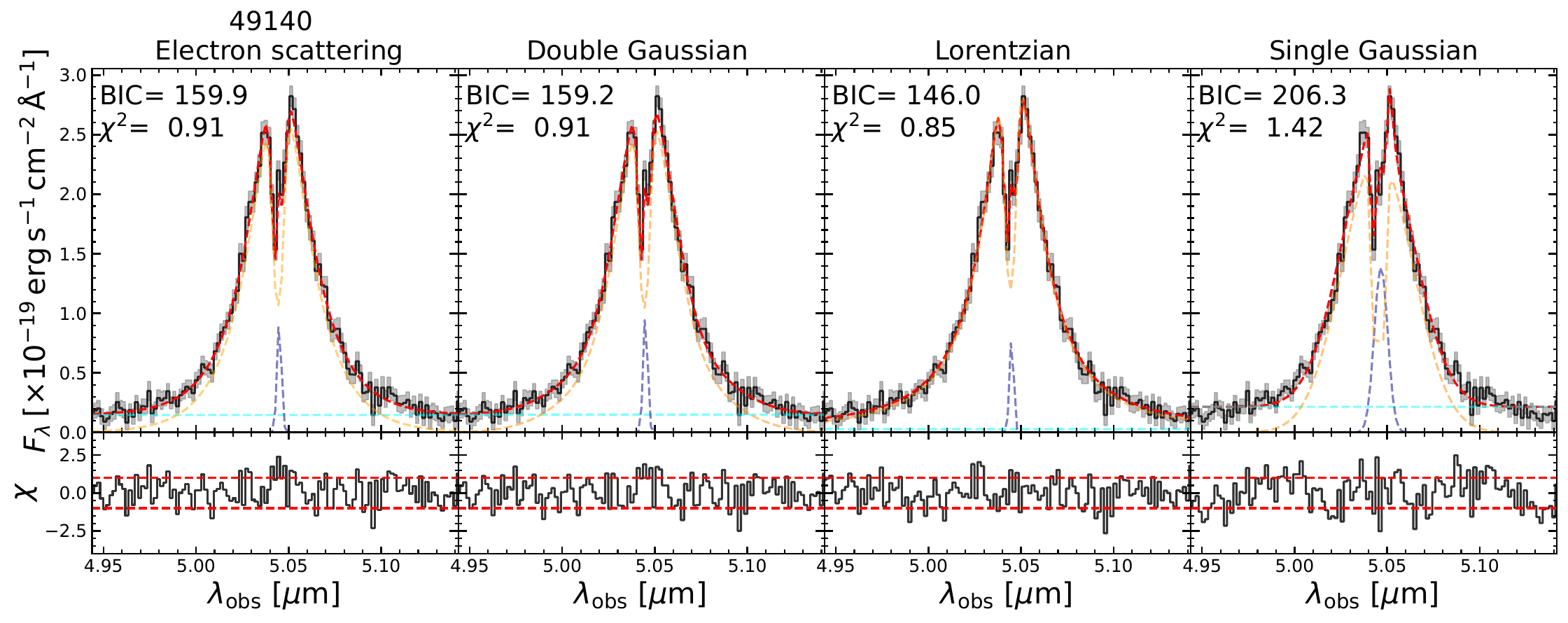}
    \includegraphics[width=0.42\paperwidth]{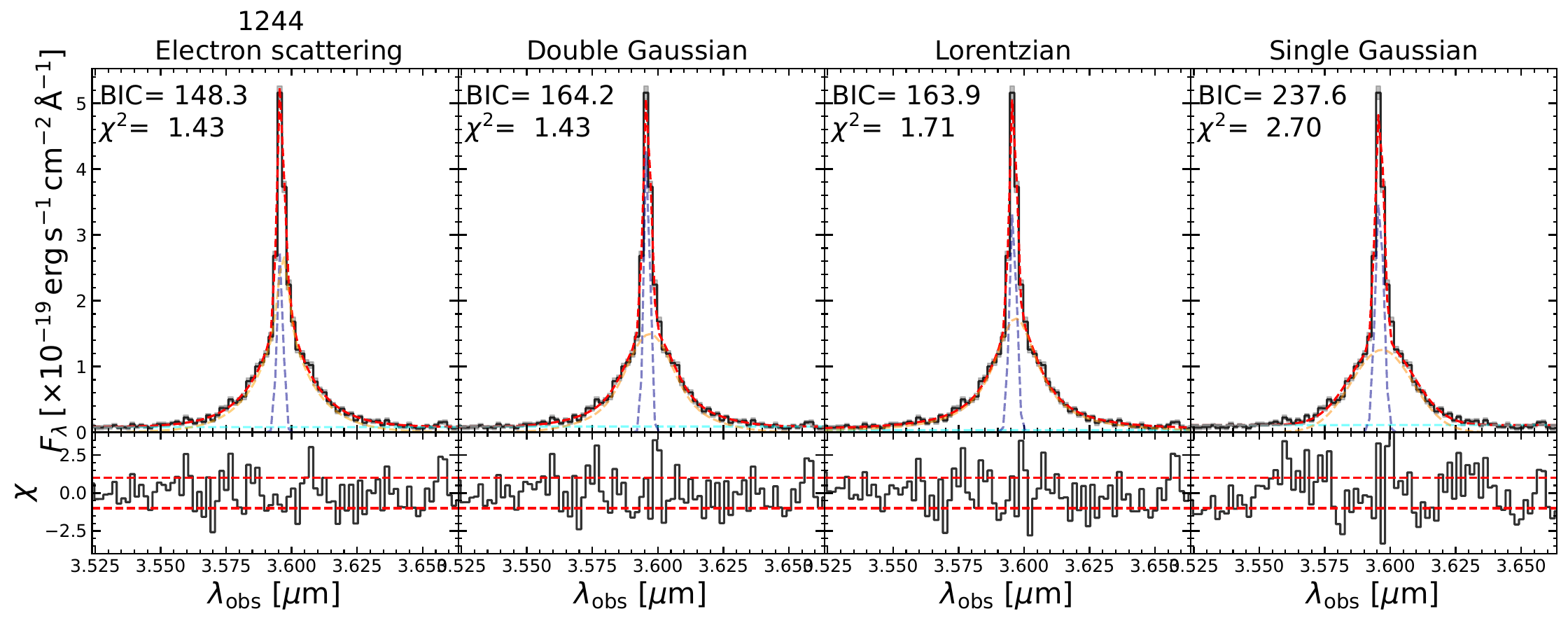}
    \includegraphics[width=0.42\paperwidth]{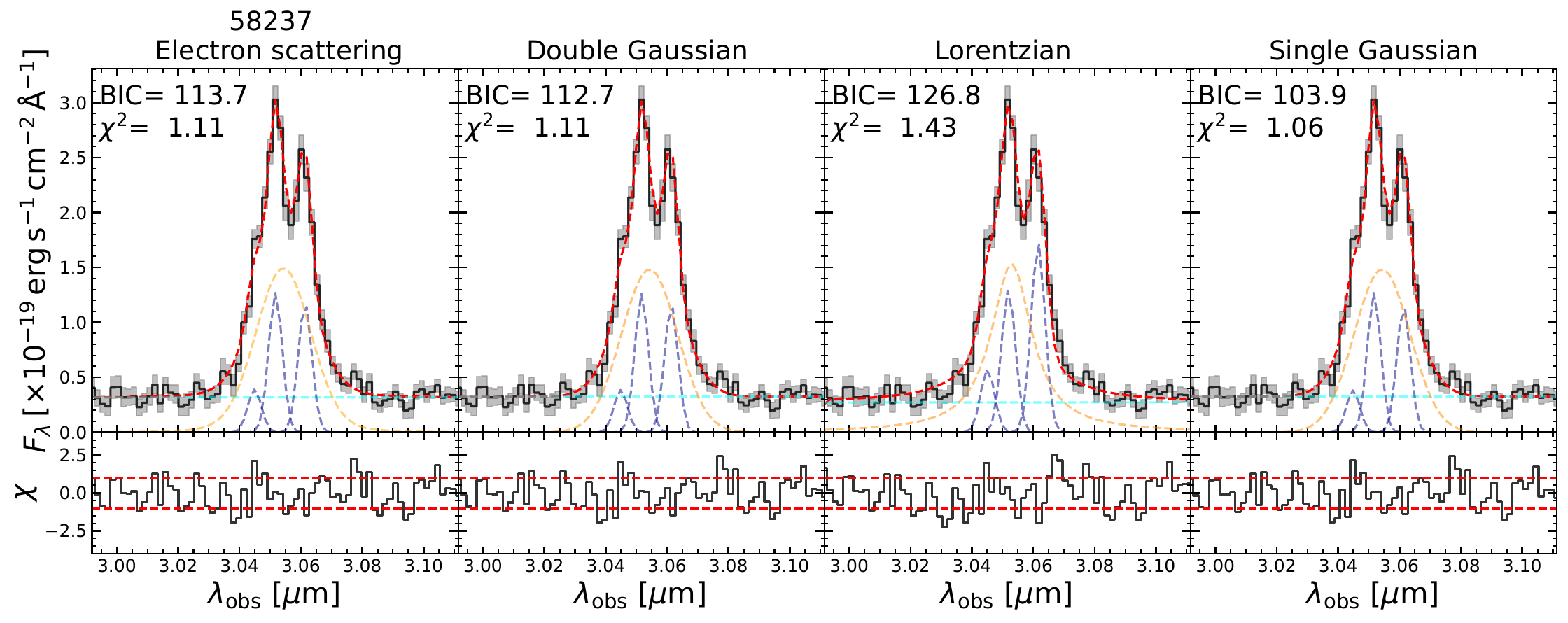}
    \includegraphics[width=0.42\paperwidth]{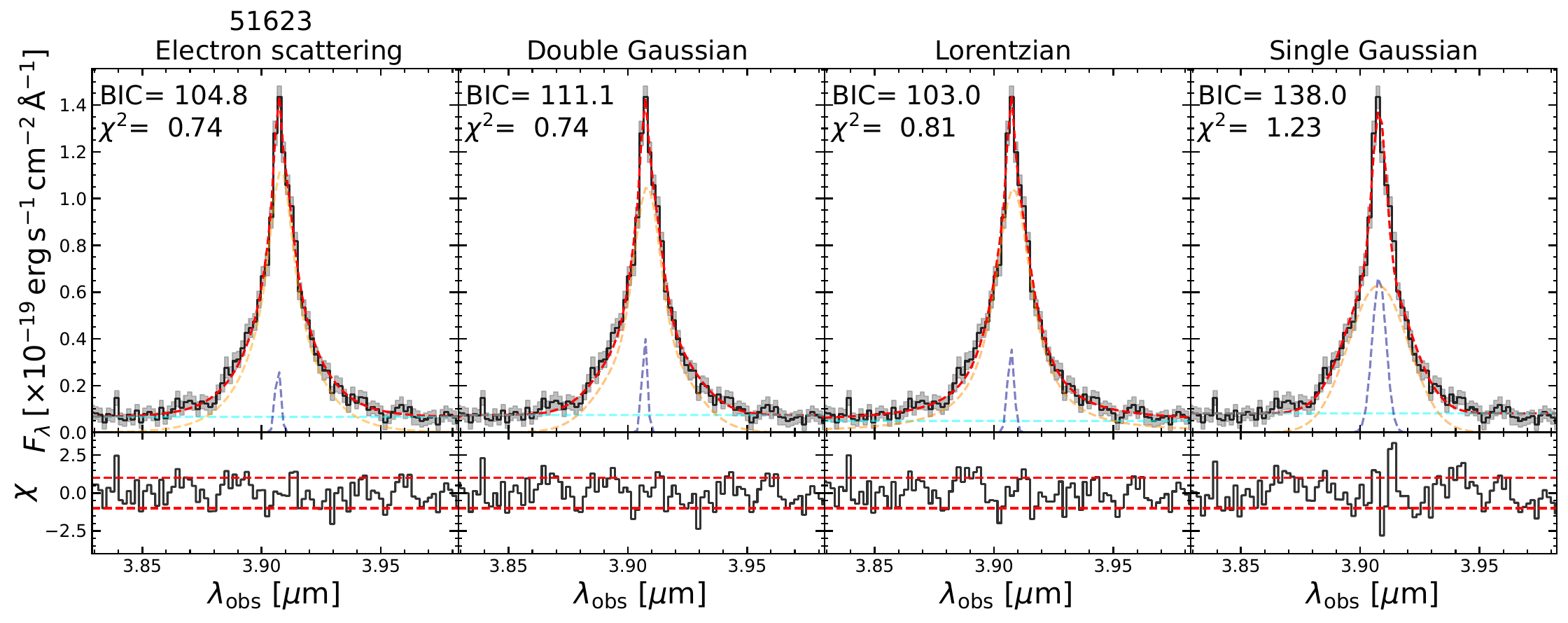}
    \includegraphics[width=0.42\paperwidth]{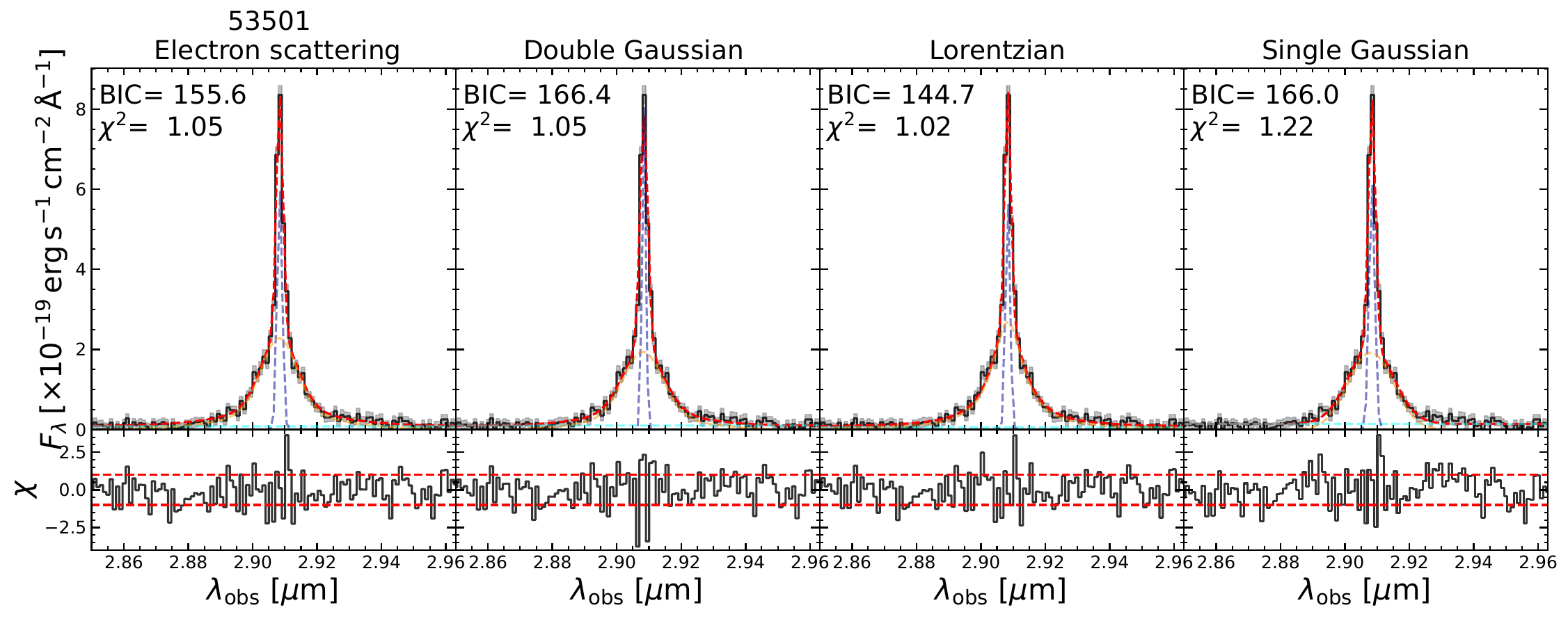}
    \includegraphics[width=0.42\paperwidth]{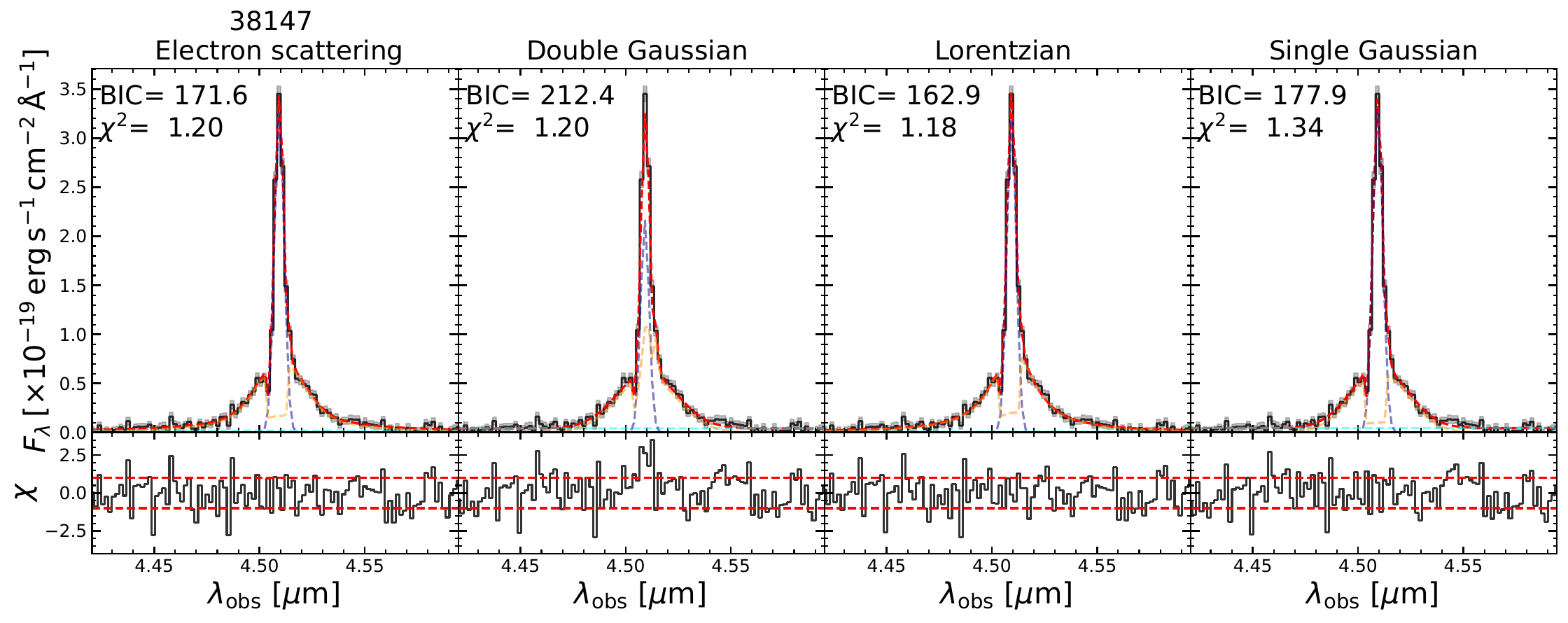}
    \includegraphics[width=0.42\paperwidth]{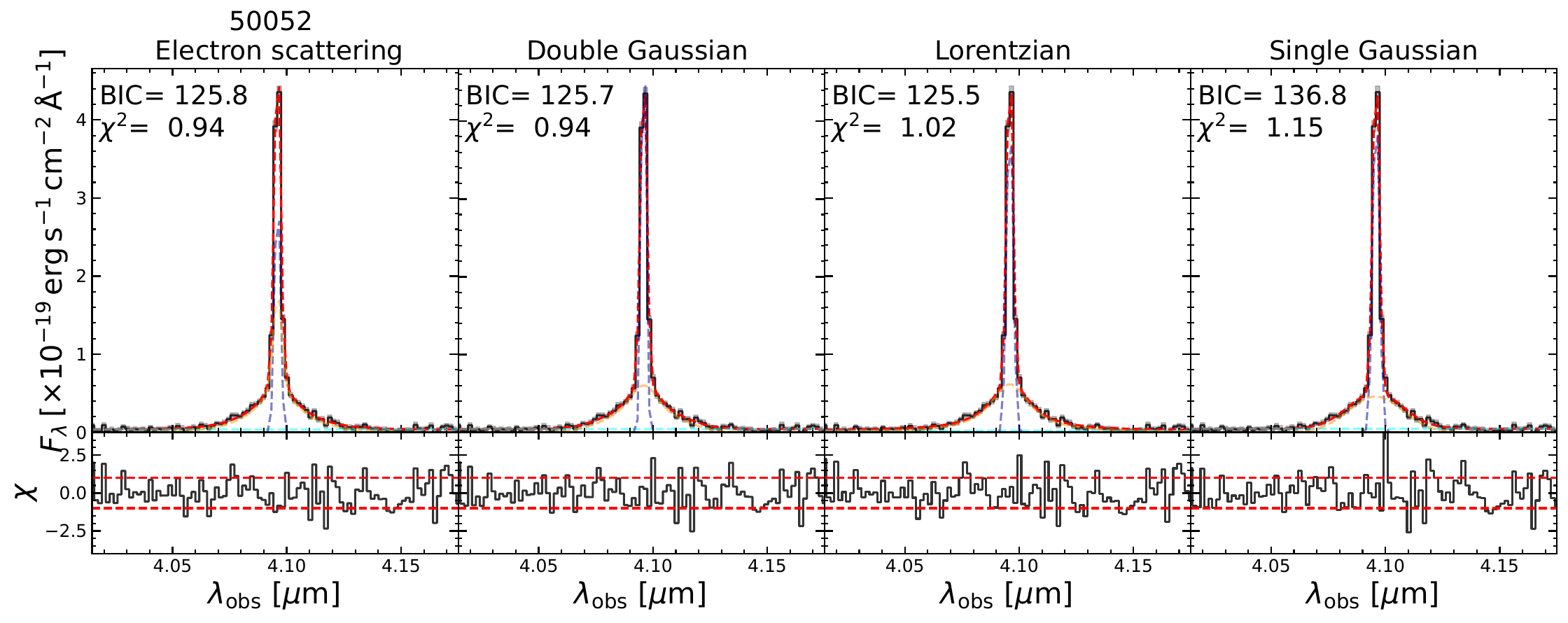}
    \includegraphics[width=0.42\paperwidth]{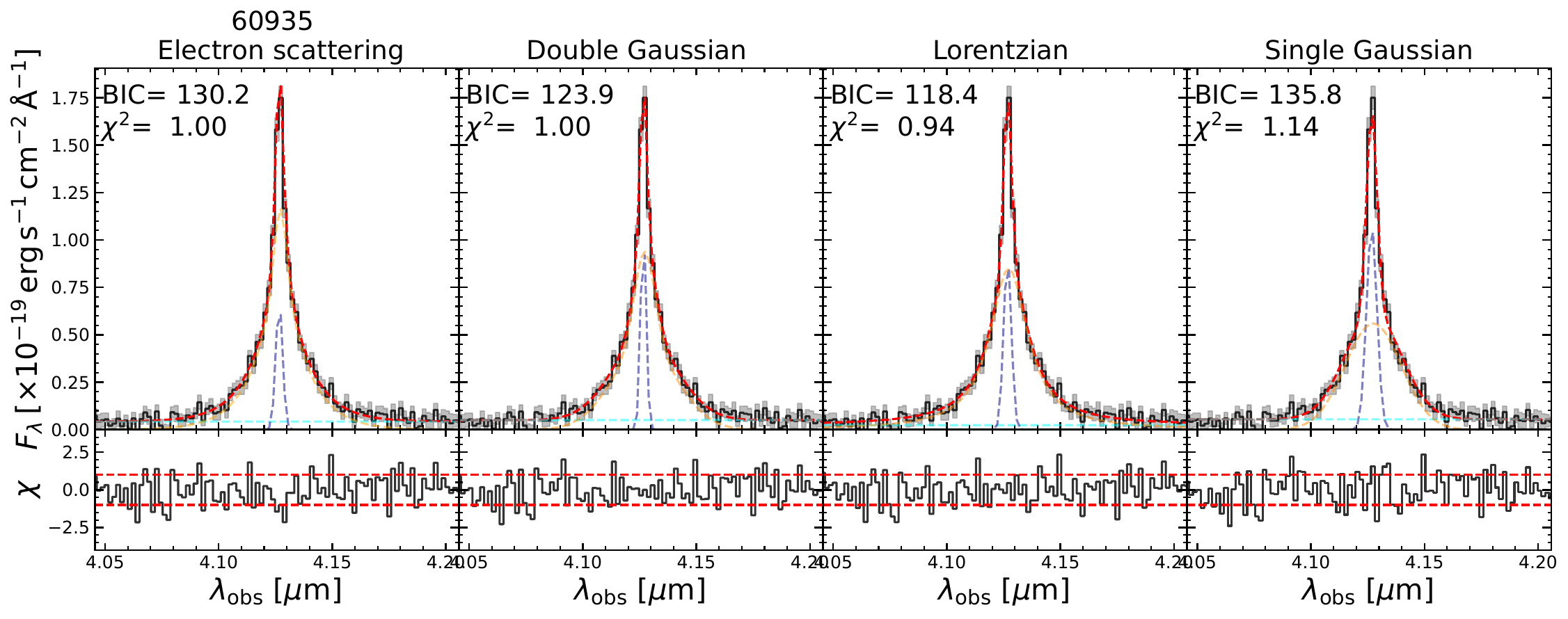}
    \caption{Example of the model fitting used in this work. From left to right: Electron scattering model, double Gaussian model, Lorentzian model and single Gaussian model. The data are shown as a black solid line with the best-fit model shown as a red dashed line. The BLR model is shown as an orange dashed line, including any absorption, and the narrow \Halpha component as a blue dashed line. We show the $\chi$ residuals in the bottom panel for each model.}
    \label{fig.a1_obj}
\end{figure*}

\begin{figure*}
    \centering
    \includegraphics[width=0.42\paperwidth]{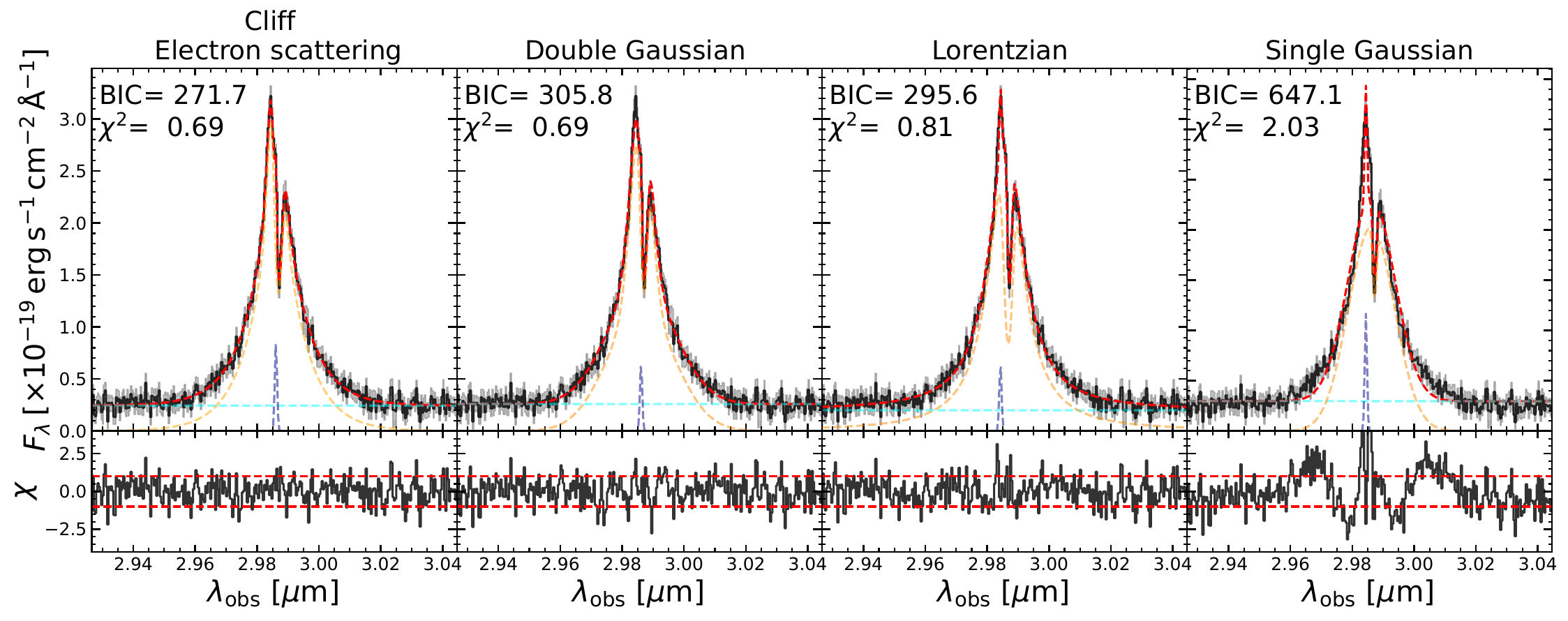}
    \includegraphics[width=0.42\paperwidth]{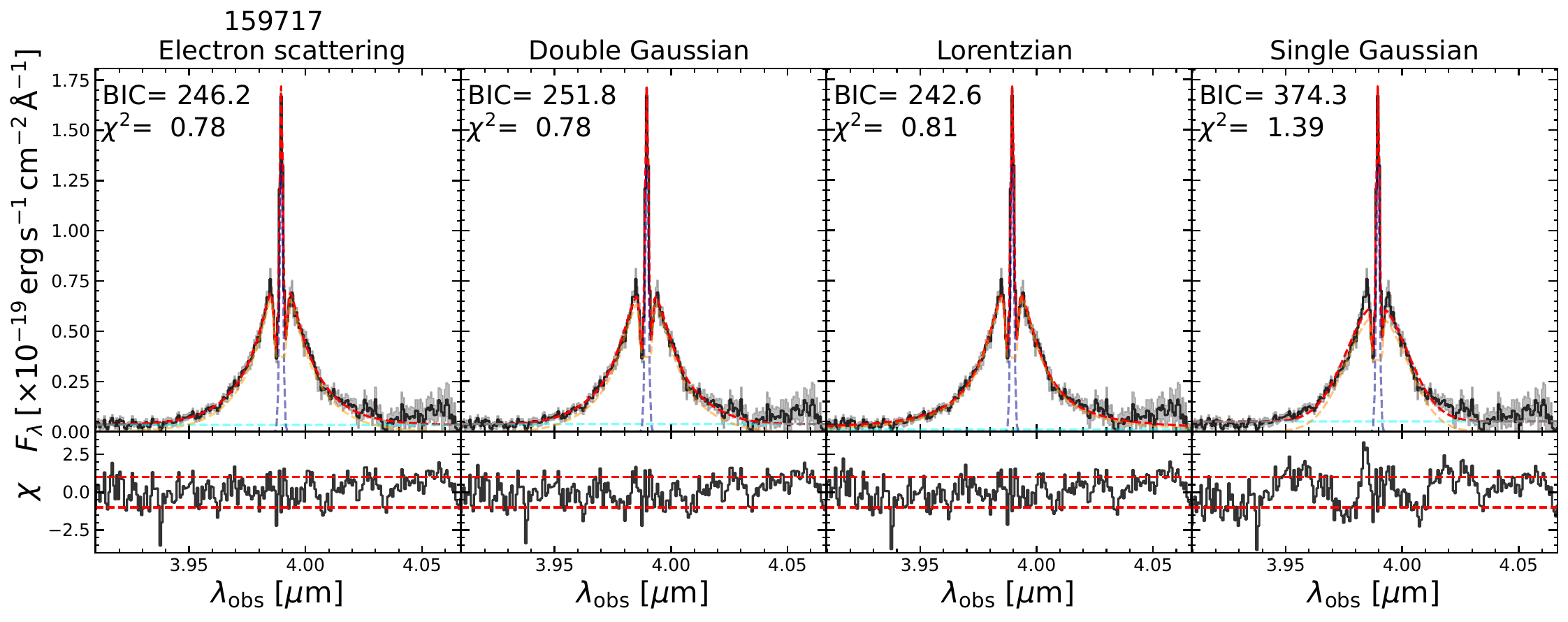}
    \includegraphics[width=0.42\paperwidth]{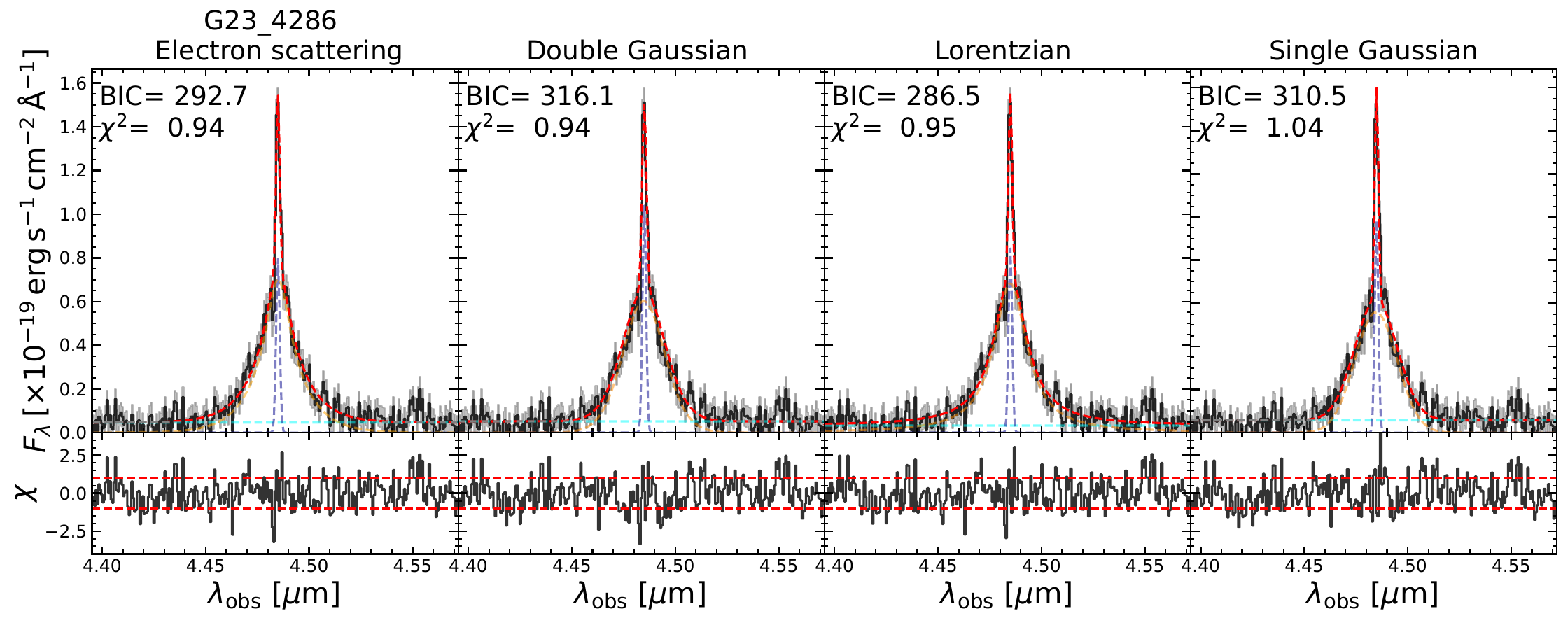}
    \includegraphics[width=0.42\paperwidth]{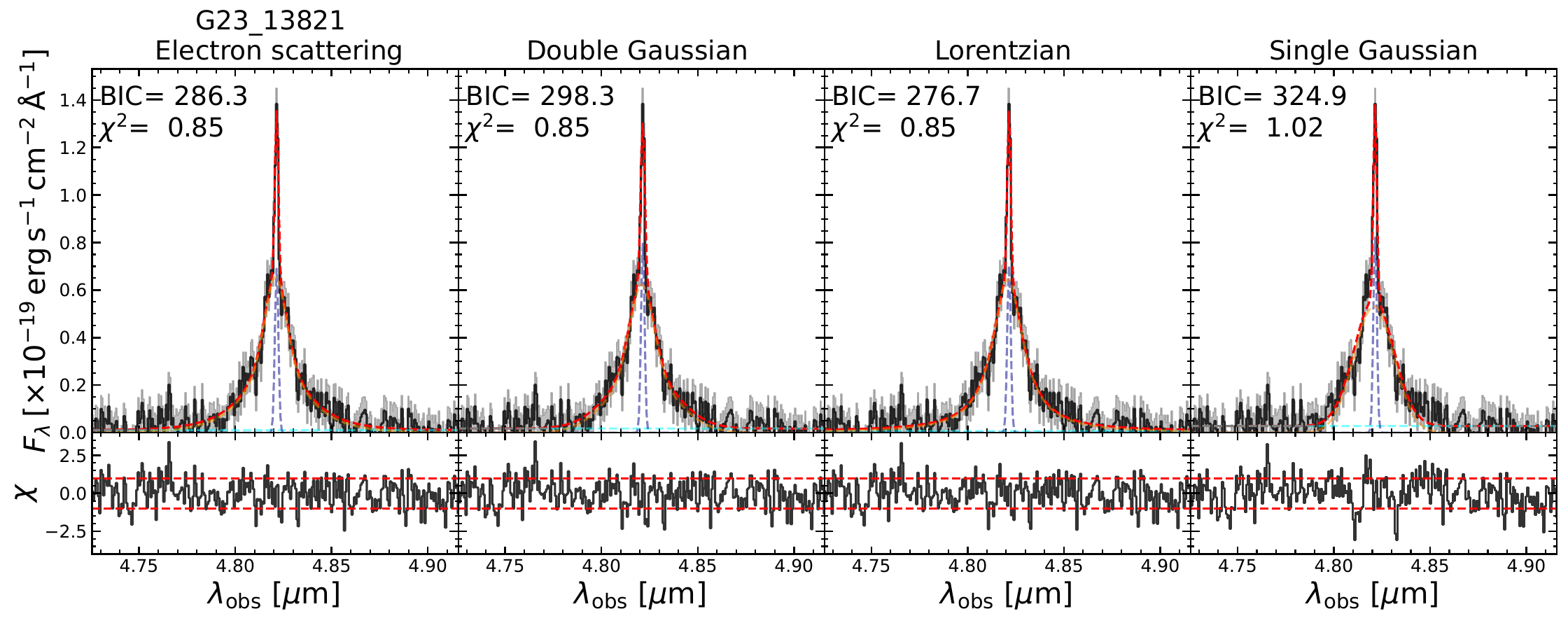}
    \includegraphics[width=0.42\paperwidth]{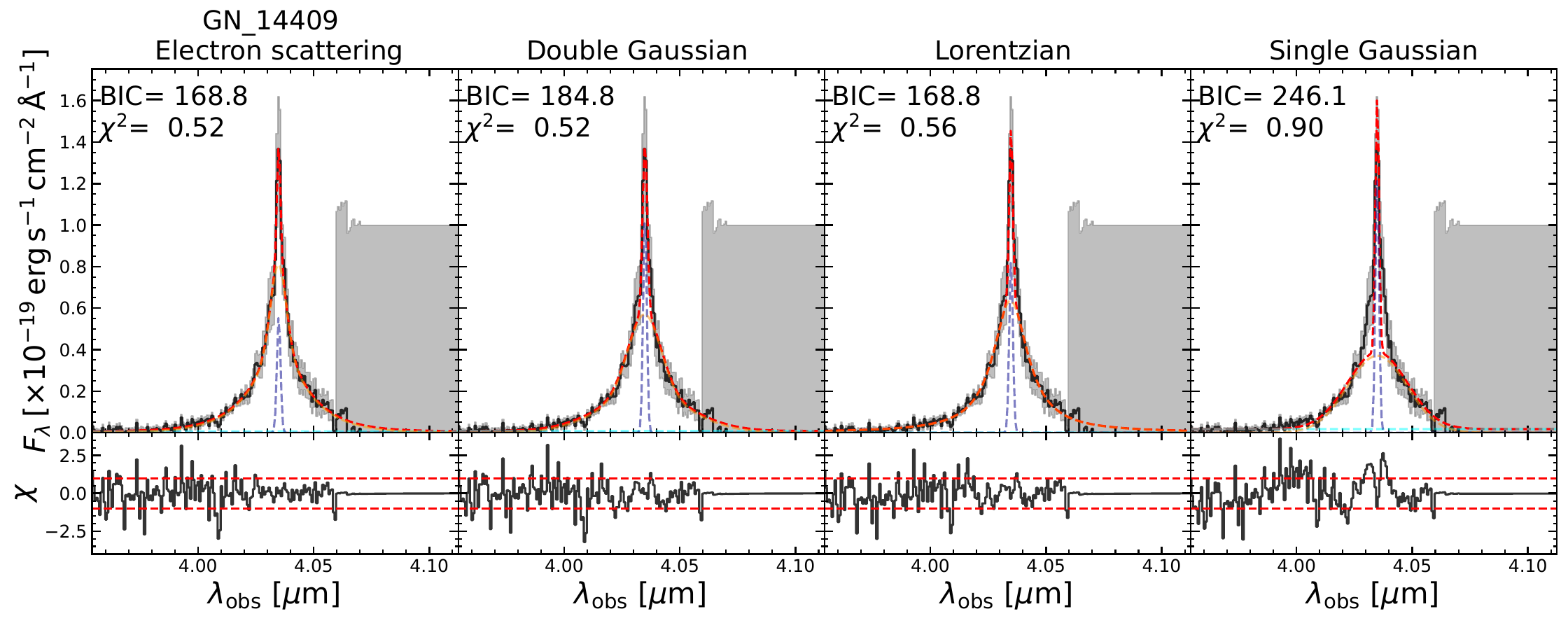}
    \includegraphics[width=0.42\paperwidth]{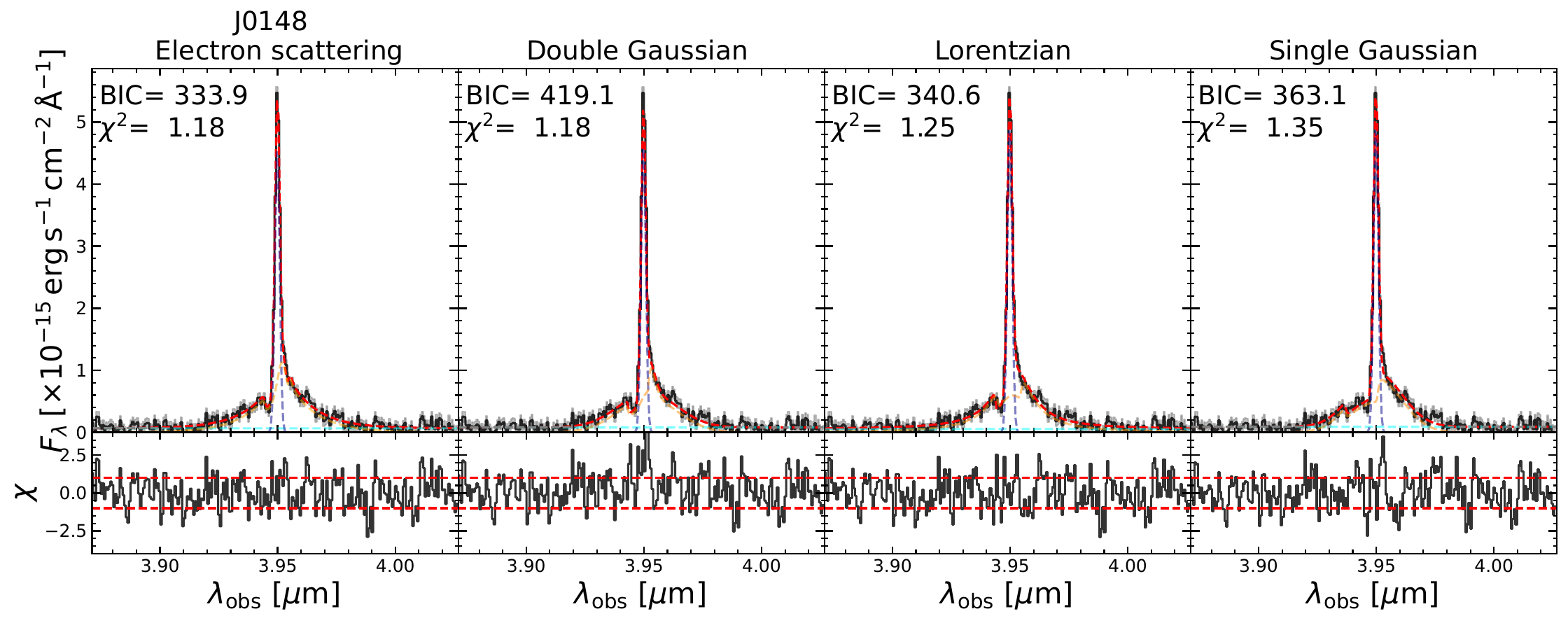} 
    \includegraphics[width=0.42\paperwidth]{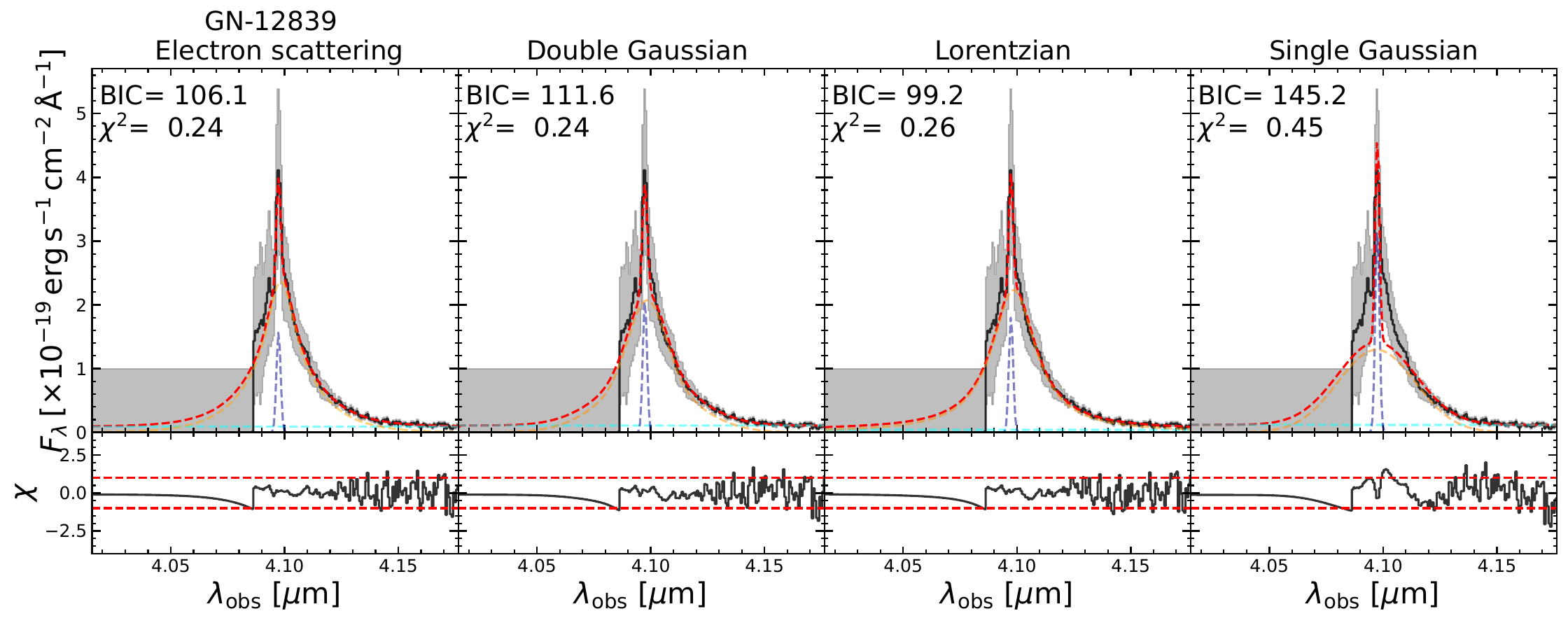}
    \includegraphics[width=0.42\paperwidth]{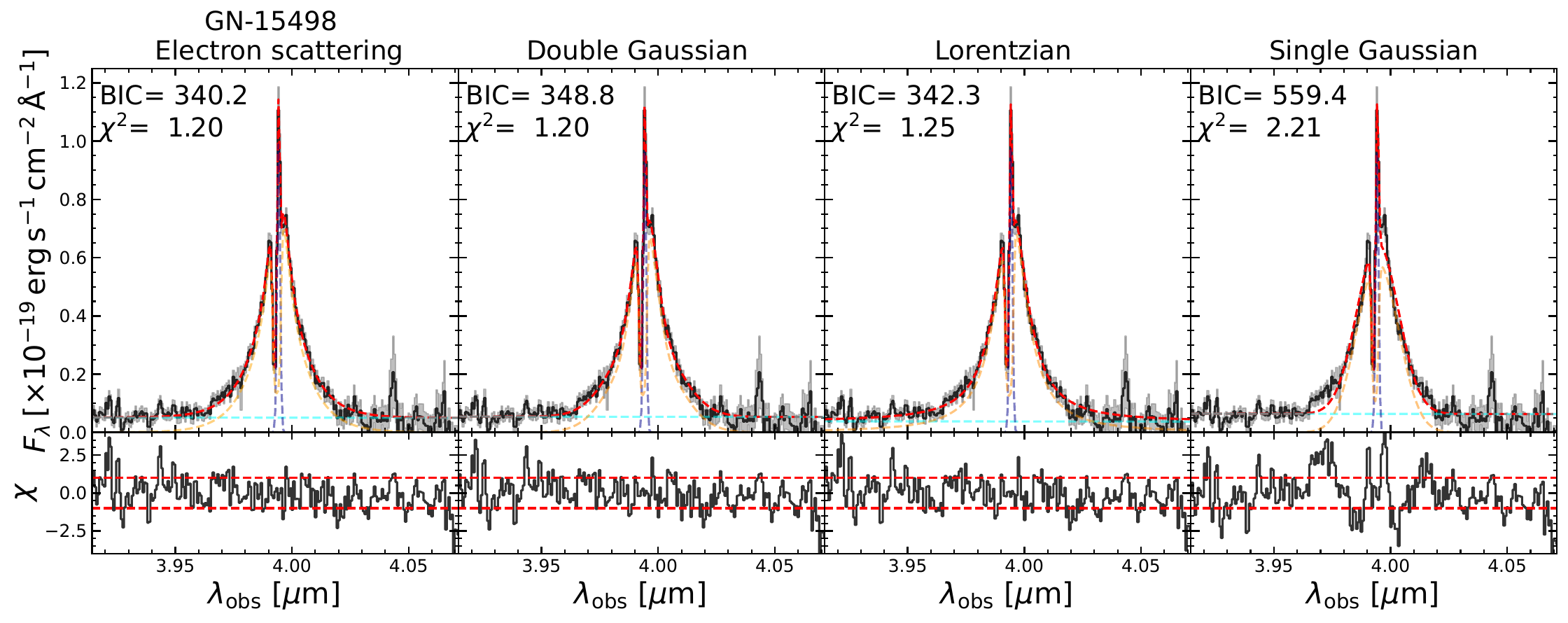}
    \includegraphics[width=0.42\paperwidth]{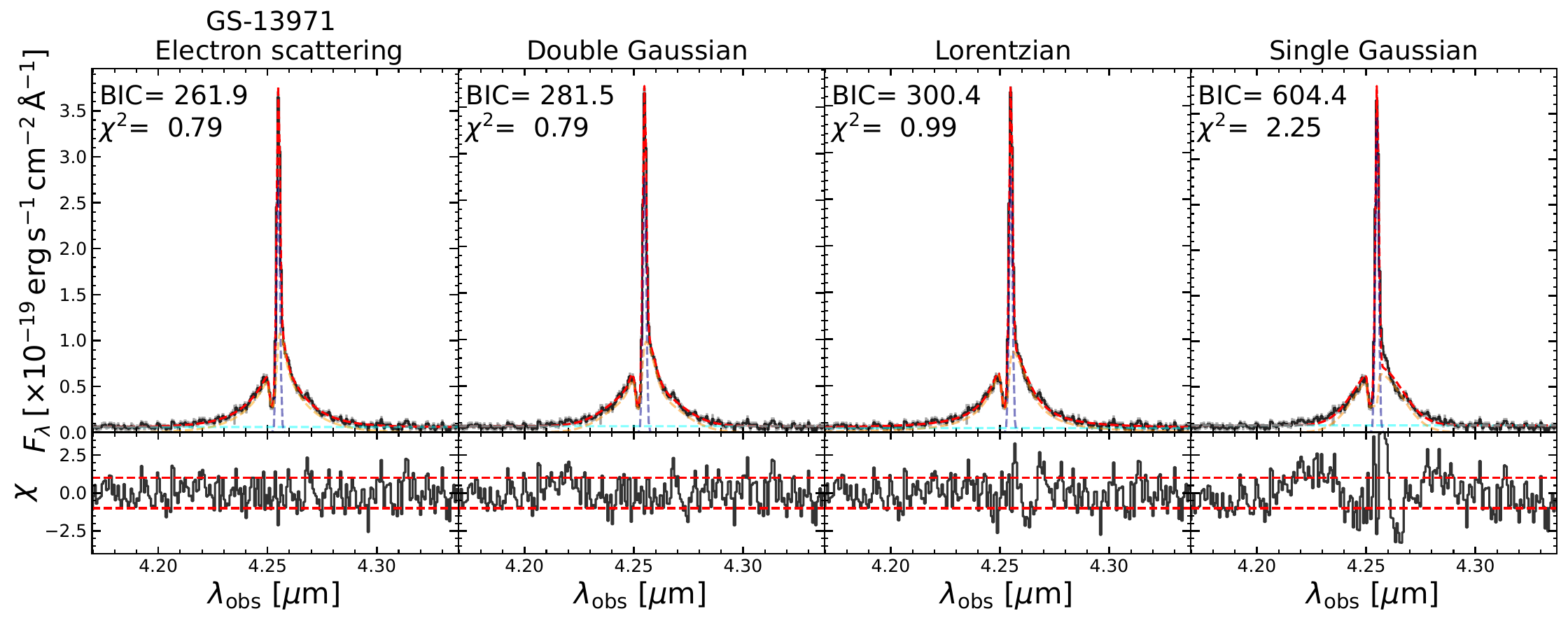}
    \includegraphics[width=0.42\paperwidth]{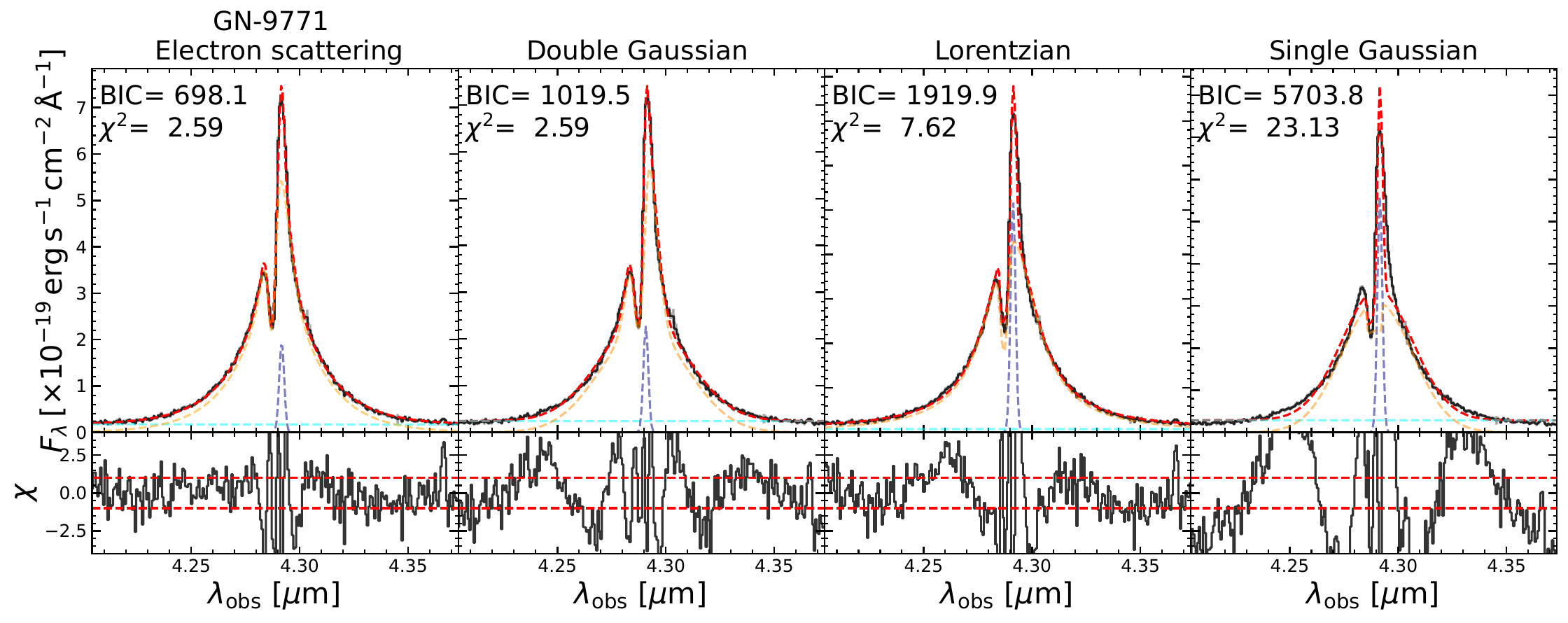}
    \includegraphics[width=0.42\paperwidth]{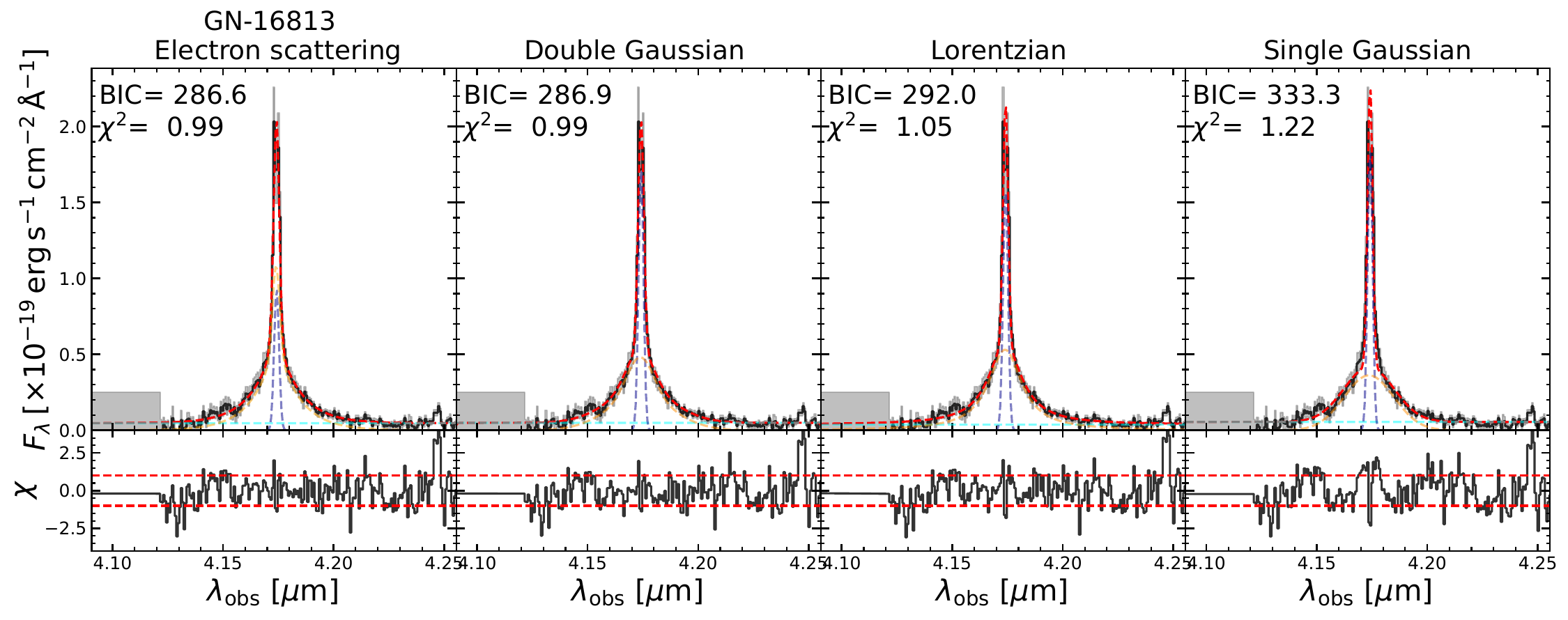}
    \includegraphics[width=0.42\paperwidth]{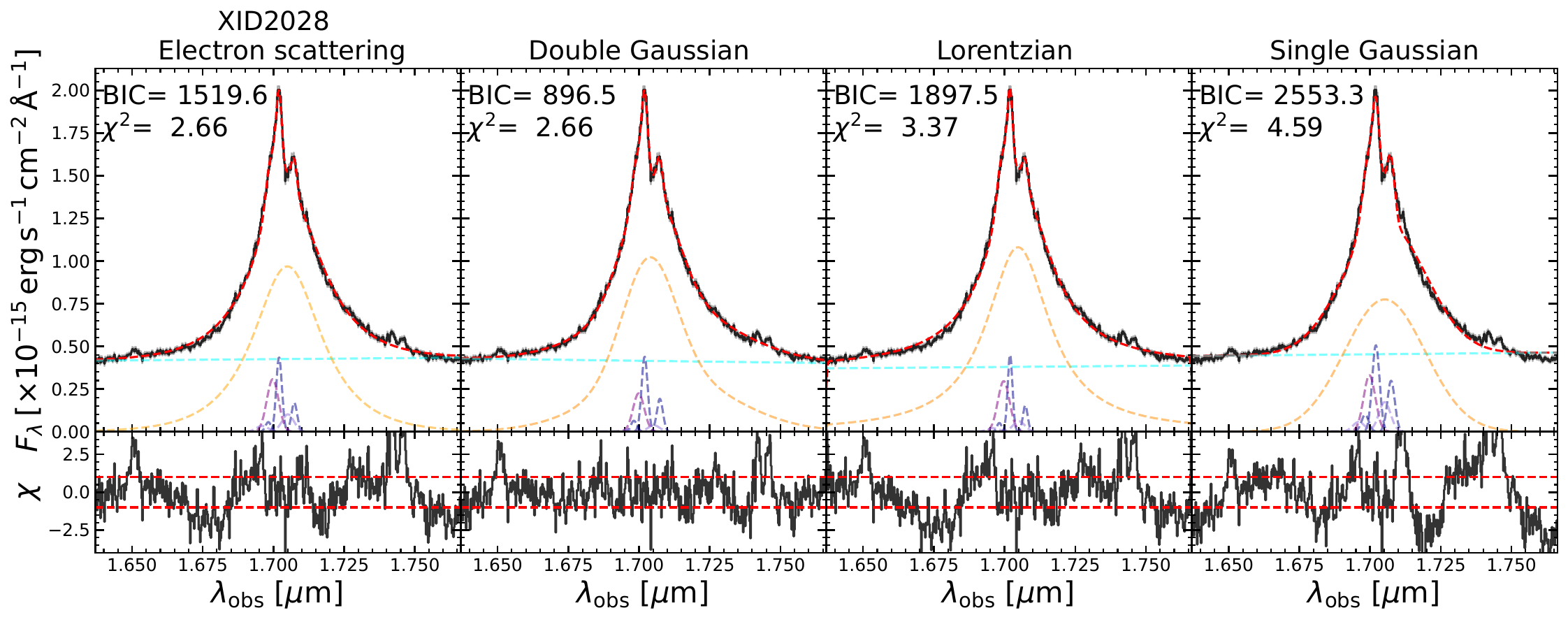}
    \caption{Continued from Fig.~\ref{fig.a1_obj}.}
    \label{fig.a2_obj}
\end{figure*}

\begin{figure*}
    \centering
    \includegraphics[width=0.42\paperwidth]{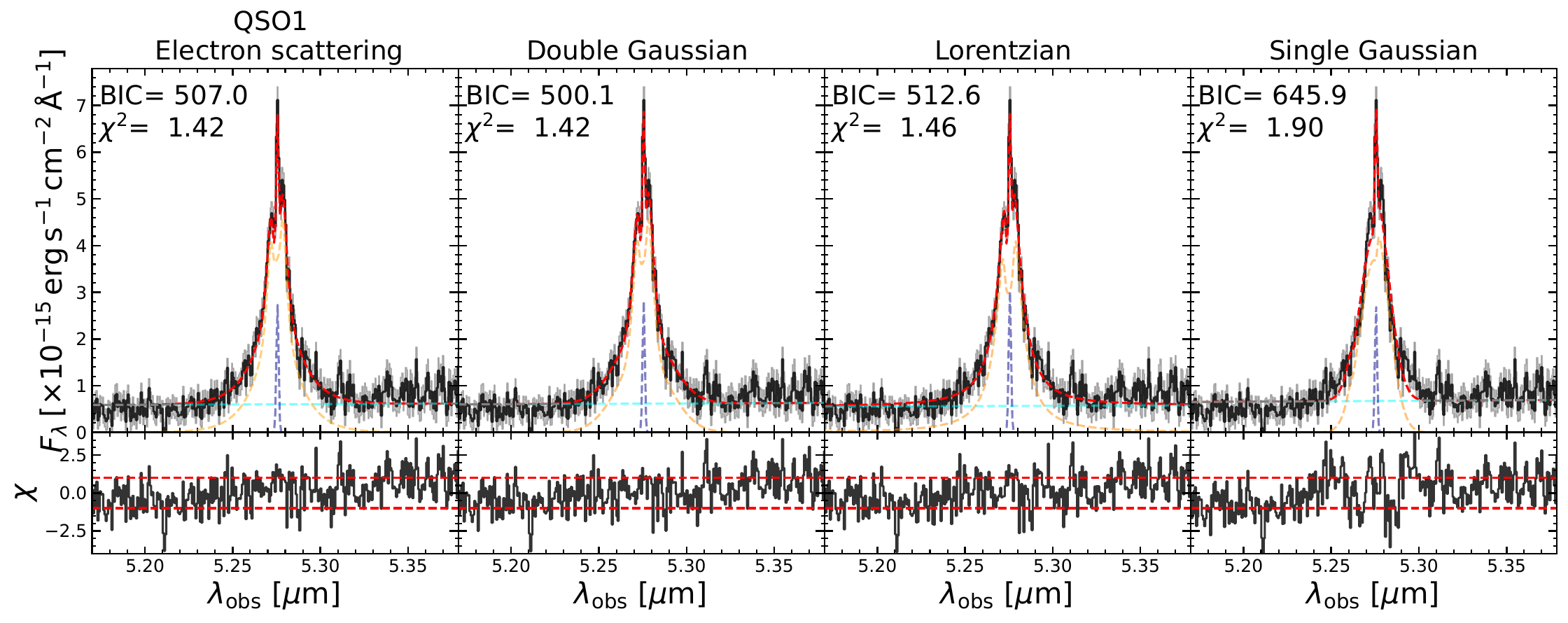}
    \includegraphics[width=0.42\paperwidth]{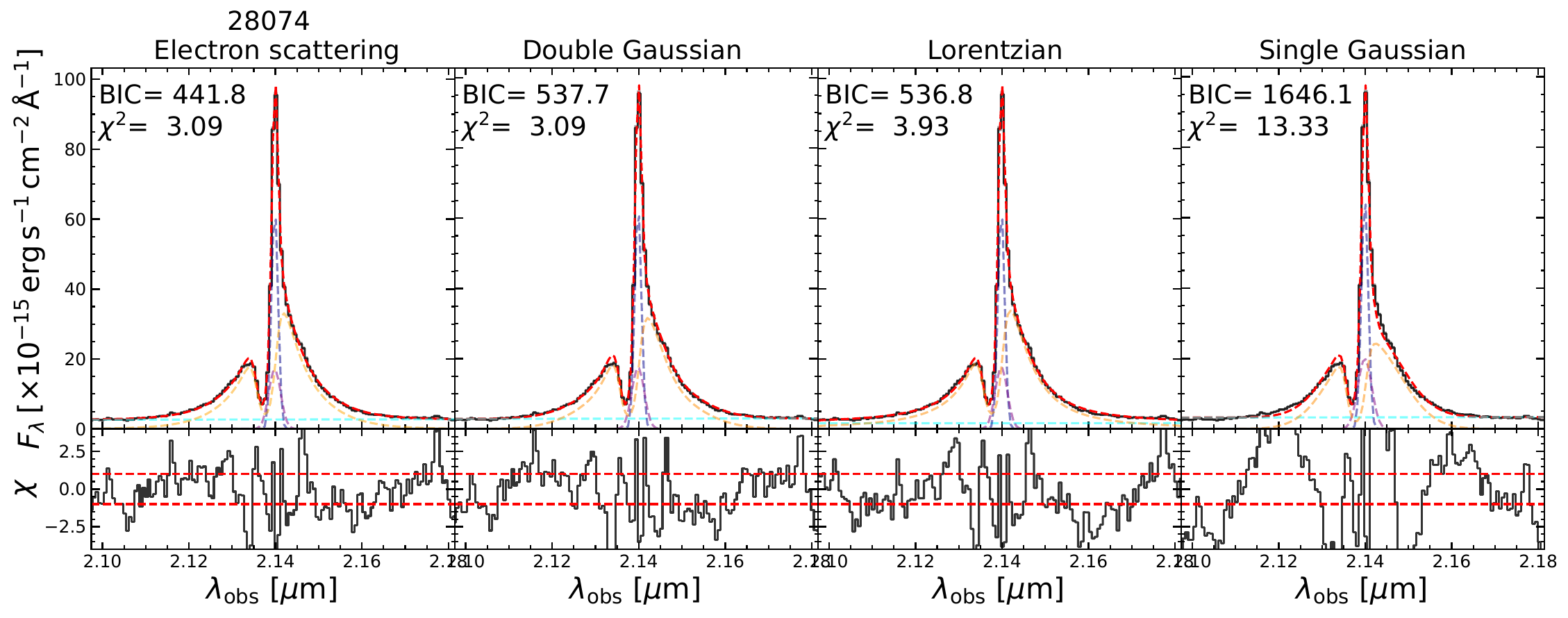}
    \includegraphics[width=0.42\paperwidth]{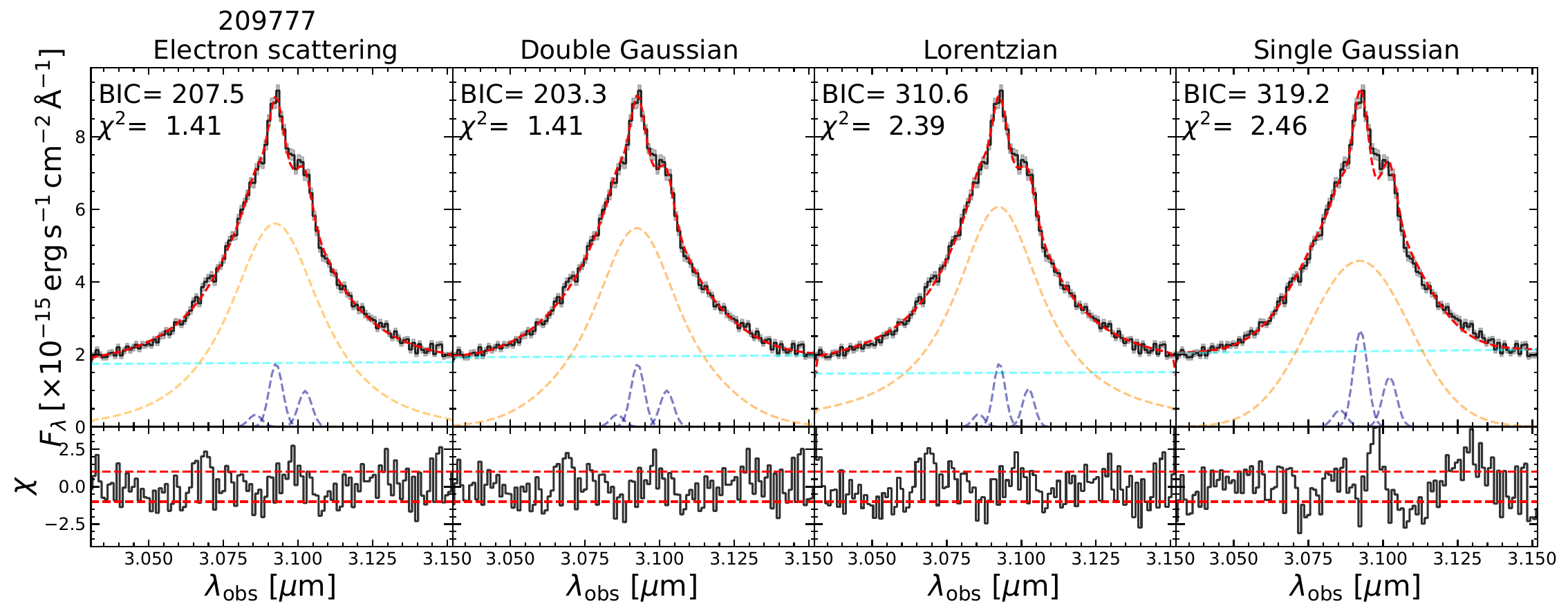}
    \includegraphics[width=0.42\paperwidth]{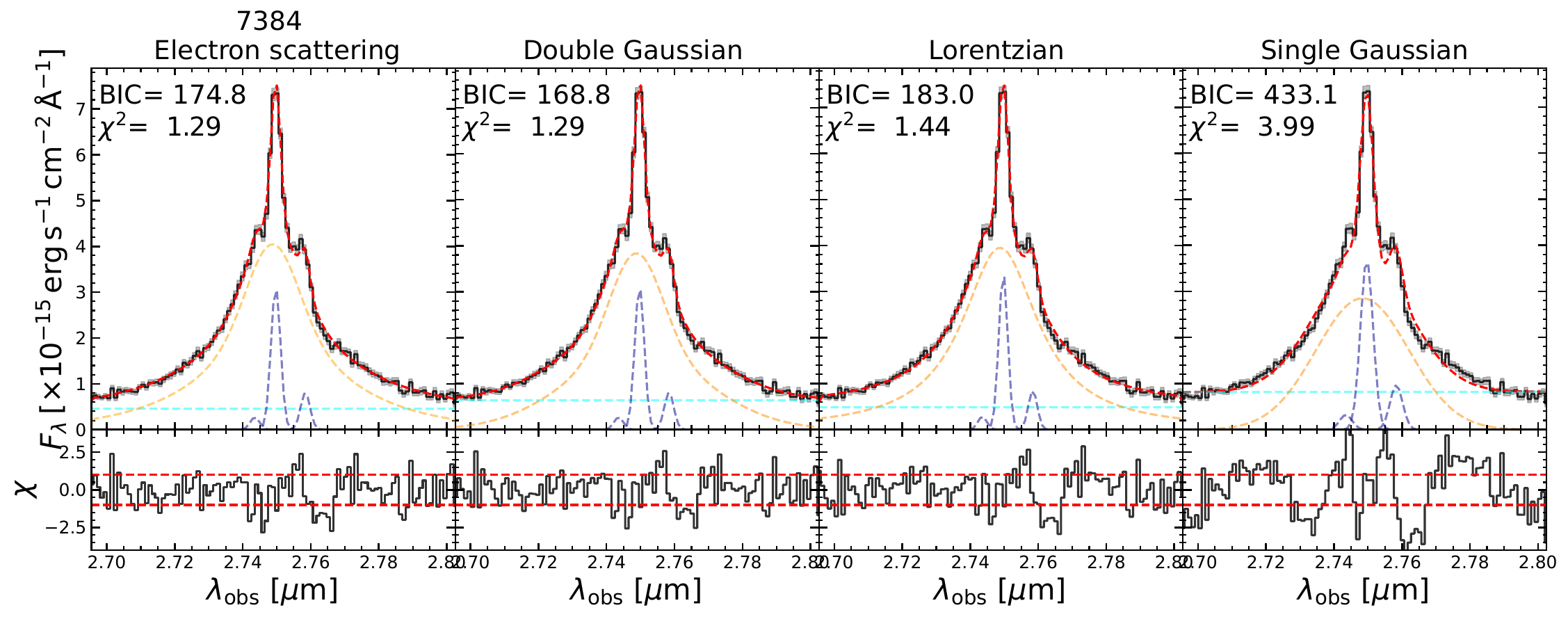}
    \includegraphics[width=0.42\paperwidth]{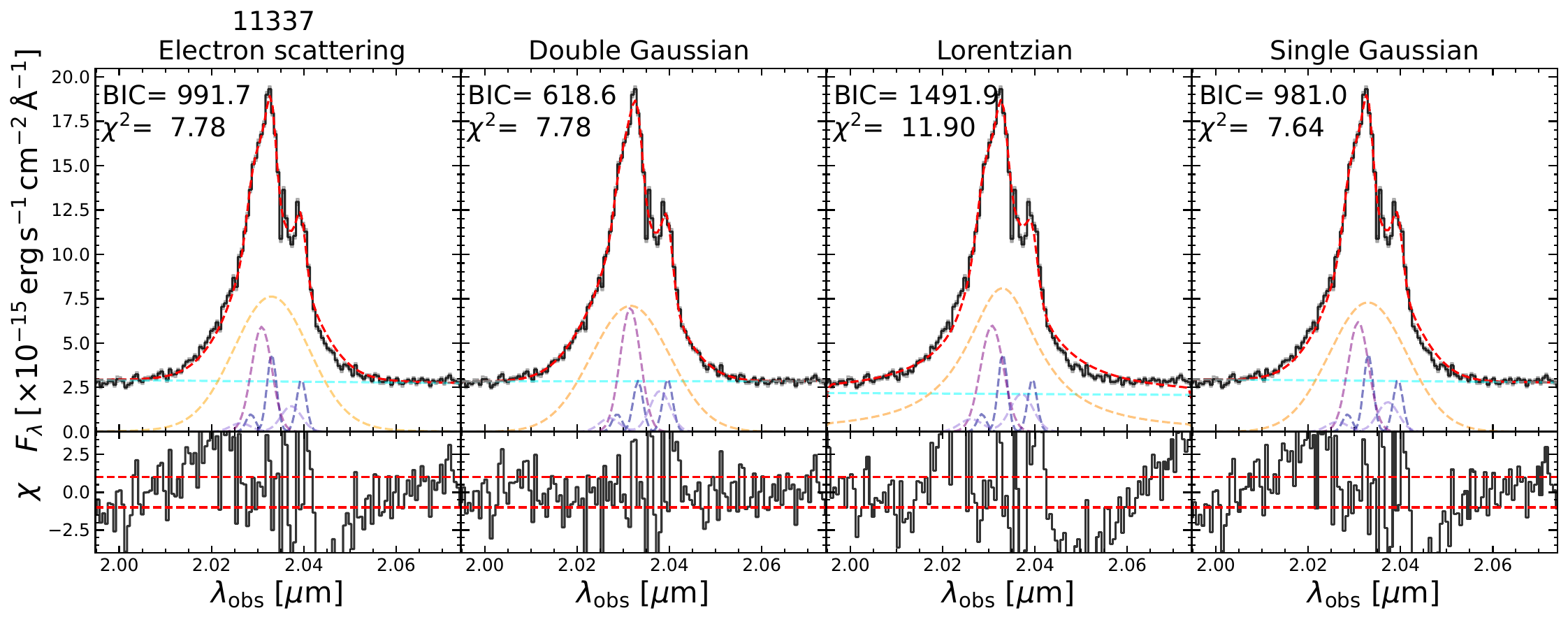}
    \includegraphics[width=0.42\paperwidth]{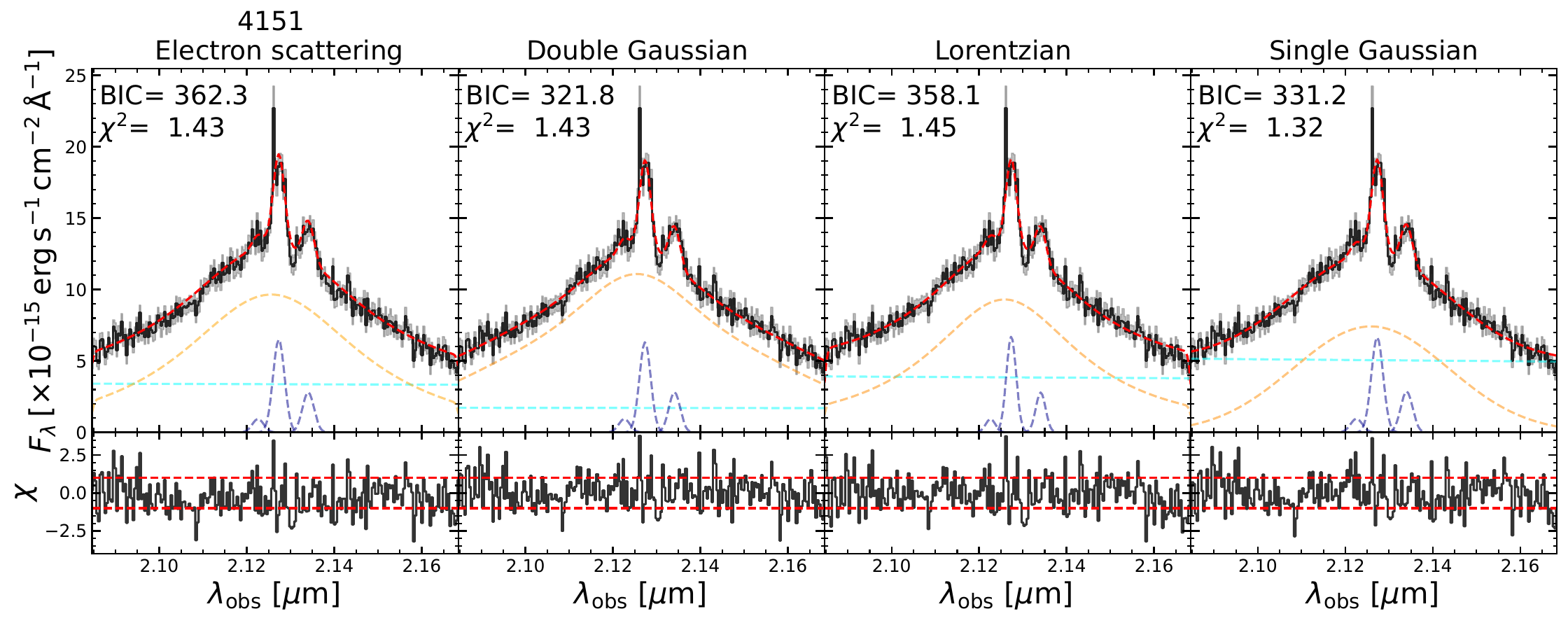} 
    \includegraphics[width=0.42\paperwidth]{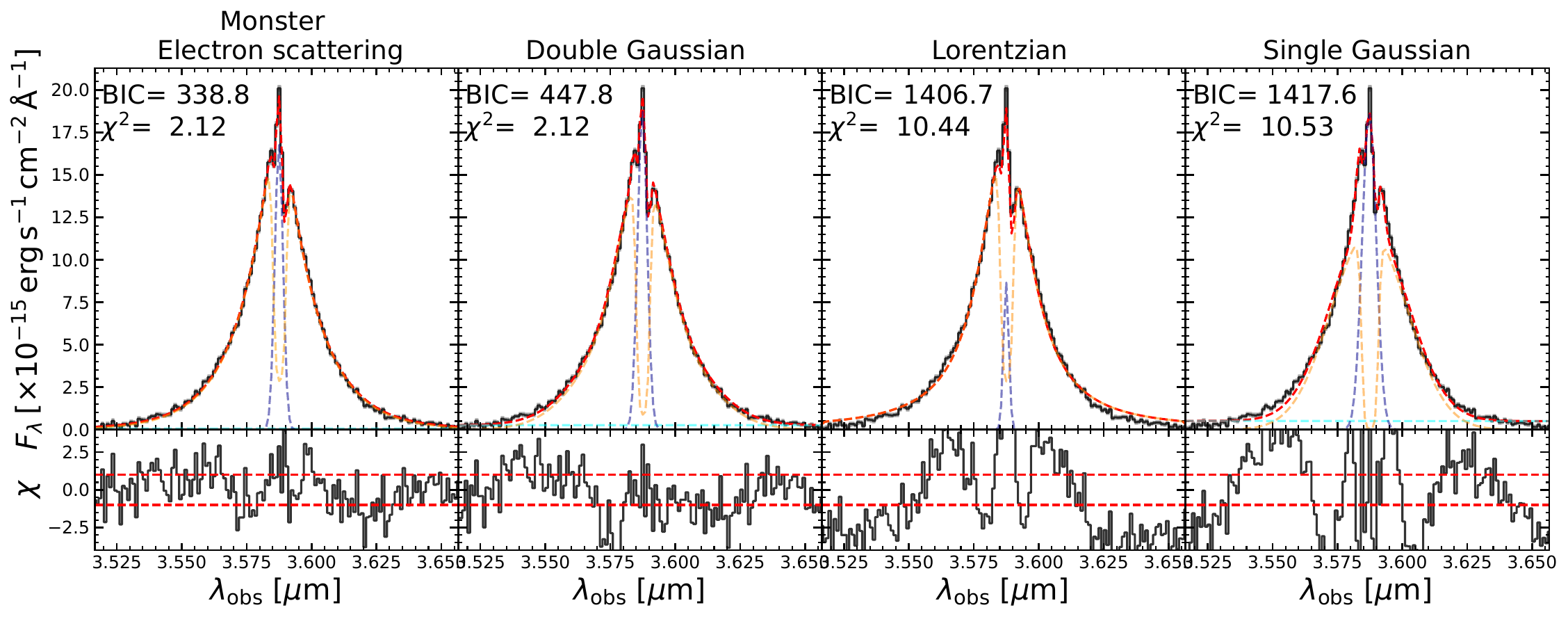}
    \caption{Continued from Fig.~\ref{fig.a2_obj}.}
    \label{fig.a3_obj}
\end{figure*}

\begin{figure*}
    \centering
    \includegraphics[width=0.7
    \paperwidth]{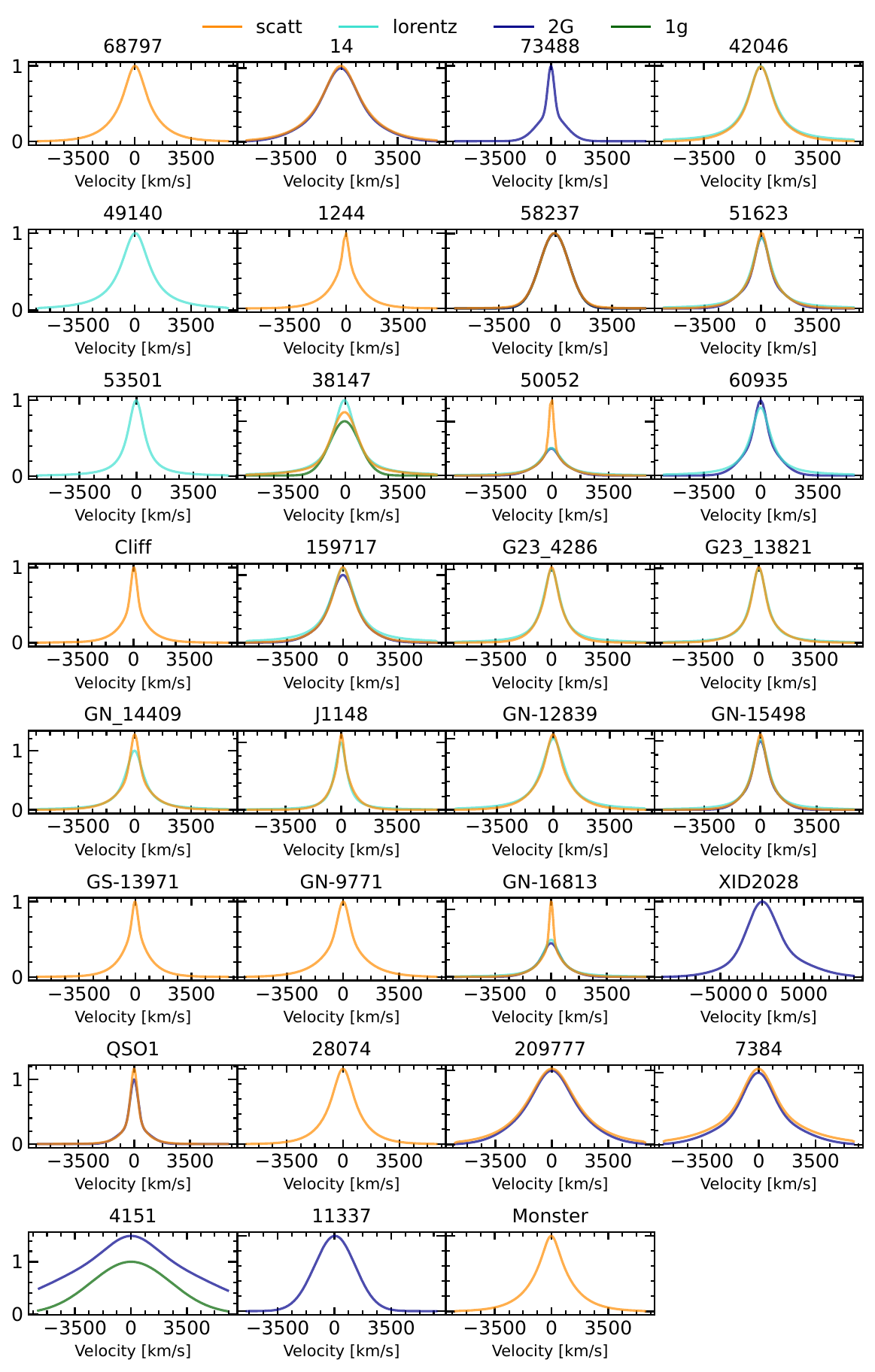}
    \caption{Showcase of all broad \Halpha profiles with $\delta$BIC$<$10 for each source in our sample.}
    \label{fig.BLR_best_comp}
\end{figure*}

\begin{figure*}
    \centering
    \includegraphics[width=0.7
    \paperwidth]{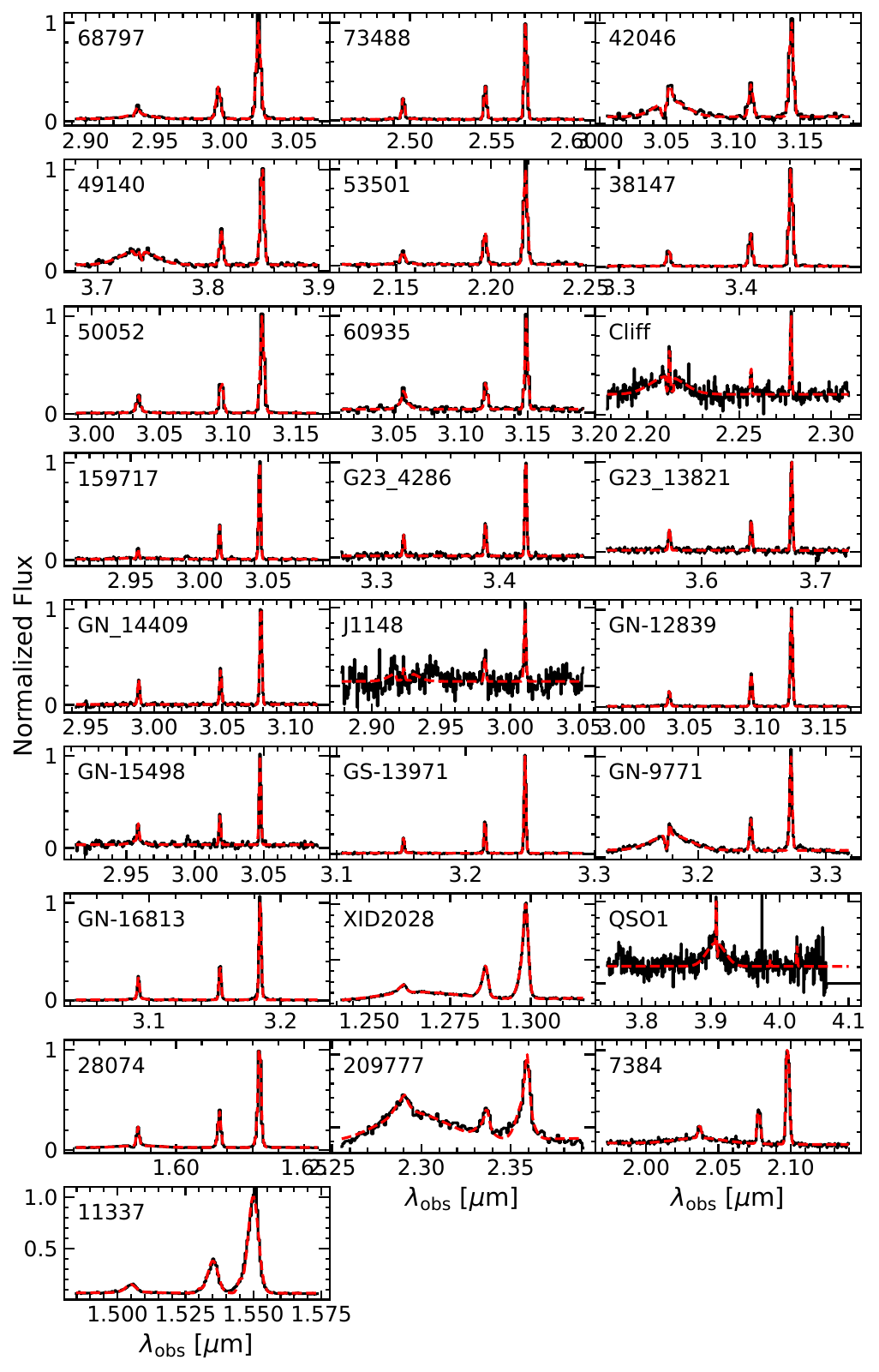}
    \caption{\OIIIall+\Hbeta spectra for sources in our sample with spectral coverage of the emission lines.}
    \label{fig.OIII_spec}
\end{figure*}

\section*{Appendix B: Results of the spectral fitting}

 In Table~\ref{tab.prop} we present the estimated bolometric luminosities and \MBH from \S~\ref{s.derived_property}.

\begin{table*}
    \caption{List of targets in our sample with derived physical properties: ID, best fit, \MBH and bolometric luminosities for each fitted model. }
    \centering
    \begin{tabular}{@{}lccccccccccc@{}} 
\hline
\hline
ID & Best fit & \MBH$_{\rm Lorentz}$ & L$_{\rm Bol, Lorentz}$ & \MBH$_{\rm 2G}$ &  L$_{\rm Bol, 2G}$ & \MBH$_{\rm exp, vir}$ & L$_{\rm Bol, exp, vir}$ & \MBH$_{\rm exp,scatt}$ & L$_{\rm Bol, exp, scatt}$ & \MBH$_{\rm 1G}$ & L$_{\rm Bol, 1G}$\\
   \\
\hline
68797 & \escats & $8.1^{+0.01}_{-0.01}$ & $45.9^{+0.00}_{-0.00}$ & $8.1^{+0.01}_{-0.02}$ & $45.8^{+0.00}_{-0.00}$ & $8.0^{+0.02}_{-0.02}$ & $45.8^{+0.00}_{-0.00}$ & $7.3^{+0.03}_{-0.02}$ & $45.8^{+0.00}_{-0.00}$ & $8.3^{+0.01}_{-0.01}$ & $45.7^{+0.00}_{-0.00}$\\ [6pt]
14 & 2G & $8.2^{+0.02}_{-0.02}$ & $45.5^{+0.01}_{-0.01}$ & $8.2^{+0.03}_{-0.03}$ & $45.5^{+0.01}_{-0.01}$ & $8.2^{+0.03}_{-0.03}$ & $45.5^{+0.01}_{-0.01}$ & $7.5^{+0.14}_{-0.30}$ & $45.5^{+0.01}_{-0.01}$ & $8.3^{+0.01}_{-0.01}$ & $45.4^{+0.01}_{-0.01}$\\ [6pt]
73488 & 2G & $6.9^{+0.03}_{-0.80}$ & $44.7^{+0.01}_{-0.05}$ & $6.5^{+0.04}_{-0.05}$ & $44.7^{+0.01}_{-0.01}$ & $6.5^{+0.06}_{-0.06}$ & $44.7^{+0.01}_{-0.01}$ & $6.0^{+0.06}_{-0.06}$ & $44.7^{+0.01}_{-0.01}$ & $7.3^{+0.02}_{-0.02}$ & $44.6^{+0.01}_{-0.01}$\\ [6pt]
42046 & Lorentzian & $7.9^{+0.03}_{-0.03}$ & $45.8^{+0.01}_{-0.01}$ & $7.9^{+0.05}_{-0.03}$ & $45.7^{+0.01}_{-0.01}$ & $7.9^{+0.07}_{-0.09}$ & $45.8^{+0.01}_{-0.01}$ & $7.5^{+0.15}_{-0.24}$ & $45.8^{+0.01}_{-0.01}$ & $8.2^{+0.02}_{-0.01}$ & $45.6^{+0.01}_{-0.01}$\\ [6pt]
49140 & Lorentzian & $8.2^{+0.03}_{-0.03}$ & $46.0^{+0.01}_{-0.01}$ & $8.2^{+0.04}_{-0.04}$ & $46.0^{+0.02}_{-0.01}$ & $8.2^{+0.04}_{-0.05}$ & $46.0^{+0.01}_{-0.01}$ & $7.9^{+0.12}_{-0.33}$ & $46.0^{+0.01}_{-0.01}$ & $8.3^{+0.02}_{-0.01}$ & $45.9^{+0.01}_{-0.01}$\\ [6pt]
1244 & \escats & $7.6^{+0.03}_{-0.03}$ & $45.1^{+0.01}_{-0.01}$ & $7.6^{+0.04}_{-0.04}$ & $45.1^{+0.01}_{-0.01}$ & $7.0^{+0.30}_{-0.19}$ & $45.1^{+0.01}_{-0.02}$ & $6.0^{+0.21}_{-0.13}$ & $45.1^{+0.01}_{-0.02}$ & $7.8^{+0.02}_{-0.02}$ & $45.0^{+0.01}_{-0.01}$\\ [6pt]
58237 & 1G & $7.3^{+0.13}_{-0.12}$ & $44.8^{+0.03}_{-0.03}$ & $7.6^{+0.04}_{-0.04}$ & $44.8^{+0.04}_{-0.04}$ & $7.6^{+0.03}_{-0.04}$ & $44.8^{+0.03}_{-0.03}$ & $7.5^{+0.04}_{-0.05}$ & $44.8^{+0.03}_{-0.03}$ & $7.6^{+0.04}_{-0.03}$ & $44.8^{+0.03}_{-0.03}$\\ [6pt]
51623 & Lorentzian & $7.3^{+0.05}_{-0.03}$ & $45.0^{+0.01}_{-0.01}$ & $7.2^{+0.06}_{-0.06}$ & $44.9^{+0.01}_{-0.01}$ & $7.2^{+0.07}_{-0.07}$ & $45.0^{+0.01}_{-0.01}$ & $6.6^{+0.16}_{-0.24}$ & $45.0^{+0.01}_{-0.01}$ & $7.7^{+0.03}_{-0.03}$ & $44.8^{+0.01}_{-0.01}$\\ [6pt]
53501 & Lorentzian & $7.1^{+0.06}_{-0.06}$ & $44.8^{+0.01}_{-0.01}$ & $7.3^{+0.05}_{-0.05}$ & $44.7^{+0.04}_{-0.03}$ & $7.3^{+0.06}_{-0.10}$ & $44.8^{+0.04}_{-0.03}$ & $7.2^{+0.07}_{-0.19}$ & $44.8^{+0.04}_{-0.03}$ & $7.3^{+0.04}_{-0.04}$ & $44.7^{+0.01}_{-0.01}$\\ [6pt]
38147 & \escats & $6.9^{+0.12}_{-0.09}$ & $45.4^{+0.05}_{-0.05}$ & $7.5^{+0.16}_{-1.16}$ & $45.2^{+0.22}_{-0.07}$ & $7.7^{+0.05}_{-0.08}$ & $45.1^{+0.03}_{-0.03}$ & $7.5^{+0.06}_{-0.13}$ & $45.1^{+0.03}_{-0.03}$ & $7.7^{+0.03}_{-0.05}$ & $45.0^{+0.01}_{-0.01}$\\ [6pt]
50052 & Lorentzian & $7.4^{+0.06}_{-0.05}$ & $44.9^{+0.01}_{-0.01}$ & $7.3^{+0.11}_{-0.13}$ & $44.8^{+0.02}_{-0.02}$ & $6.3^{+0.63}_{-0.21}$ & $44.9^{+0.05}_{-0.10}$ & $5.8^{+0.32}_{-0.18}$ & $44.9^{+0.05}_{-0.10}$ & $7.5^{+0.04}_{-0.04}$ & $44.7^{+0.01}_{-0.01}$\\ [6pt]
60935 & Lorentzian & $7.4^{+0.06}_{-0.05}$ & $45.0^{+0.01}_{-0.01}$ & $7.1^{+0.17}_{-0.13}$ & $44.9^{+0.02}_{-0.02}$ & $7.0^{+0.22}_{-0.28}$ & $45.0^{+0.02}_{-0.02}$ & $6.2^{+0.47}_{-0.50}$ & $45.0^{+0.02}_{-0.02}$ & $7.6^{+0.04}_{-0.04}$ & $44.8^{+0.01}_{-0.02}$\\ [6pt]
Cliff & \escats & $7.2^{+0.02}_{-0.05}$ & $44.9^{+0.01}_{-0.01}$ & $7.0^{+0.05}_{-0.05}$ & $44.9^{+0.01}_{-0.01}$ & $6.6^{+0.15}_{-0.11}$ & $44.9^{+0.02}_{-0.01}$ & $6.1^{+0.11}_{-0.08}$ & $44.9^{+0.02}_{-0.01}$ & $7.4^{+0.02}_{-0.02}$ & $44.8^{+0.00}_{-0.00}$\\ [6pt]
159717 & Lorentzian & $7.5^{+0.02}_{-0.02}$ & $45.0^{+0.01}_{-0.01}$ & $7.5^{+0.03}_{-0.03}$ & $44.9^{+0.01}_{-0.01}$ & $7.5^{+0.07}_{-0.15}$ & $44.9^{+0.01}_{-0.01}$ & $6.6^{+0.44}_{-0.46}$ & $44.9^{+0.01}_{-0.01}$ & $7.7^{+0.01}_{-0.01}$ & $44.8^{+0.00}_{-0.01}$\\ [6pt]
G23\_4286 & Lorentzian & $7.2^{+0.05}_{-0.04}$ & $45.0^{+0.01}_{-0.01}$ & $7.3^{+0.07}_{-0.08}$ & $44.9^{+0.02}_{-0.02}$ & $7.1^{+0.09}_{-0.13}$ & $44.9^{+0.02}_{-0.02}$ & $6.4^{+0.36}_{-0.57}$ & $44.9^{+0.02}_{-0.02}$ & $7.4^{+0.03}_{-0.03}$ & $44.9^{+0.01}_{-0.01}$\\ [6pt]
G23\_13821 & Lorentzian & $7.2^{+0.06}_{-0.05}$ & $45.1^{+0.01}_{-0.01}$ & $7.2^{+0.07}_{-0.08}$ & $45.0^{+0.02}_{-0.02}$ & $7.1^{+0.08}_{-0.07}$ & $45.0^{+0.03}_{-0.02}$ & $6.7^{+0.14}_{-0.20}$ & $45.0^{+0.03}_{-0.02}$ & $7.4^{+0.04}_{-0.04}$ & $44.9^{+0.01}_{-0.01}$\\ [6pt]
GN\_14409 & \escats & $7.2^{+0.04}_{-0.04}$ & $44.8^{+0.01}_{-0.01}$ & $7.2^{+0.12}_{-0.15}$ & $44.8^{+0.02}_{-0.02}$ & $6.8^{+0.18}_{-0.22}$ & $44.8^{+0.02}_{-0.02}$ & $6.4^{+0.23}_{-0.40}$ & $44.8^{+0.02}_{-0.02}$ & $7.6^{+0.04}_{-0.03}$ & $44.7^{+0.01}_{-0.02}$\\ [6pt]
J1148 & Lorentzian & $7.0^{+0.06}_{-0.06}$ & $45.2^{+0.02}_{-0.02}$ & $7.1^{+0.05}_{-0.06}$ & $45.1^{+0.02}_{-0.03}$ & $6.9^{+0.18}_{-0.08}$ & $45.2^{+0.02}_{-0.04}$ & $6.0^{+0.88}_{-0.19}$ & $45.2^{+0.02}_{-0.04}$ & $7.3^{+0.04}_{-0.03}$ & $45.0^{+0.02}_{-0.02}$\\ [6pt]
GN-12839 & Lorentzian & $7.7^{+0.05}_{-0.05}$ & $45.5^{+0.03}_{-0.03}$ & $7.7^{+0.11}_{-0.13}$ & $45.4^{+0.02}_{-0.03}$ & $7.6^{+0.13}_{-0.20}$ & $45.4^{+0.03}_{-0.03}$ & $6.8^{+0.39}_{-0.40}$ & $45.4^{+0.03}_{-0.03}$ & $8.1^{+0.04}_{-0.04}$ & $45.3^{+0.03}_{-0.03}$\\ [6pt]
GN-15498 & \escats & $7.1^{+0.02}_{-0.02}$ & $44.8^{+0.01}_{-0.01}$ & $7.0^{+0.02}_{-0.02}$ & $44.7^{+0.01}_{-0.01}$ & $7.0^{+0.05}_{-0.06}$ & $44.8^{+0.01}_{-0.01}$ & $6.4^{+0.13}_{-0.16}$ & $44.8^{+0.01}_{-0.01}$ & $7.3^{+0.02}_{-0.02}$ & $44.7^{+0.01}_{-0.01}$\\ [6pt]
GS-13971 & \escats & $7.3^{+0.03}_{-0.03}$ & $45.0^{+0.01}_{-0.01}$ & $7.0^{+0.12}_{-0.06}$ & $45.0^{+0.01}_{-0.01}$ & $6.9^{+0.08}_{-0.10}$ & $45.0^{+0.01}_{-0.01}$ & $6.0^{+0.19}_{-0.25}$ & $45.0^{+0.01}_{-0.01}$ & $7.5^{+0.02}_{-0.02}$ & $44.9^{+0.01}_{-0.01}$\\ [6pt]
GN-9771 & \escats & $8.0^{+0.01}_{-0.01}$ & $45.9^{+0.00}_{-0.00}$ & $7.5^{+0.02}_{-0.02}$ & $45.9^{+0.00}_{-0.00}$ & $7.7^{+0.01}_{-0.01}$ & $45.9^{+0.00}_{-0.00}$ & $7.2^{+0.01}_{-0.00}$ & $45.9^{+0.00}_{-0.00}$ & $8.3^{+0.00}_{-0.00}$ & $45.8^{+0.00}_{-0.00}$\\ [6pt]
GN-16813 & \escats & $7.1^{+0.05}_{-0.05}$ & $44.8^{+0.01}_{-0.01}$ & $7.1^{+0.08}_{-0.10}$ & $44.7^{+0.02}_{-0.01}$ & $6.2^{+0.52}_{-0.27}$ & $44.8^{+0.05}_{-0.04}$ & $5.7^{+0.30}_{-0.14}$ & $44.8^{+0.05}_{-0.04}$ & $7.4^{+0.04}_{-0.04}$ & $44.6^{+0.01}_{-0.01}$\\ [6pt]
XID2028 & 2G & $7.9^{+0.01}_{-0.01}$ & $44.0^{+0.00}_{-0.00}$ & $7.9^{+0.01}_{-0.01}$ & $43.9^{+0.00}_{-0.00}$ & $7.9^{+0.02}_{-0.01}$ & $43.9^{+0.00}_{-0.00}$ & $7.4^{+0.12}_{-0.11}$ & $43.9^{+0.00}_{-0.00}$ & $8.0^{+0.00}_{-0.00}$ & $43.8^{+0.00}_{-0.00}$\\ [6pt]
QSO1 & 2G & $7.1^{+0.05}_{-0.05}$ & $45.8^{+0.01}_{-0.01}$ & $7.0^{+0.07}_{-0.07}$ & $45.7^{+0.02}_{-0.02}$ & $7.0^{+0.18}_{-0.15}$ & $45.8^{+0.03}_{-0.02}$ & $6.6^{+0.26}_{-0.14}$ & $45.8^{+0.03}_{-0.02}$ & $7.4^{+0.15}_{-0.04}$ & $45.7^{+0.03}_{-0.05}$\\ [6pt]
GS-3073 & \escats & $7.7^{+0.20}_{-0.20}$ & $45.4^{+0.08}_{-0.10}$ & $8.0^{+0.20}_{-0.20}$ & $45.3^{+0.08}_{-0.10}$ & $7.4^{+0.20}_{-0.20}$ & $45.4^{+0.08}_{-0.10}$ & $6.8^{+0.20}_{-0.20}$ & $45.4^{+0.08}_{-0.10}$ & $8.2^{+0.20}_{-0.20}$ & $45.2^{+0.08}_{-0.10}$\\ [6pt]
28074 & \escats & $7.8^{+0.33}_{-0.02}$ & $45.6^{+0.00}_{-0.09}$ & $7.8^{+0.34}_{-0.02}$ & $45.5^{+0.14}_{-0.00}$ & $7.8^{+0.57}_{-0.02}$ & $45.5^{+0.01}_{-0.09}$ & $7.1^{+1.02}_{-0.05}$ & $45.5^{+0.01}_{-0.09}$ & $8.1^{+0.03}_{-0.08}$ & $45.4^{+0.02}_{-0.04}$\\ [6pt]
209777 & 2G & $8.5^{+0.02}_{-0.02}$ & $45.7^{+0.00}_{-0.00}$ & $8.4^{+0.02}_{-0.02}$ & $45.6^{+0.01}_{-0.01}$ & $8.4^{+0.02}_{-0.02}$ & $45.6^{+0.01}_{-0.01}$ & $7.8^{+0.10}_{-0.10}$ & $45.6^{+0.01}_{-0.01}$ & $8.5^{+0.01}_{-0.01}$ & $45.5^{+0.00}_{-0.01}$\\ [6pt]
7384 & 2G & $8.1^{+0.02}_{-0.02}$ & $45.2^{+0.01}_{-0.01}$ & $8.0^{+0.03}_{-0.03}$ & $45.2^{+0.01}_{-0.01}$ & $8.1^{+0.04}_{-0.03}$ & $45.2^{+0.03}_{-0.02}$ & $7.7^{+0.05}_{-0.06}$ & $45.2^{+0.03}_{-0.02}$ & $8.2^{+0.02}_{-0.02}$ & $45.1^{+0.01}_{-0.01}$\\ [6pt]
4151 & 2G & $8.8^{+0.01}_{-0.02}$ & $45.3^{+0.01}_{-0.01}$ & $9.1^{+0.11}_{-0.16}$ & $45.5^{+0.05}_{-0.08}$ & $8.9^{+0.06}_{-0.05}$ & $45.4^{+0.03}_{-0.03}$ & $8.4^{+0.14}_{-0.12}$ & $45.4^{+0.03}_{-0.03}$ & $8.7^{+0.03}_{-0.03}$ & $45.2^{+0.02}_{-0.02}$\\ [6pt]
11337 & 2G & $7.9^{+0.02}_{-0.02}$ & $45.0^{+0.01}_{-0.01}$ & $7.9^{+0.01}_{-0.01}$ & $44.8^{+0.01}_{-0.01}$ & $7.9^{+0.02}_{-0.01}$ & $44.8^{+0.02}_{-0.02}$ & $7.9^{+0.04}_{-0.03}$ & $44.8^{+0.02}_{-0.02}$ & $7.9^{+0.01}_{-0.01}$ & $44.8^{+0.01}_{-0.01}$\\ [6pt]
Monster & \escats & $8.1^{+0.01}_{-0.01}$ & $45.9^{+0.00}_{-0.00}$ & $8.1^{+0.01}_{-0.01}$ & $45.9^{+0.00}_{-0.00}$ & $8.0^{+0.06}_{-0.05}$ & $45.9^{+0.00}_{-0.01}$ & $6.9^{+0.23}_{-0.13}$ & $45.9^{+0.00}_{-0.01}$ & $8.3^{+0.01}_{-0.01}$ & $45.8^{+0.01}_{-0.00}$\\ [6pt]
\hline
\end{tabular}  
    \label{tab.prop}
\end{table*}

\bsp	
\label{lastpage}
\end{document}